\pdfoutput=1
\documentclass{aa}
\usepackage{graphicx}
\usepackage{txfonts}
\usepackage{xcolor}

\usepackage{adjustbox}

\usepackage{microtype}
\usepackage{cancel}
\newlength\figwidth
\newlength\figstampcolsep
\newcommand\BowshockFig[1]{
  \includegraphics[width=\figwidth, clip, trim=10 10 10 10]
  {#1}
}
\newcommand\raiselabel[1]{\raisebox{0.9\figwidth}[-0.5\figwidth]{#1}}

\newcommand\BowshockFigg[1]{
  \includegraphics[width=0.48\figwidth, clip, trim=10 13 10 10]
  {#1}
}
\newcommand\raiselabell[1]{\raisebox{0.45\figwidth}[-0.5\figwidth]{#1}}

\newcommand\oii{[\ion{O}{ii}]}
\newcommand\oiii{[\ion{O}{iii}]}
\newcommand\nii{[\ion{N}{ii}]}

\newcommand\sii{[\ion{S}{ii}]}
\newcommand\hei{[\ion{He}{i}]}
\newcommand\heii{[\ion{He}{ii}]}
\newcommand\ha{\ensuremath{\mathrm{H}\alpha}}

\newcommand\hii{\ion{H}{ii}}
\usepackage{ulem,xpatch}

\xpatchcmd{\sout}
  {\bgroup}
  {\bgroup}
  {}{}

\begin{document}

   \title{J-PLUS: Tools to identify compact planetary nebulae in the Javalambre and southern photometric local universe surveys
    }

    \author{L. A. Gutiérrez-Soto
          \inst{1},
          D. R. Gon\c calves \inst{1}, 
                  S. Akras\inst{1},
                  A. Cortesi\inst{2},
                  C. López-Sanjuan\inst{3},
          M. A. Guerrero\inst{4},
                S. Daflon\inst{5},
                M. Borges Fernandes\inst{5},
                C. Mendes de Oliveira\inst{2},
                A. Ederoclite\inst{2},
                L. Sodré Jr\inst{2},
                C. B. Pereira\inst{5},
                A. Kanaan\inst{6},
                A. Werle\inst{2,6},
                H. Vázquez Ramió\inst{7}, 
                     J. S. Alcaniz\inst{5},
                     R. E. Angulo\inst{7},
                     A. J. Cenarro\inst{3},
                     D. Cristóbal-Hornillos\inst{7},
                     R. A. Dupke\inst{5,8,9},
                     C. Hernández-Monteagudo\inst{3},
                     A. Marín-Franch\inst{7},
                     M. Moles\inst{7},
                     J. Varela\inst{3},
                     T. Ribeiro\inst{10,11},
                     W. Schoenell\inst{12},                   
                        A. Alvarez-Candal\inst{5},
                        L. Galbany\inst{13,14},
                        F. M. Jim\'enez-Esteban\inst{15,16},
                        R. Logro\~no-Garc\'ia\inst{7}
                         and 
                         D. Sobral\inst{17}
}  
   \institute{Observatório do Valongo, Universidade Federal do Rio de Janeiro,
              Ladeira Pedro Antonio 43, 20080-090 Rio de Janeiro, Brazil\\
              \email{lgutierrez@astro.ufrj.br} \and
              Instituto de Astronomia, Geofísica e Ci\^encias Atmosféricas, Universidade de S\~ao Paulo, 05508-090 São Paulo, Brazil \and
              Centro de Estudios de Física del Cosmos de Aragón (CEFCA), Unidad Asociada al CSIC, Plaza San Juan 1, E-44001 Teruel, Spain \and
              Instituto de Astrof\'isica de Andaluc\'ia, IAA-CSIC, Granada, Spain \and
              Observat\'orio Nacional, Rua Gal. Jos\'e Cristino 77, 20921-400, Rio de Janeiro, RJ, Brazil \and
              Departamento de F\'isica, Universidade Federal de Santa Catarina, Florian\'opolis, SC, 88040-900, Brazil \and
              Centro de Estudios de F\'isica del Cosmos de Arag\'on, Plaza San Juan 1, 44001 Teruel, Spain \and
              University of Michigan, Dept. Astronomy,1085 S. University Ann Arbor, MI 48109, USA. \and
              University of Alabama, Dept. of Phys. \& Astronomy, Gallalee Hall, Tuscaloosa, AL 35401, USA \and
              NOAO, P.O. Box 26732, Tucson, AZ 85726 \and
              Departamento de F\'isica, Universidade Federal de Sergipe, Av. Marechal Rondon, S/N, 49000-000 S\~ao Crist\'ov\~ao, SE, Brazil \and
Departamento de Astronomia, Instituto de F\'isica, Universidade Federal do Rio Grande do Sul (UFRGS), Av. Bento Gon\c{c}alves 9500, Porto Alegre, RS, Brazil \and 
 PITT PACC, Department of Physics and Astronomy, University of Pittsburgh, Pittsburgh, PA 15260, USA. \and
              Departamento de F\'isica Te\'orica y del Cosmos, Universidad de Granada, E-18071 Granada, Spain \and
              Departmento de Astrof\'{\i}sica, Centro de Astrobiolog\'{\i}a (CSIC-INTA), ESAC Campus, Camino Bajo del Castillo s/n, E-28692 Villanueva de la Ca\~nada, Madrid, Spain \and
              Spanish Virtual Observatory, Spain \and             
              Department of Physics, Lancaster University, Lancaster, LA1 4YB, UK 
              }
             
   \date{Received ?, 2019; accepted ?, 2019}

 \abstract
  {From the approximately $\sim$3,500 planetary nebulae (PNe) discovered in our Galaxy, only 14  are known to be members of the Galactic halo. Nevertheless, a systematic search for halo PNe has never been performed.} 
{In this study, we present new photometric diagnostic tools to identify compact PNe in the Galactic halo by making use of the novel 12-filter system projects, J-PLUS (Javalambre Photometric Local Universe Survey) and S-PLUS (Southern-Photometric Local Universe Survey).}
{We reconstructed the IPHAS (Isaac Newton Telescope (INT) Photometric \ha{} Survey of the Northern Galactic Plane) diagnostic diagram and propose four new ones using i) the J-PLUS and S-PLUS synthetic photometry for a grid of photo-ionisation models of halo PNe, ii) several observed halo PNe, as well as iii) a number of other emission-line objects that resemble PNe. All colour-colour diagnostic diagrams are validated using two known halo PNe observed by J-PLUS during the scientific verification phase and the first data release (DR1) of S-PLUS and the DR1 of J-PLUS.}
{By applying our criteria to the DR1s ($\sim$1,190~deg$^2$), we identified one PN candidate. However, optical follow-up spectroscopy proved it to be a  H~{\sc ii} region belonging to the UGC 5272 galaxy. Here, we also discuss the PN and two H~{\sc ii} galaxies recovered by these selection criteria. Finally, the cross-matching with the most updated PNe catalogue (HASH) helped us to highlight the potential of these surveys, since we recover all the known PNe in the observed area.}
{The tools here proposed to identify PNe and separate them from their emission-line contaminants proved to be very efficient thanks to the combination of many colours, even when applied --like in the present work-- to an automatic photometric search that is limited to compact PNe.}


\titlerunning{Tools for compact PNe in J-PLUS and S-PLUS}
\authorrunning{L. A. Guti\'errez-Soto, D. R. Gon\c calves, S. Akras, et al.}

\maketitle

\section{Introduction}
\label{sec:into}

\indent Planetary nebulae (PNe) represent the final stage of the evolution of low-to-intermediate mass stars ($0.8$-$8.0 \mathrm{M}_{\odot}$), when the ejected asymptotic giant branch (AGB) and post-AGB material is ionised by the UV radiation field of the hot and luminous descendant. Due to the high temperature of the central star, a typical PN spectrum shows a great variety of emission lines: not only the recombination lines of the Balmer series and He, but also collisionally excited lines from several elements like O, N, S, Ne, and Ar (e.g. Figure 1 of \citealp{Kwitter:2001}). Given the wide range of temperatures of the central stars (50-250)$\times10^3$K, some PNe are lower-excitation nebulae, and their spectral characteristics can be very similar to those of H~{\sc ii} regions \citep{Osterbrock:2006}. 

The AGB and post-AGB ejecta enrich the interstellar medium (ISM) with heavy elements  that were produced during the evolution of the progenitor star. Thus, PNe provide important information about the production and evolution of chemical elements in the local Universe, and even more locally in the Milky Way (for reviews, see \citealp*{Magrini:2012} and \citealp{Goncalves:2019}).

\indent In the recent past, several surveys have contributed to the discovery of new PNe. The AAO/UKST \ha{} Survey \citep{Parker:2005}, later called Macquarie/AAO/Strasbourg H{$\alpha$} Planetary Nebula Catalogue (MASH catalogue, \citealp{parker:2006}), Southern \ha{} Sky Survey Atlas (SHASSA; \citealp{Gaustad:2001}), Wisconsin \ha{} Mapper (WHAM; \citealp{Haffner:2003}), and the Isaac Newton Telescope (INT) Photometric \ha{} Survey of the Northern Galactic Plane (IPHAS; \citealp{Drew:2005}), in which a number of PN candidates have been found in an automated manner \citep{Viironen:2009a, Viironen:2009b, Sabin:2010}. Small, private surveys have also contributed to the discovery of new Galactic PNe \citep[e.g.][]{Boumis:2003, Boumis:2006}. The most recent compilation of PNe is published in the Hong Kong/AAO/Strasbourg/\ha{} PN database (HASH; \citealp{Parker:2016}). To date, HASH contains 2401 true, 447 likely, and 692 possible Galactic PNe. More genuine PNe candidates have been recently reported in the UKIRT Wide-Field Imaging Survey (UWISH2) of the Northern Galactic Plane \citep{Gledhill:2018} and in the CORNISH catalogue \citep{Irabor:2018, Fragkou:2018}. 
 
So far, only 14 PNe are confirmed members of the Galactic halo \citep{Otsuka:2015}. These are low-metallicity objects, since they come from the oldest population of the Galaxy, with $\log$(O/H) + 12 < 8.1 \citep{Peimbert:1978}. Their height above the Galactic plane and their kinematics are additional criteria to locate them in the halo. Thus, halo PNe (hPNe) show large vertical distances from the Galactic plane, ~$\sim 7.2~\mathrm{kpc}$ \citep{Otsuka:2015}. The understanding of the chemical evolution of the Galaxy can be improved by studying the old and metal-poor PNe located in the halo of our Galaxy, since these stars were born in the earlier phases of Galactic evolution \citep{Otsuka:2015}.

The Javalambre Photometric Local Universe Survey (J-PLUS\footnote{\url{https://www.j-plus.es}}, \citealp{Cenarro:2019}) and the Southern-Photometric Local Universe Survey (S-PLUS\footnote{\url{http://www.splus.iag.usp.br}}, \citealp{Mendes:2019}) provide observations of the Galactic halo covering both northern and southern celestial hemispheres in a systematic way with twin telescopes using the same set of multi-band filters. In addition to the \ha{} filter, which is already vastly applied to systematically searching for \ha{} emitters such as PNe and symbiotic stars (SySt), the telescopes offer 11 more filters. In this work, we present and discuss new colour-colour diagnostic diagrams based on the available filters from J-PLUS and S-PLUS projects to identify halo planetary nebulae. Our main goal is to validate these tools, in addition to applying them to very first J-PLUS and S-PLUS data covering about 1,190 deg$^2$ of the sky. We show that the J-PLUS and S-PLUS filter configuration provide a characterisation of the whole optical spectra of the sources and their potential to such an end with the completion of the surveys, thus opening new horizons for the search of PNe and SySt and other emission line systems.

Our paper is organised as follows: in Section \ref{sec:obser}, we summarise the observations related to J-PLUS and S-PLUS projects, as well as important information of the first data release for each survey. In Section \ref{sec:syn}, the synthetic photometry of the different emission-line sources are explored. Section~\ref{sec:colour} describes our selection method and the new diagnostic diagrams. Section~\ref{sec:valid} presents the validation of the colour-colour diagrams, using known hPNe observed by J-PLUS, and in terms of sources we recovered within the DR1 area, the PN candidate selected with our methodology. In Section~\ref{sec:hash} the results obtained after cross-match the J-PLUS and S-PLUS data with the HASH
catalogue are discussed. In Section~\ref{sec:Dis}, we provide a general discussion and conclusions.

\section{Observations}
\label{sec:obser}

\indent J-PLUS is a multi-filter imaging survey, which is being carried out from the Observatorio Astrof\'isico de Javalambre (OAJ, \citealp{Cenarro:2014}), using the 83\,cm Javalambre Auxiliary Survey Telescope (JAST/T80) and the T80Cam camera. This survey, which maps the northern sky, has a southern counterpart, the S-PLUS. The latter uses a twin telescope, the T80-South, located at the Cerro Tololo Interamerican Observatory (CTIO), in Chile. Both telescopes are equipped with cameras that provide a $\sim$2.0 deg$^2$ field of view (FoV) of the sky, and each covers an area of around 8,500 $\mathrm{deg}^{2}$ in total. Their filter systems are made up of 12 filters spanning the optical range from 3,000 to 10,000\AA, approximately. Although the prime goal of the J-PLUS survey is to perform the photometric calibration for the Javalambre Physics of the Accelerating Universe Astrophysical Survey (J-PAS; \citealp{Benites:2015}), it also provides a rich set of photometric data allowing the investigation of many classes of objects in the Galactic halo and in the local universe. For example, given their multi-filter configuration, they can be used to search for emission-line sources as done by previous narrow-band surveys.  

J-PLUS and S-PLUS filter sets are formed of seven narrow-band filters; \textit{J}0378, \textit{J}0395, \textit{J}0410, \textit{J}0430, \textit{J}0515, \textit{J}0660 and \textit{J}0861. They also include five broad-band Sloan Digital Sky Survey (SDSS) like filters ($u, g, r, i~\mathrm{and}~z$, \citealp{Fukugita:1996}), which are designed to detect the continuum of the sources.  Table~3  and Table~2 of  J-PLUS and S-PLUS Papers 0 \citep{Cenarro:2019, Mendes:2019} specify the characteristics of the filters of the surveys, and 
the features targeted by the narrow-band ones. The data used here come from J-PLUS scientific verification data, the first J-PLUS data release, and the first S-PLUS data release. 

The J-PLUS survey is growing at $\sim$1,000~deg$^2$/year. Given the present rate of completion of the S-PLUS tiles, we anticipate that 4,500~deg$^2$ may be covered by the end of 2020, and the full survey should be completed by the middle of 2023.

\subsection{J-PLUS scientific verification data}
\label{sec:svd}

These scientific verification data (SVD) comprise a group of well-known objects to test and challenge the scientific capabilities of J-PLUS \citep{Cenarro:2019}. The J-PLUS SVD published observations of the galaxy clusters A2589 and A2593, M101, M49, the Arp313 triplet of galaxies, and a few nearby galaxies, including NGC~4470 and the Coma cluster. The Galactic PNe H~4-1 \citep{Miller:1969} and PNG 135.9+55.9 \citep{Tovmassian:2001} were also included within these data.\\

\subsection{First J-PLUS data release}
\label{sec:obser-jplus}

The J-PLUS first data release (DR1\footnote{\url{https://www.j-plus.es/datareleases/data_release_dr1}}) is composed of 511 fields observed with the 12 narrow- and broad-band filters, and it covered 1,022 deg$^2$ of the sky. DR1 is based on images collected by the JAST/T80 telescope from November 2015 to January 2018. Around nine million objects are included in this first data release. The limiting magnitudes (5$\sigma$, 3$''$ aperture) in each filter are presented in Table~4 of \citet{Cenarro:2019}. The median point spread function (PSF) in the DR1 r-band images is 1.1$''$. The detection of the sources was done in the $r$-band using {\sc SExtractor} \citep{Bertin:1996}, and the photometry in the 12 J-PLUS bands at the position of the detected sources was made using the aperture defined in the r-band image. Different types of apertures were used to perform the photometry: i) circular aperture photometry ({\sc SExtractor's MAG\_APER}) with apertures of different sizes (3$''$ and 6$''$), ii) isophotal photometry ({\sc SExtractor's MAG\_ISO}), iii) Kron photometry ({\sc SExtractor's MAG\_AUTO}, iv) Petrosian photometry ({\sc SExtractor's MAG\_PETRO}), v) photometry after degrading the images of all the bands to the worst PSF, vi) photometry after convolving all the images with a Gaussian kernel of $\sigma$=1.5$'',$ and vii) photometry applying a PSF correction as that applied in \citep{Molino:2014}. In the present work, we use the 6$''$ aperture. This choice follows the automatic method used by IPHAS photometry to find compact PNe, with a small angular diameter (typically $\leq$5$''$, \citealp{Viironen:2009b}). The 6$''$-aperture chosen also follows \citet*{Akras:2019c}, who found that PNe with angular sizes larger than $6''$ exhibit very low ($r - \ha{}$) colour index.  

\subsection{First S-PLUS data release}
\label{sec:obser-splus}

The S-PLUS DR1\footnote{\url{https://datalab.noao.edu/splus/}} includes 80 fields of the Stripe-82 area, a rectangular area with coordinates 4$\mathrm{^{h}}$ $\mathrm{<RA}$ $<20$$^{h}$ and -1.26$^{\circ}$$<\mathrm{Dec}<$1.26$^{\circ}$. These fields were observed during the scientific validation process of the survey. DR1 contains about one million sources, and the limiting magnitude of each filter is presented in Table~8 of the S-PLUS presentation paper \citep{Mendes:2019}. Sources were detected using {\sc SExtractor}.  Photometry was performed by adopting three types of aperture: i) a circular aperture of 3$''$ in diameter ({\sc SExtractor's APER}), ii)  Kron photometry ({\sc SExtractor's AUTO}), and iii) petrosian photometry ({\sc SExtractor's PETRO}). Similarly to the case of J-PLUS, we use the circular aperture of 3$''$. This aperture is 
the best within the three possibilities, since -- in addition to the arguments given in Section~\ref{sec:obser-jplus} -- it ensures that large objects with high surface brightness, like emission-line galaxies or H~{\sc ii} regions, are not recovered (with this automated method). 

The tools we use to identify PNe in the two surveys are  synthetic photometry and colour-colour diagrams, which we describe in detail in the following sections.

\section{Synthetic photometry}
\label{sec:syn}

In order to recover the synthetic photometry of PNe and their contaminants, we follow the procedure developed by \citet{Aparicio:2010}, who characterised the ALHAMBRA photometric system. The magnitudes of any photometric system are defined by a set of AB magnitudes \citep{Oke:1983}:

\begin{equation}
  \label{eq:AB-magn}
 \mathrm{AB}_{\nu} = -2.5 \log f_{\nu} - 48.6,
\end{equation}
\noindent
where the constant is obtained from the equation 
\begin{equation}
  \label{eq:conts}
  48.6 = -2.5 \log F_{0} 
\end{equation}
\noindent  $F_{0} = 3.63 \times 10^{-20}~\mathrm{erg~s^{-1}~cm^{-2}~Hz^{-1}}$ being the flux of Vega at $\lambda = 5,500$~\AA. The term $f_{\nu}$ is the flux per unit of frequency [$\mathrm{erg~s^{-1}~cm^{-2}~Hz^{-1}}$]. 

The synthetic photometry is obtained from the convolution of the  observed or modelled spectra with the transmission function or pass-band of a specific photometric system. The transmission function accounts for the transmission filter, the reflectivity of the telescope mirror, the transmission of the camera optics and the quantum efficiency of the detector used \citep{Bessell:2005}. By considering these characteristics of the photometric system, Eq.~\ref{eq:AB-magn} turns to
\begin{equation}
  \label{eq:AB-magn-banpass-lenght}
   \mathrm{AB}_{\lambda} = -2.5 \log \frac{1} {c} \frac{\int f_{\lambda} S_{\lambda} \lambda d \lambda} {\int  S_{\lambda} d\lambda/\lambda}  - 48.60,
\end{equation}
\noindent
where $c$ is the light speed, $S_{\nu}$ is the transmission function, and the units of $f_{\lambda}$ are [$\mathrm{erg~s^{-1}~cm^{-2}}$~\AA$^{-1}$]. Therefore, Eq.~\ref{eq:AB-magn-banpass-lenght} makes it possible to estimate the synthetic photometry (AB magnitude) in a specific photometric system.

\begin{figure}
\setlength\tabcolsep{\figstampcolsep}
\centering

\begin{tabular}{l l}
 
 \includegraphics[width=0.45\linewidth, trim=50 100 100 8]{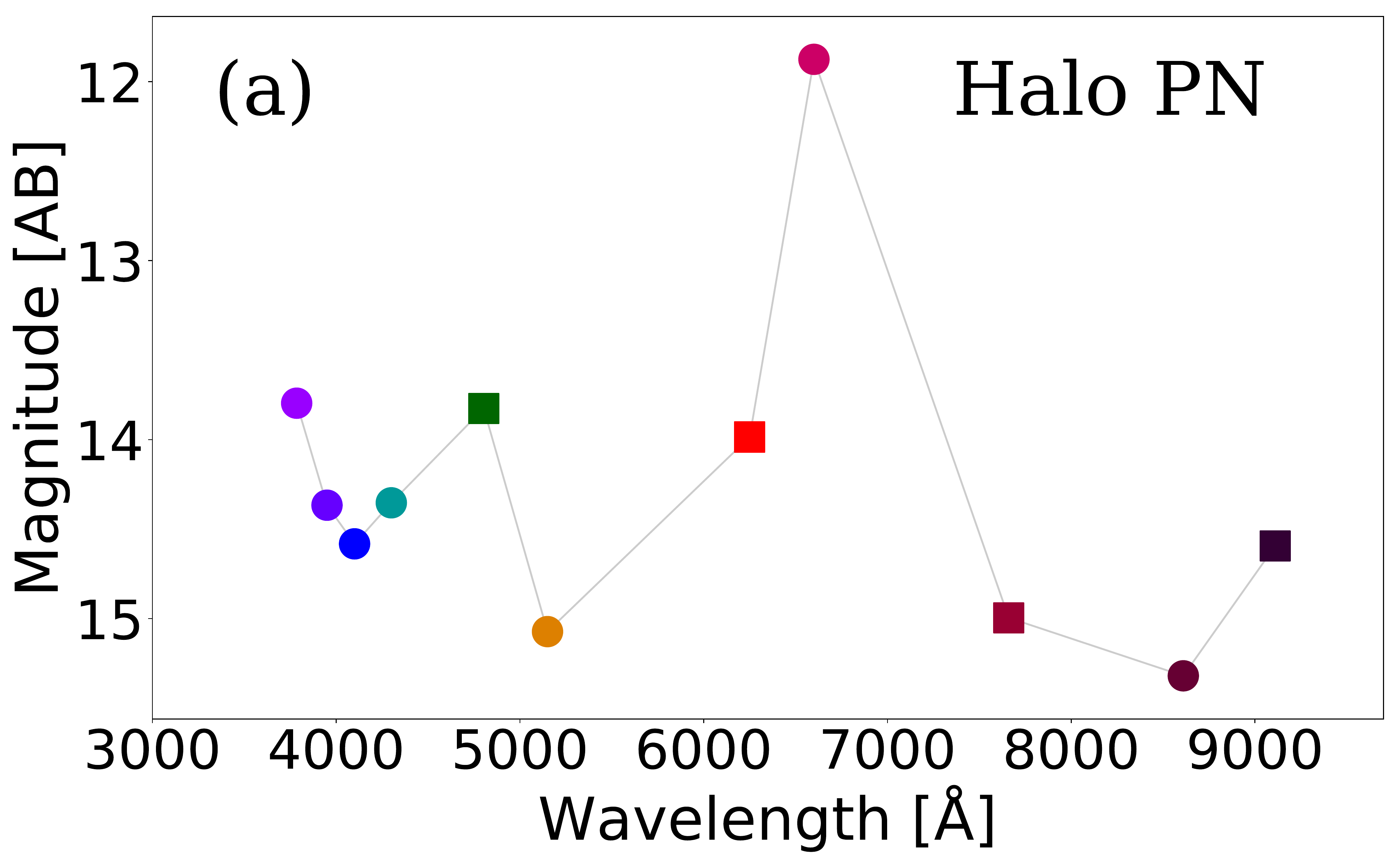} & \includegraphics[width=0.48\linewidth, trim=60 90 10 8,clip]{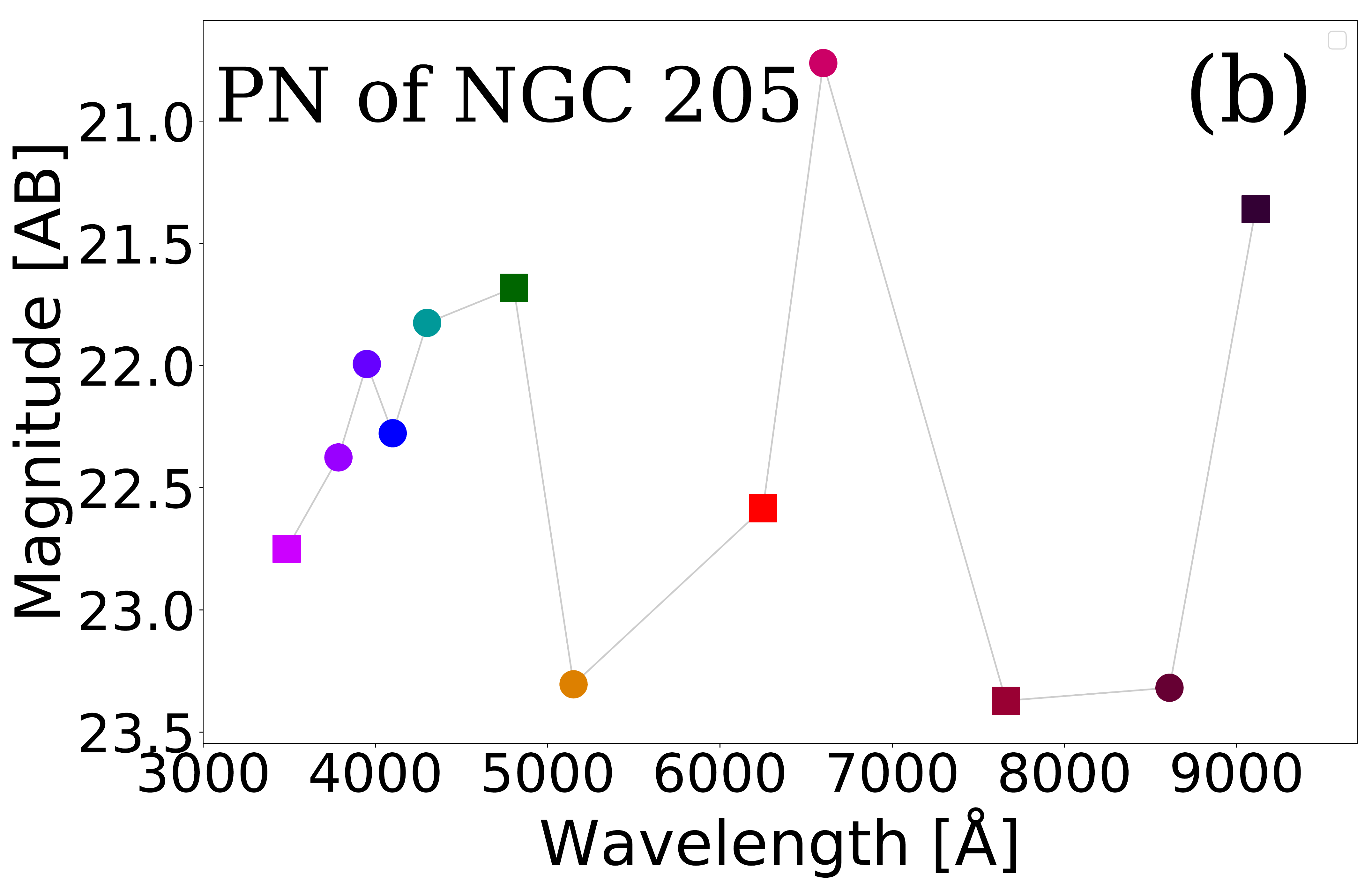}  \\  

\includegraphics[width=0.45\linewidth, trim=80 100 70 8]{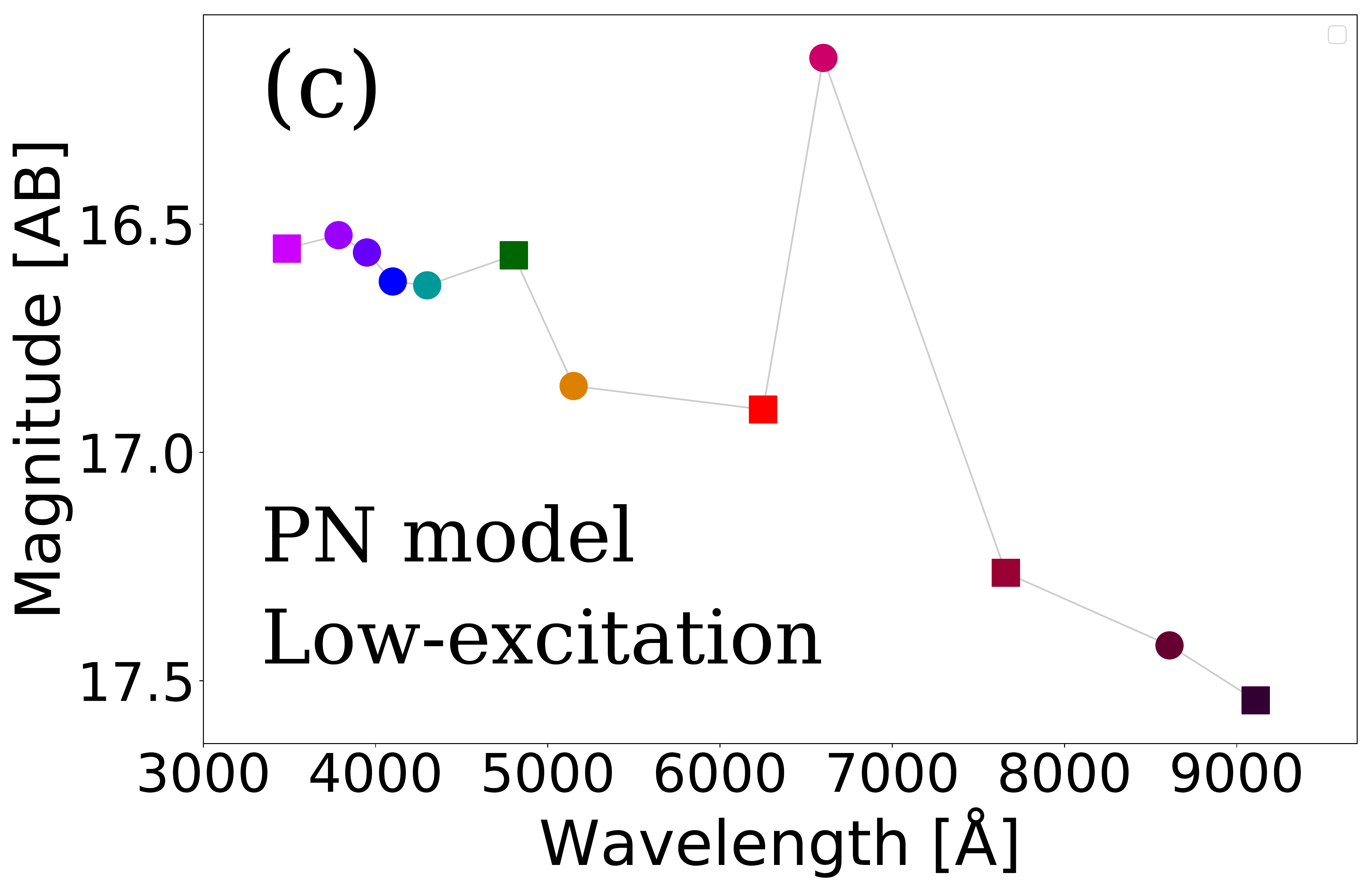} & \includegraphics[width=0.48\linewidth, trim=65 90 10 10, clip]{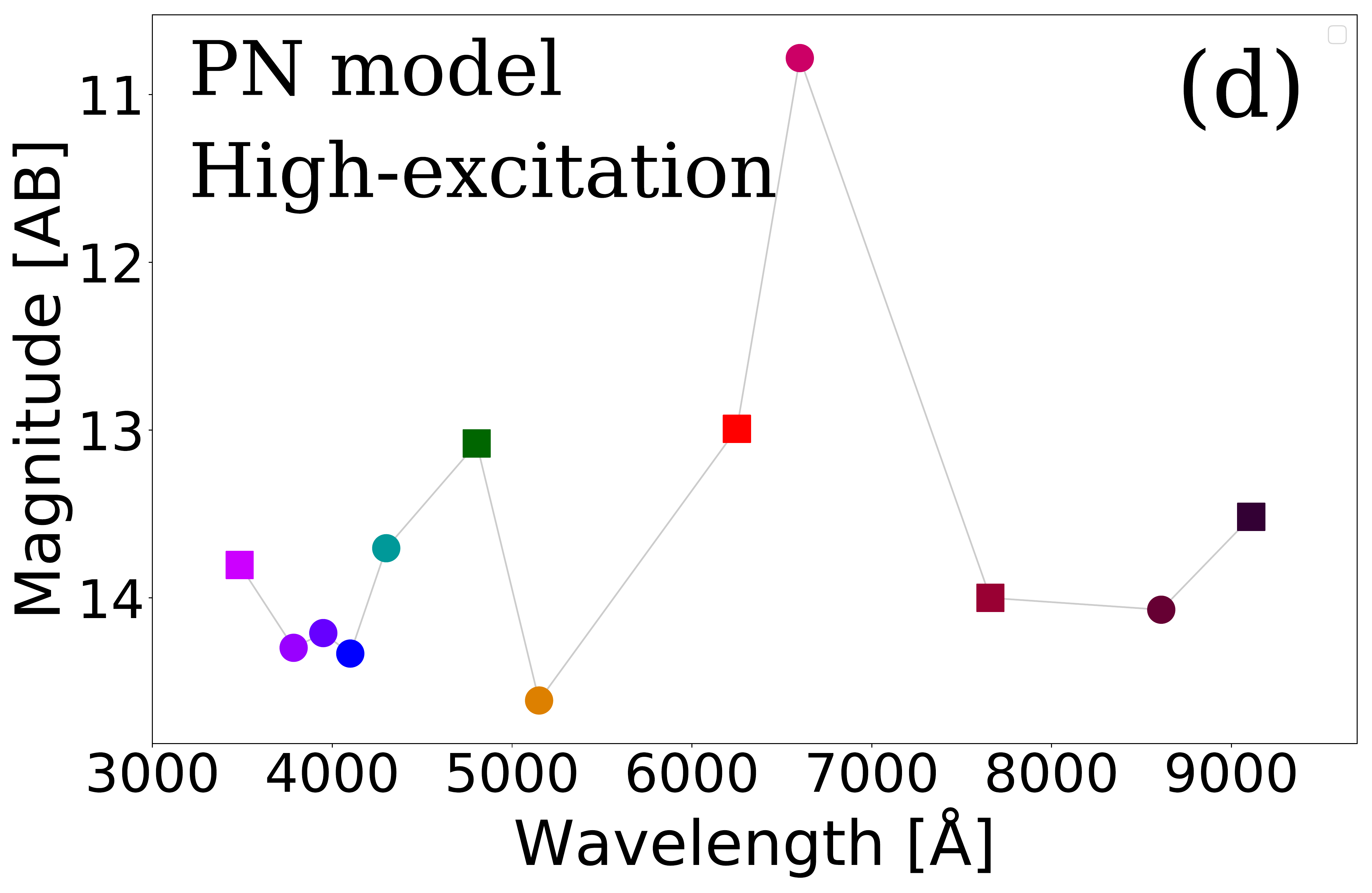} \\

\includegraphics[width=0.45\linewidth, trim=50 100 100 8]{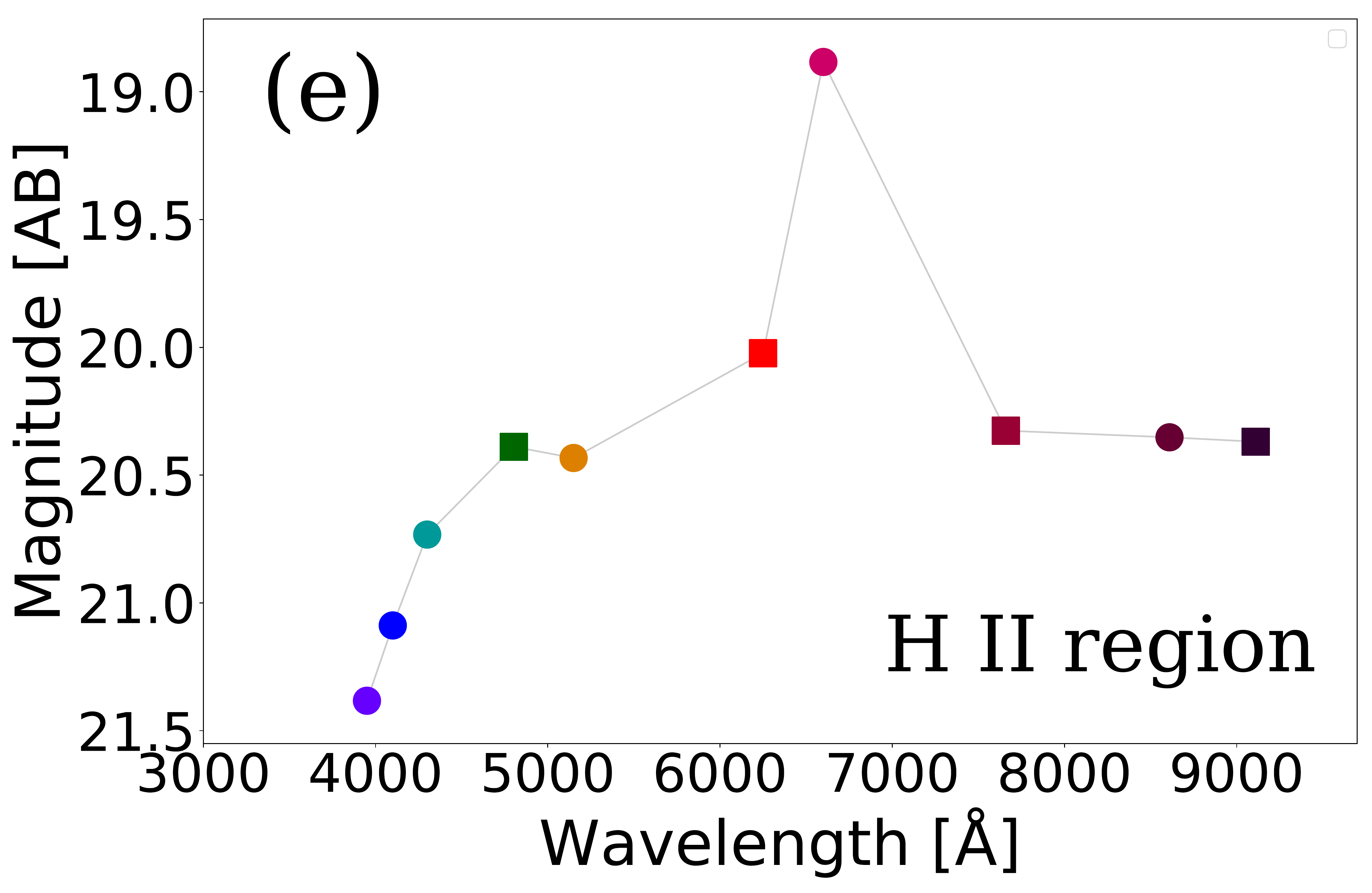} & \includegraphics[width=0.48\linewidth, trim=65 90 10 10, clip]{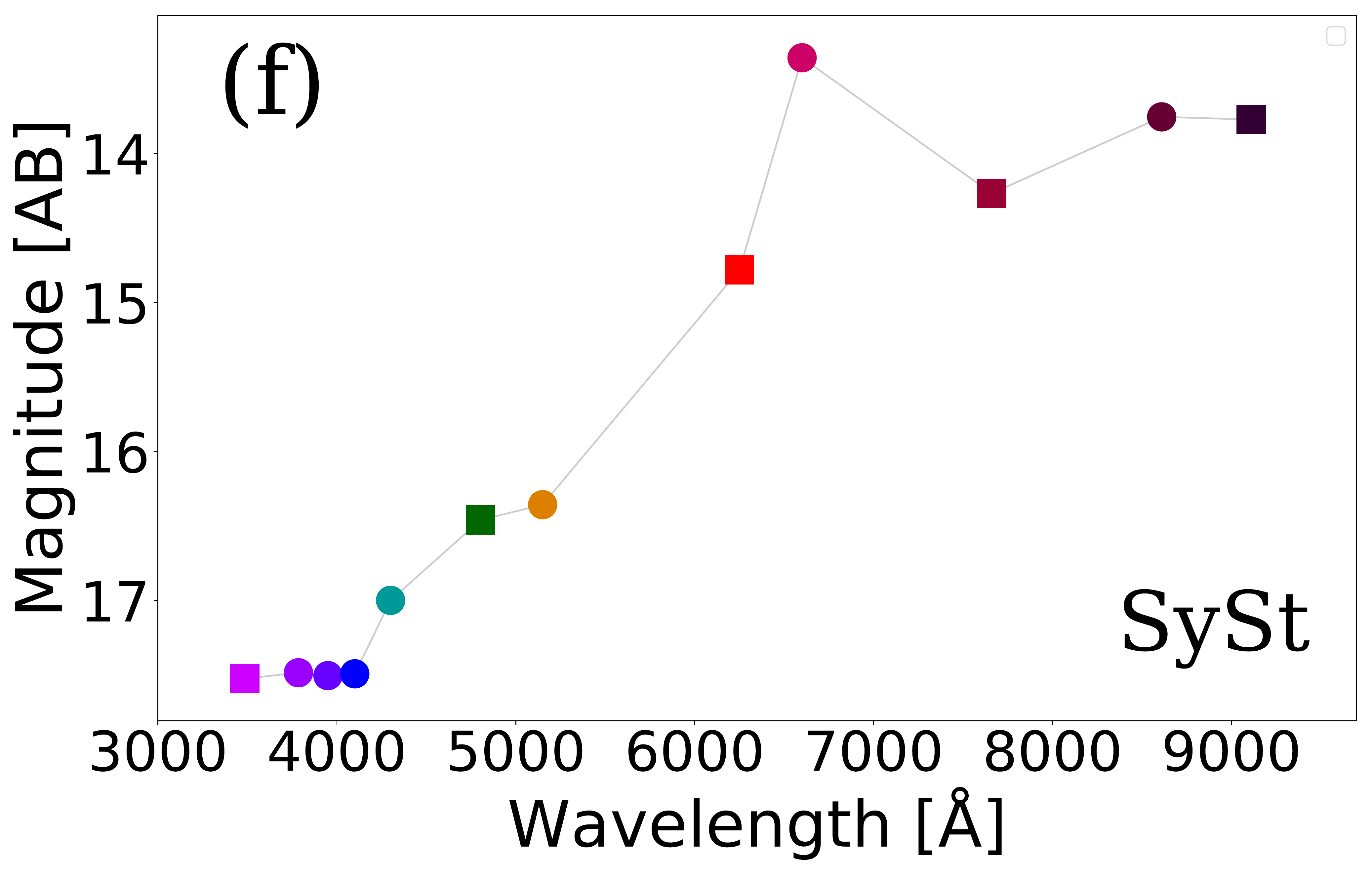} \\

\includegraphics[width=0.45\linewidth, trim=50 100 100 8]{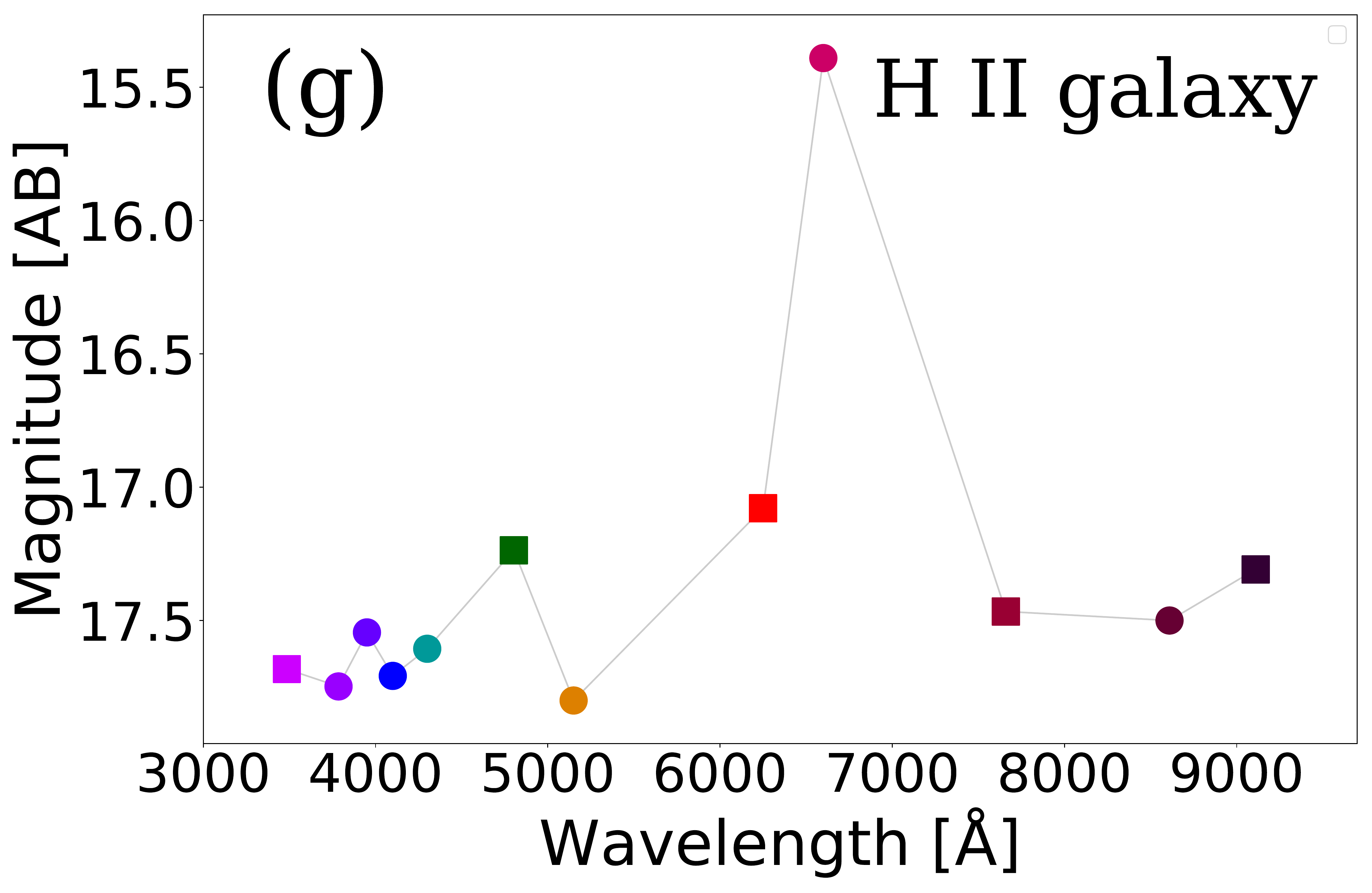}& \includegraphics[width=0.483\linewidth, trim=60 90 10 10, clip]{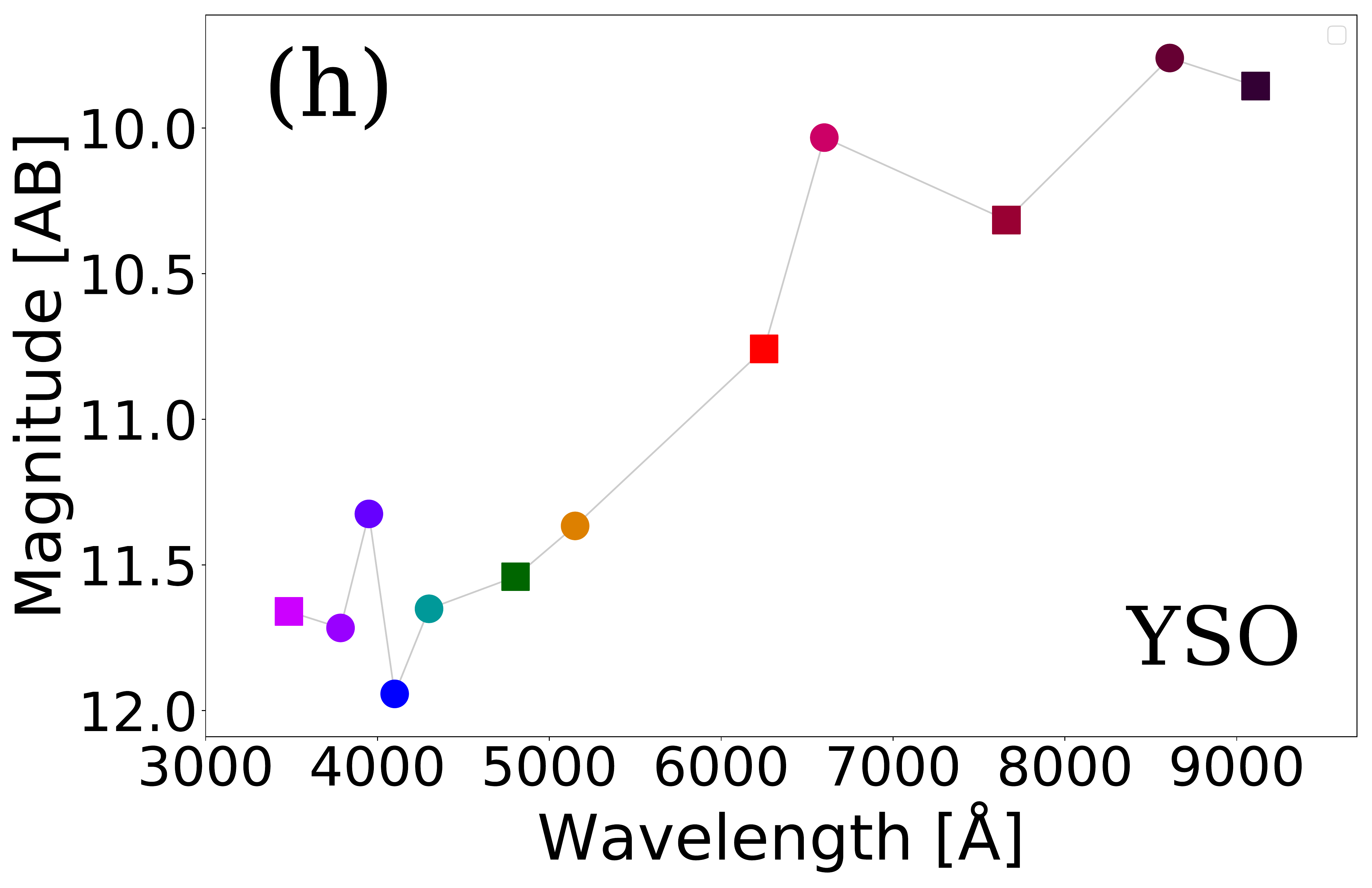}\\  \includegraphics[width=0.45\linewidth, trim=50 100 100 8]{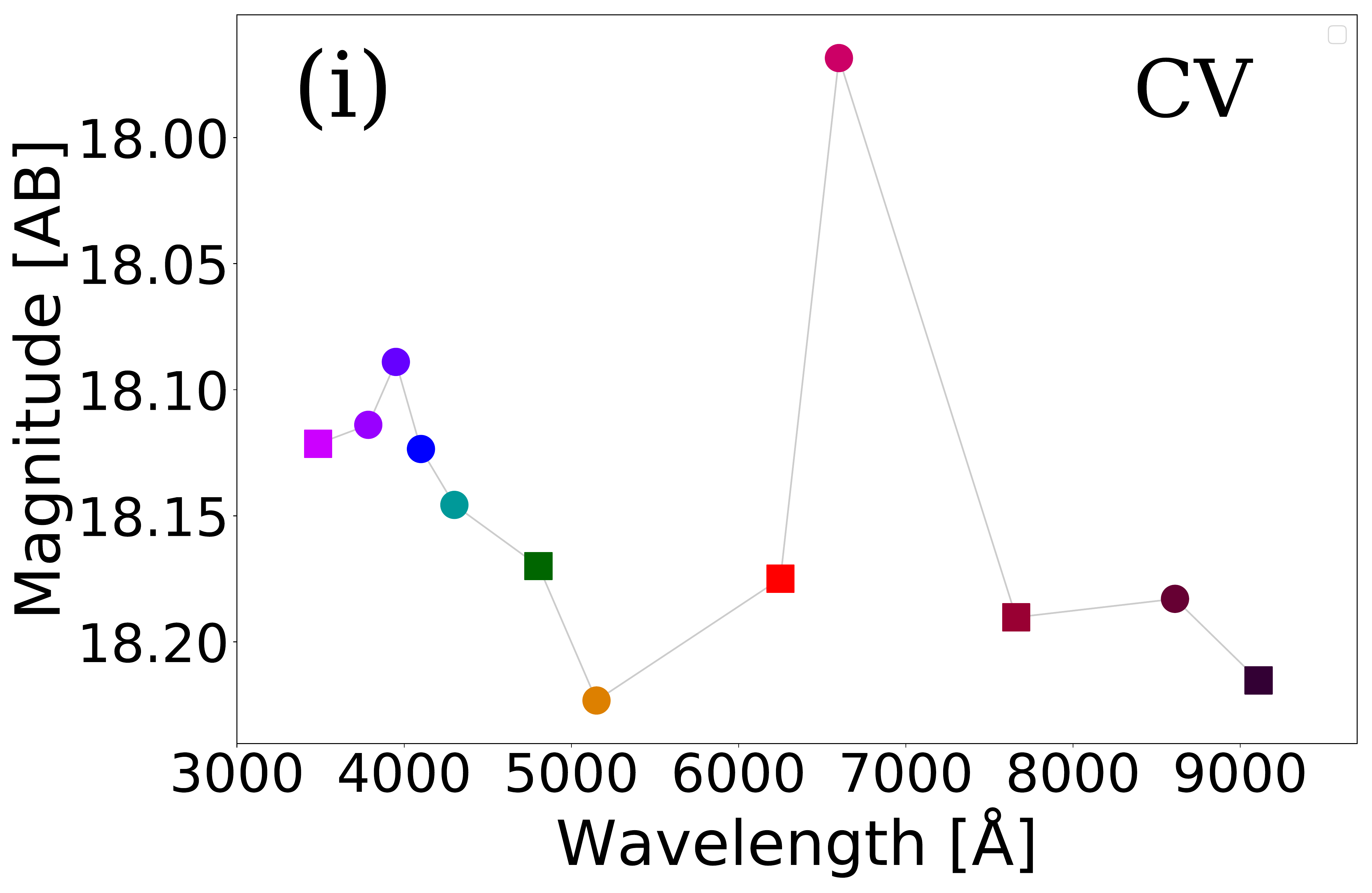} &

\includegraphics[width=0.48\linewidth, trim=60 90 10 8, clip]{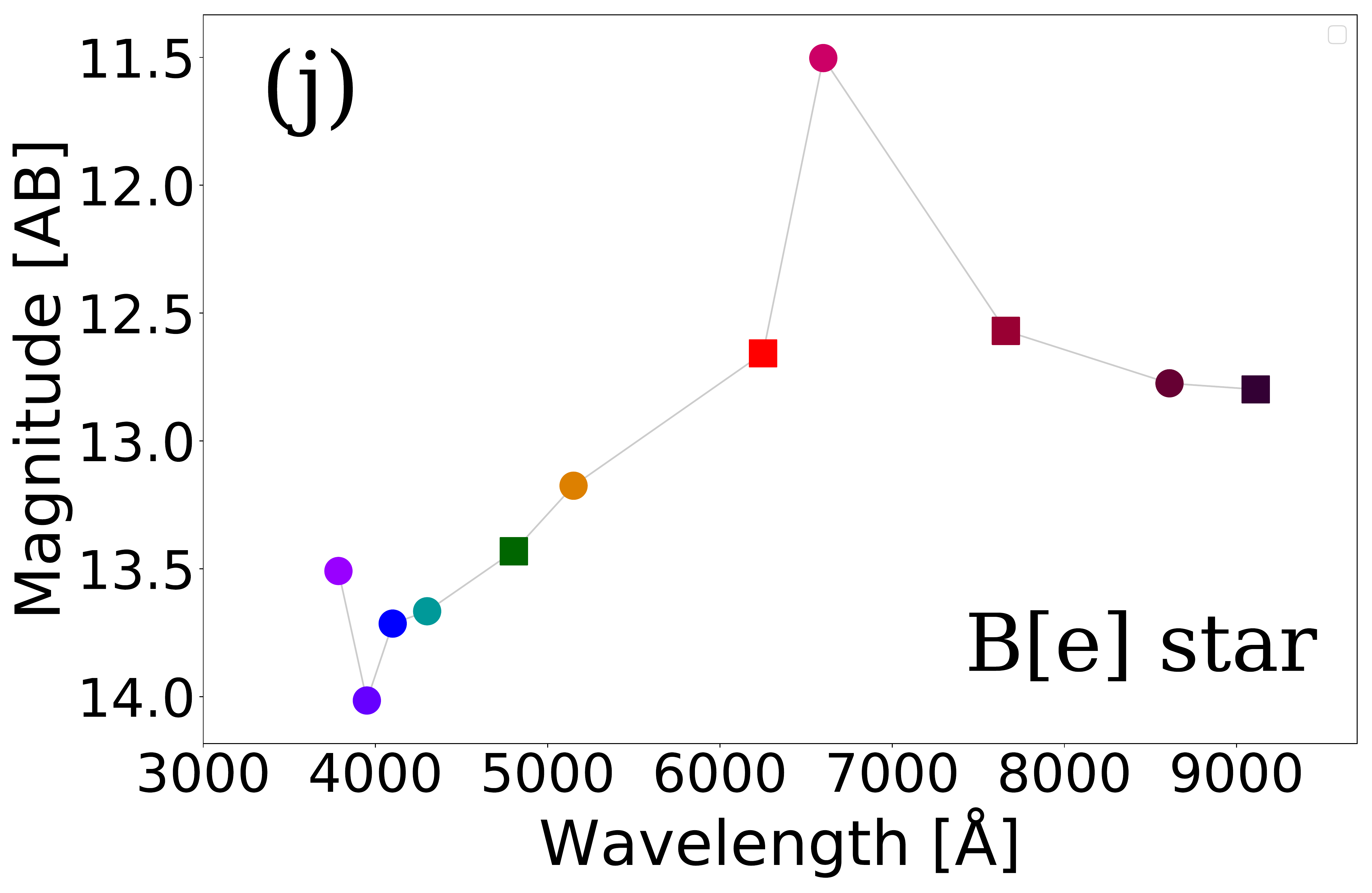} \\
\includegraphics[width=0.45\linewidth, trim=50 100 100 8]{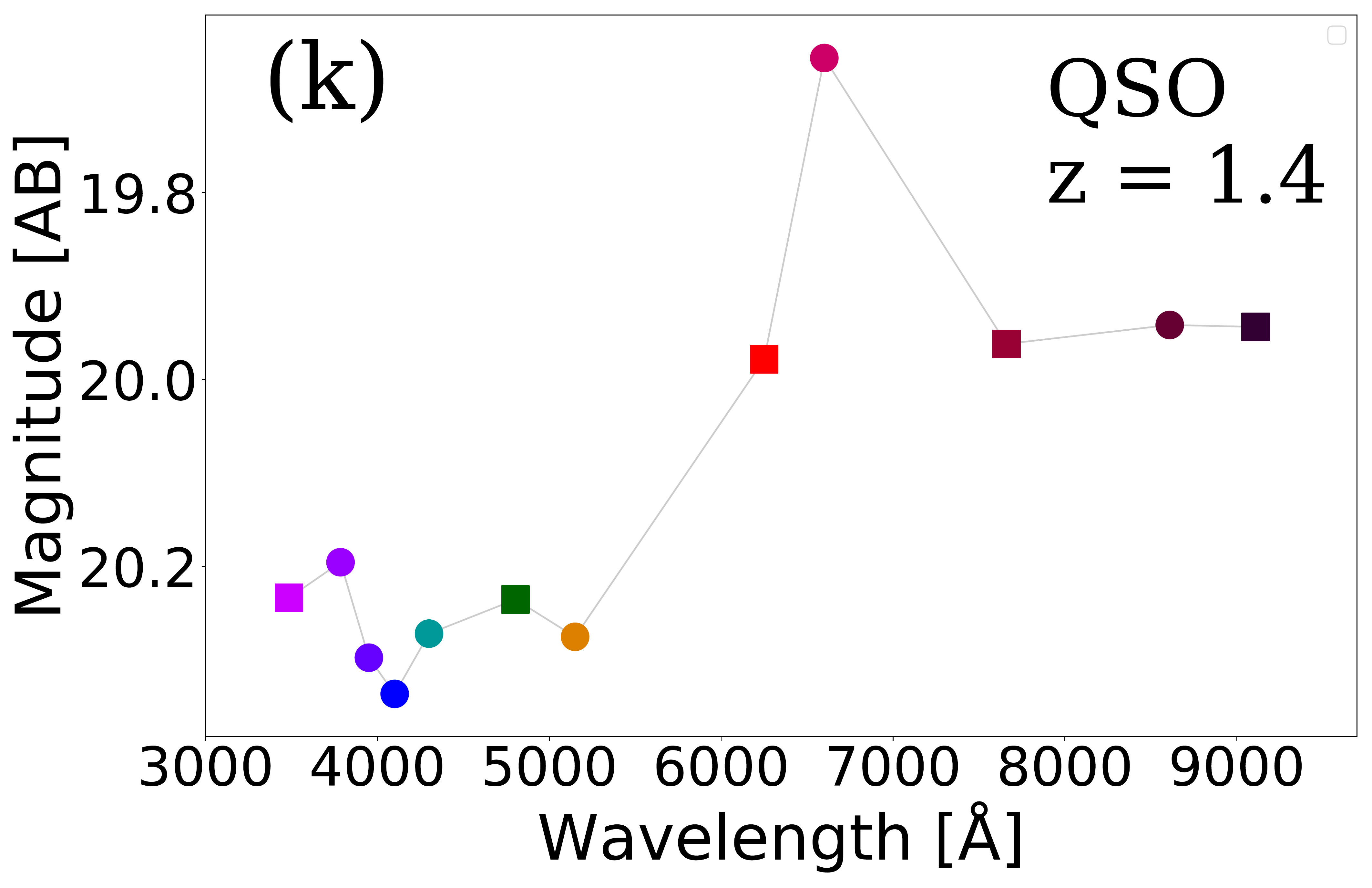}&
\includegraphics[width=0.48\linewidth, trim=60 90 10 8, clip]{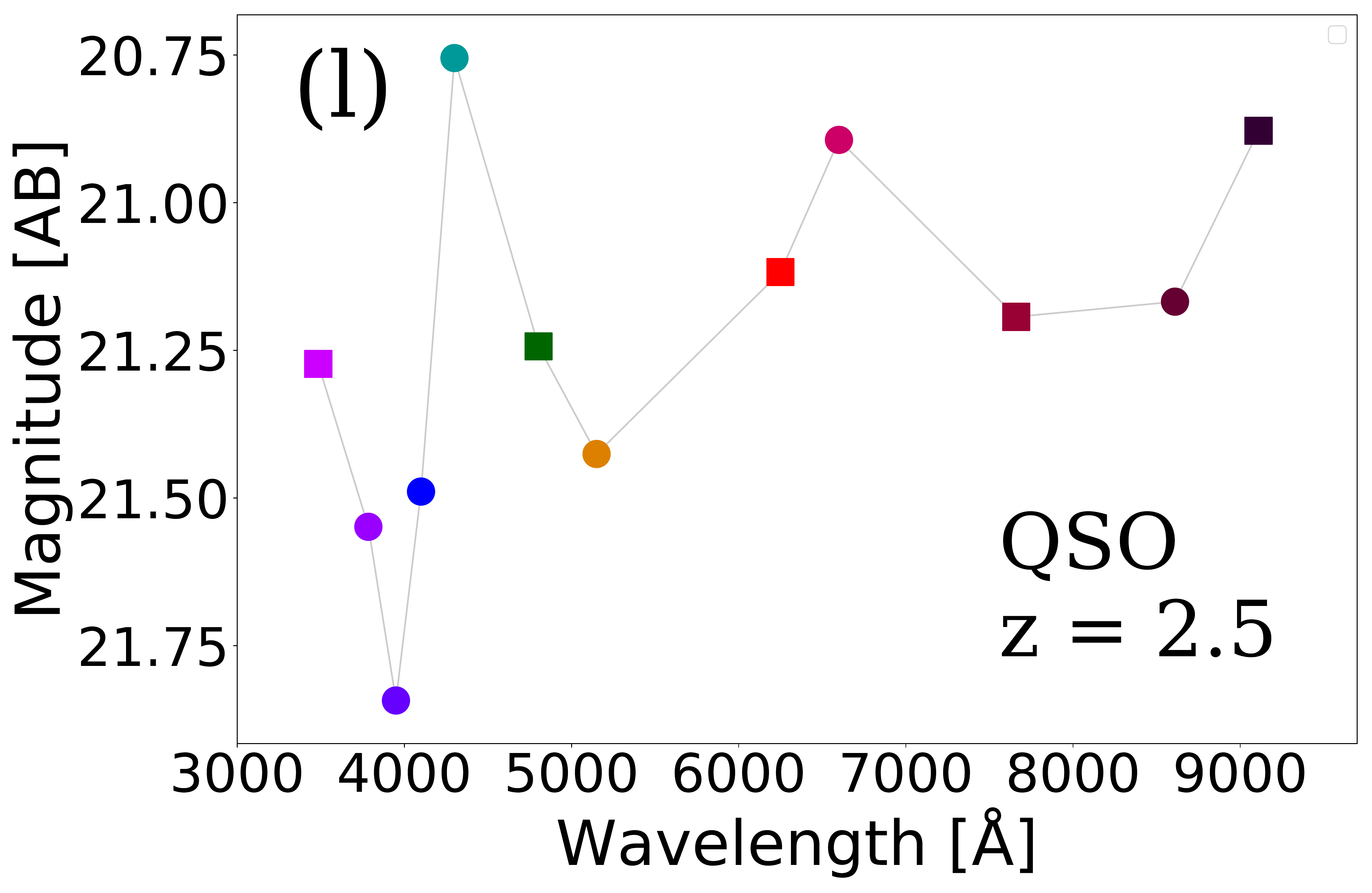} \\
\includegraphics[width=0.45\linewidth, trim=50 10 100 8]{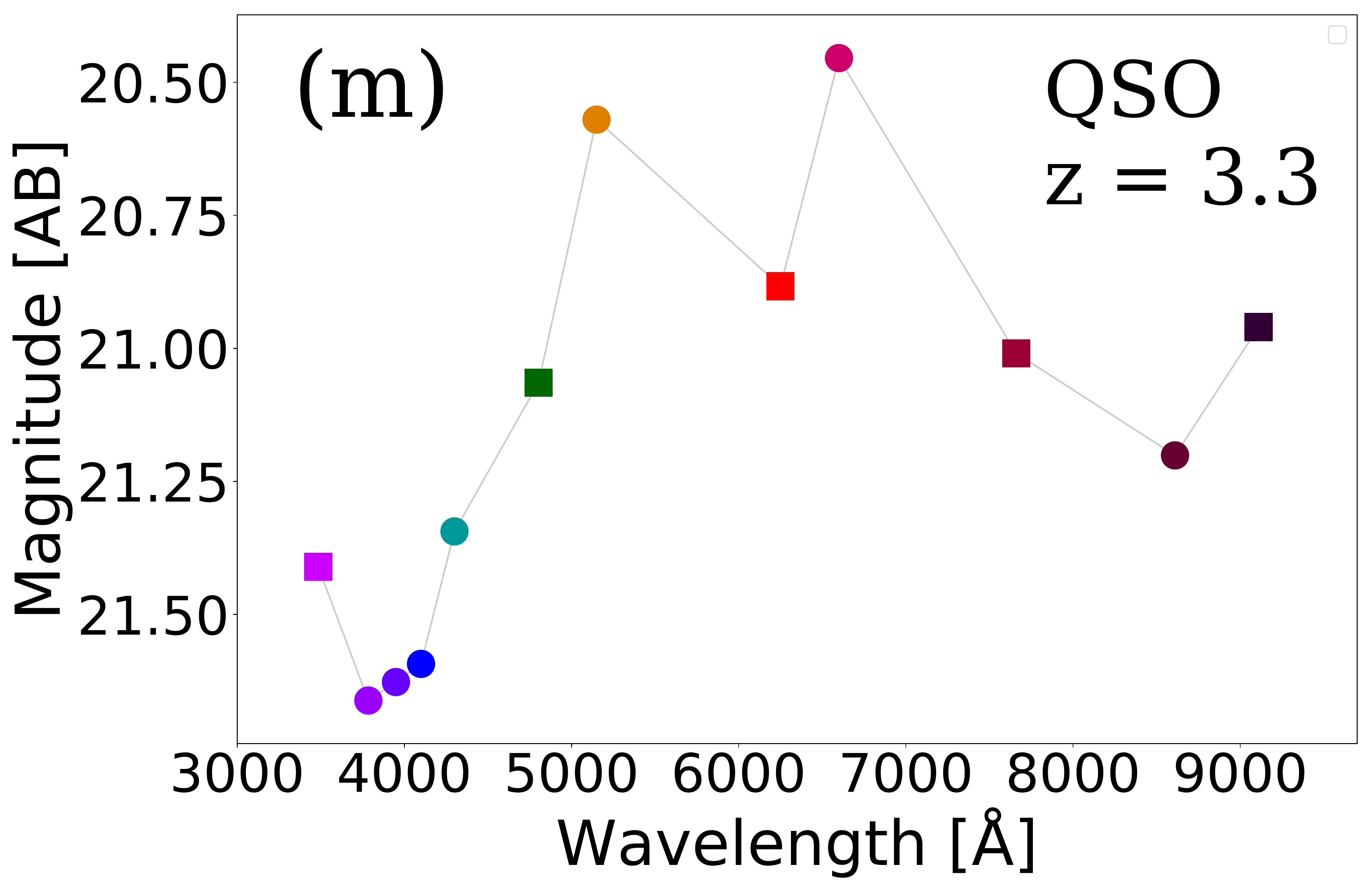}&
\includegraphics[width=0.48\linewidth, trim=65 10 10 4, clip]{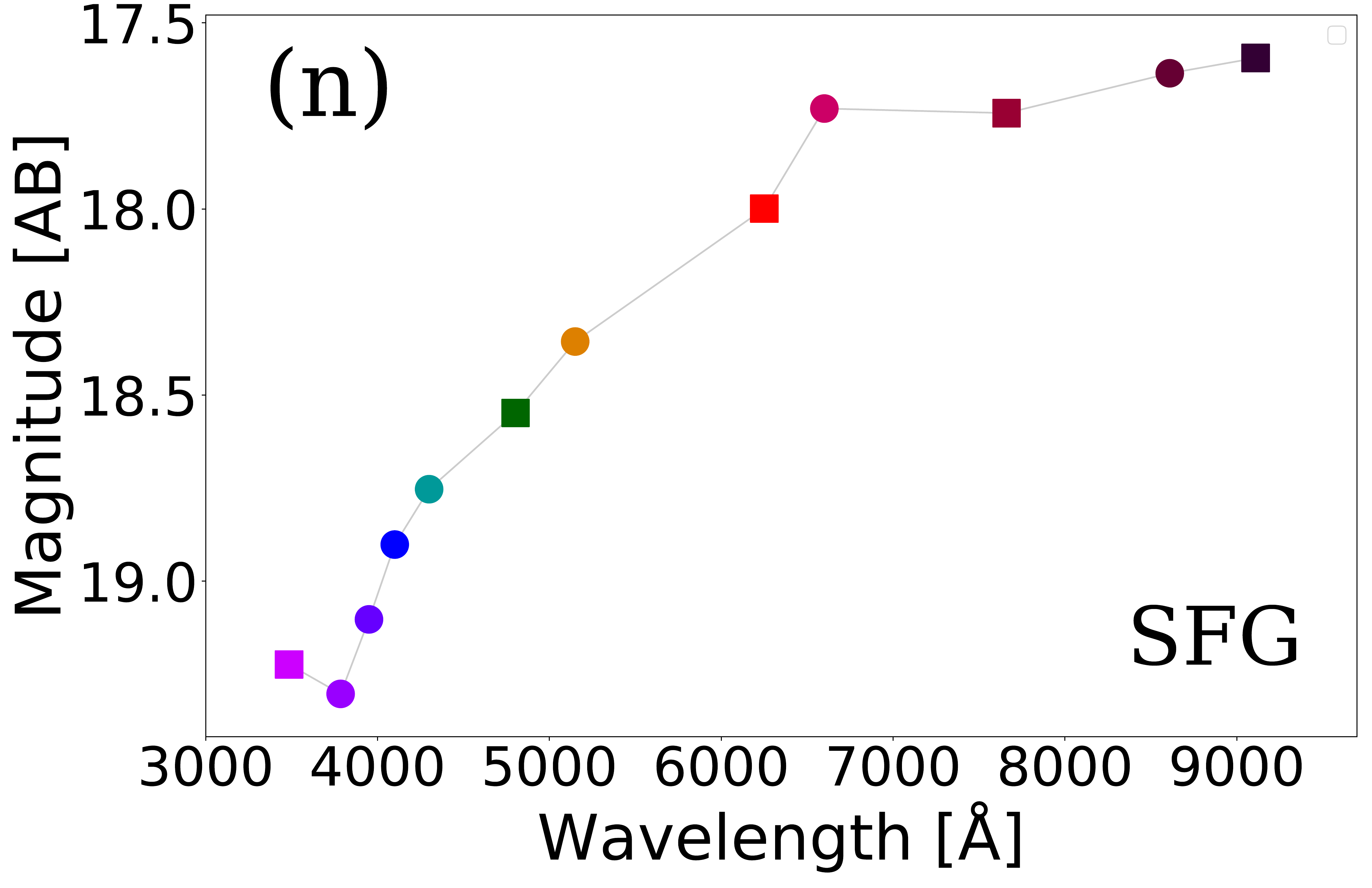}

  \end{tabular}
  
  \caption{J-PLUS and S-PLUS synthetic photo-spectra of several emission-line objects. The DdDm~1 PN $(a)$. A PN in the dwarf galaxy NGC~205 $(b)$. Two {\sc cloudy} modelled PNe with 
  T$_{\rm{eff}}$=$60 \times 10^3$~K  and $200 \times 10^3$~K, and H density of 1000~cm$^{-3}$ and 3000~cm$^{-3}$, which concern the abundances of the hPN 
  DdDm~1 and L=10,000~L$_{\odot}$, $(c, d)$. A H~{\sc ii} region in the nearby galaxy NGC~55 $(e)$. The SySt LMC~1 $(f)$. The \hii{} galaxy Mrk 1318 $(g)$ from SDSS. The YSO RU~Lup $(h)$. A CV $(i)$ from SDSS. A B[e] star $(j)$. QSOs with redshift of 1.4 $(k)$, $z = 2.5$ $(l),$ and $z = 3.3$ $(m),$ where the strong emission in the \textit{J}0660 filter, is caused by the Mg ~{\sc ii} $\lambda2798$,  C~{\sc iii}] $\lambda1909$ and C~{\sc IV} $\lambda1550$  emission lines, respectively. And, a star-forming galaxy $(n)$. For this object, the \ha{} line is responsible for the \textit{J}0660 magnitude. Squares represent the SDSS-like broad-band filters. From left to right they are $u, g, r, i,~ \mathrm{and}~ z$. Circles are the narrow-band filters, \textit{J}0378, \textit{J}0395, \textit{J}0410, \textit{J}0515, \textit{J}0660, and \textit{J}0861, from left to right.}
  \label{fig:phot-spect}
\end{figure}

The synthetic photometry or photo-spectra  of several emission-line sources -- PNe, SySt, cataclysmic variables (CVs), quasi-stellar objects (QSO), extragalactic H~{\sc ii} regions, young stellar objects (YSOs), B[e] stars, and star-forming galaxies (SFGs) -- were obtained through the convolution of the theoretical transmission curves of the J-PLUS and S-PLUS systems with the available optical spectra, as in Eq.~\ref{eq:AB-magn-banpass-lenght}.

The same procedure was also applied to a grid of photoionisation models for hPNe to obtain their synthetic magnitudes. For the hPN modelling, we adopted the photoionisation code, \textsc{cloudy} \citep{Ferland:2013}. The initial parameters used to compute the models reflect the typical properties of the PN population located in the Galactic halo. They represent different sets of nebular abundances, with electron densities of 1,000~$\mathrm{cm^{-3}}$, 3,000~$\mathrm{cm^{-3}}$ and 6,000~$\mathrm{cm^{-3}}$, of a spherically symmetric nebula with a radius of $2.7''$, at distance of 10~kpc.  
Central star (black body) effective temperatures from 50,000~K to 250,000~K, in steps of 10$\times 10^3$~K, and luminosities of 500, 1,000, 5,000 and 10,000~L$_{\odot}$ were considered. We reddened the modelled spectra applying the reddening curve by \citet{Fitzpatrick:1999}, and  using two different colour excesses, E(B - V) = 0.1 and 0.2, following the J-PLUS and S-PLUS average extinction of E(B - V) = 0.1 \citep{Cenarro:2019, Mendes:2019}. 

Figure~\ref{fig:phot-spect} comprises all synthetic photo-spectra recovered as explained above.
Panels (\textit{a}) and (\textit{b}) display the photo-spectra of the Galactic hPN DdDm~1 \citep{Kwitter:1998}, and of an extragalactic PN located in NGC~205 \citep{Goncalves:2014}, as examples. Both objects show strong \ha{} emission, with a continuum a little more intense in the blue than in the red part of the spectrum. Panels (\textit{c}) and (\textit{d}) show two \textsc{cloudy} modelled hPNe. Both have the same chemical abundance  (equivalent to that of the hPN DdDm~1), and the same luminosity of $10^4 L_{\odot}$. The  effective temperature and density of the model in panel (\textit{c}) are of $60 \times 10^3$~K and 1,000~cm$^{-3}$ and for the model in panel (\textit{d}) of $200 \times 10^3$~K and 6,000~cm$^{-3}$, respectively. So, the first model (panel \textit{c}) corresponds to a low-excitation PN. The ionising star is not hot enough to produce such an \oiii{} emission able to stand out significantly in the $g$ broad-band magnitude. The second model (panel \textit{d}) represents a high-excitation PN. In this case, the magnitude of the $g$ broad-band filter is impacted by the strong \oiii{} and \heii{} emission lines. Note that the $g$-band also includes the  H$\beta$ line, therefore it is not possible to disentangle the H$\beta$ and \oiii\ emission lines in the J-PLUS and S-PLUS photo-spectra and to quantify the level of excitation more precisely. The hPNe (modelled or observed) differ from Galactic-disc PNe in size, velocity, distance, and extinction \citep{Howard:1997, Otsuka:2015}. All the properties -- high distances implying small angular sizes, poor in metals and low extinction -- of hPNe, with the exception of the large velocities, make them occupy a specific locus in the colour-colour diagrams, significantly different to that of the disc PNe.

In addition to PNe, we also computed the photo-spectra of a number of other emission-line sources that mimic PNe. Panel (\textit{e}) of Figure~\ref{fig:phot-spect} shows the photo-spectrum of an extragalactic H~{\sc ii} region located in the dwarf galaxy NGC~55 \citep{Magrini:2017}, in which the \ha{} emission is also perceptible. The low-excitation PN (on panel \textit{c}) and the H~{\sc ii} region show very similar photo-spectra, which will make them hard to distinguish within the J-PLUS and S-PLUS catalogues. \citet{Merrett:2006} have shown that for low-luminosity objects, it is hard to distinguish PNe and H~{\sc ii} regions, using the ratio between \oiii{} and \ha{}+\nii{} emission lines. We encounter the same problem here, even using a wider range of wavelengths. We further investigate this issue, including morphological criteria, in Section~\ref{sec:valid}. With the exception of the morphology, the same difficulty occurs for \hii{} galaxies, as is shown in panel \textit{g}, for a SDSS \hii{} galaxy. At this point, it should be noted that the synthetic photometric magnitudes are below the limiting magnitudes for all the filters bluer than 4,500\AA, in the case of the extragalactic PN (panel {\it b}), and for \textit{u} and \textit{J}0378 filters in the case of the extragalactic H~{\sc ii} region (panel {\it e}).

Panel (\textit{f}) of Figure~\ref{fig:phot-spect} presents the photo-spectrum of the SySt LMC1 \citep{Munari:2002a}, displaying a clear \ha{} emission with an increasing continuum to the longer wavelengths. The resolution of the J-PLUS and S-PLUS photometry  allows to perceive undoubtedly the reddened nature of the SySt. 
As PNe, symbiotic stars are also constituted by an evolved star, in most cases a white dwarf (WD). They are interacting binary systems with a cold giant companion (red giant or Mira star) and an evolved hot star. The wind of the giant is ionized by the UV radiation from the evolved companion, thus resulting in a spectrum composed by both absorption features from the stellar photo-sphere of the giant (e.g. TiO and VO), and emission line features from the excited ions (\citealp*{Munari:2002a, Corradi:2003}; \citealp{Rodriguez:2014, kiewicz:2017, Akras:2019b, Ange2019}). The photo-spectrum of a cataclysmic variable (CV) selected from the SDSS catalogue is also presented in Figure~\ref{fig:phot-spect} (panel \textit{i}). Cataclysmic
variables are interacting binary systems, in which a white dwarf accretes gas from a main-sequence star, via Roche lobe outflow, forming an accretion disc. The Balmer lines are produced in the optically thin outer regions of the disc \citep{Williams:1980}. It is simple to notice that based on their photo-spectra, PNe of high and low excitation could be mistaken for CVs. Panel (\textit{h}) brings the convolution results for a YSO, from \citet{Alcala:2014}. An accretion disc, from where the \ha{} emission emerges, is also present in this type of (pre-main-sequence) stars. The YSO's photometry shows a continuum that is stronger in the red part of the spectrum. Panel (\textit{j}) displays a B[e] star \citep{Lamers:1998}. B[e] stars have a surrounding nebula produced by large-scale mass loss involving one or more eruptions \citep{Marston:2006}. Its photo-spectrum resembles that of both SySt and YSOs.

The J-PLUS and S-PLUS synthetic spectra of several SDSS QSOs, in specific redshift ($z$) ranges, were also included in our selection of emission-line objects. The QSOs in the redshift ranges $1.3 < z < 1.4$, $2.4 < z < 2.6,$ and $3.2 < z < 3.4$ possess features that resemble those of PNe, in other words, they could be misinterpreted as \ha{} line emission at $z=0$. At these redshift ranges, respectively, the C~{\sc iv}~1550~\AA, C~{\sc iii}]~1909~\AA~ and Mg {\sc ii}~2798~\AA~ emission lines of QSOs fall into the $J0660$ (\ha{})
  filter.  In panel \textit{(k)} of Figure~\ref{fig:phot-spect}, we have a QSO at  $z = 1.4$, in which the excess of emission in the \textit{J}0660 filter comes from the contribution of the Mg~{\sc ii}  emission line. 
 Two more QSOs are presented in Figure~\ref{fig:phot-spect}: at $z = 2.5$ (panel \textit{l}) and $z = 3.3$  (panel \textit{m}). At these redshifts, the C~{\sc iii}] and C~{\sc iv} emission lines are also detected in the J0660 band. The magnitudes of the $J0430$ band for the QSO at $z = 2.5,$ and of the $J0515$ band for the QSOs at $z = 3.3,$ indicate strong emissions, which are provided by the L$\alpha$ emission in both cases. Panel \textit{(n)} of Figure~\ref{fig:phot-spect} shows a star-forming galaxy from the SDSS. Its photo-spectrum exhibits a moderate \ha{} emission and a continuum increasing for longer wavelengths. This is because these objects are embedded in hot gas and dust. Despite the fact that QSOs have strong emission lines that can be misidentified as \ha{}, their photo spectra are only moderately similar to those of PNe. It is worth mentioning that J-PLUS and S-PLUS can characterise not only the strong emission lines, but also the continua of these objects, thus allowing us to correctly differentiate QSO from PNe, which was a problematic task in projects dedicated to the search of PNe, which were using a fewer filters. Finally, we note that SFGs and YSOs have photo-spectra that are very different from those presented in Panel \textit{a} and Panel \textit{b,} and so they do not cause much impact on our research work.

\section{Synthetic colour-colour diagrams: selection criteria}
\label{sec:colour}

\indent The photo-spectra of the emission-line sources just discussed in Figures~\ref{fig:phot-spect} -- for PNe, SySt, CVs, B[e], QSOs, extragalactic H~{\sc ii} regions, YSOs, and SFGs -- are now used to build diagnostic colour-colour diagrams. Our goal is to pinpoint the most relevant ones to discriminate PNe from the above systems. 

\subsection{IPHAS equivalent colour-colour diagram}
\label{sec:iphas}

\begin{figure}
  \includegraphics[width=\linewidth, trim=10 10 10 10]{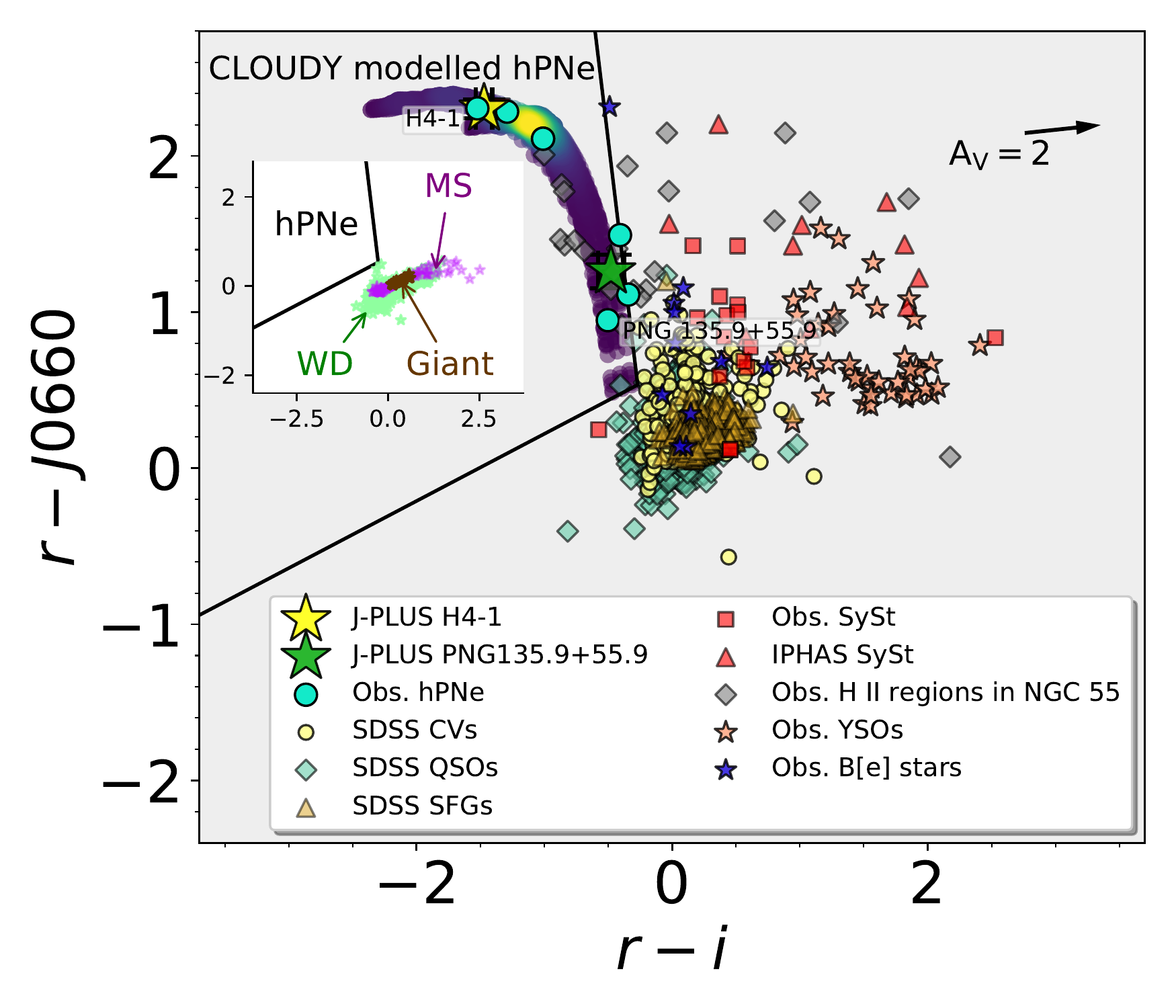}
  \caption{J-PLUS and S-PLUS ($r-J0660$) vs ($r - i$) colour-colour diagram, equivalent to IPHAS ($r' - \ha{}$) vs. ($r' - i'$). The big yellow and green stars with error-bars are the J-PLUS observations for H 4-1 and PNG 135.9+55.9, respectively. Included in the
diagrams, there are families of \textsc{cloudy} modelled hPNe spanning a range of properties (density map region). Cyan circles represent hPN  PNG 135.9+55.9 spectrum from SDSS, DdDM-1 \citep{Kwitter:1998}, NGC 2022, BB-1, H4-1 \citep{Kwitter:2003}, and MWC 574 \citep{Pereira:2007}. Grey diamonds represent H~{\sc ii} regions in NGC 55 \citep*{Magrini:2017}. Red boxes display \citet{Munari:2002a, Munari:2002b} SySt, this group also includes external SySt from NGC 205 \citep{Goncalves:2015}, IC~10 \citep{Goncalves:2008} and NGC 185 \citep{Goncalves:2012}, and red triangles correspond to IPHAS symbiotic stars \citep{Rodriguez:2014}. Yellow circles correspond to cataclysmic variables (CVs) from SDSS. Orange triangles refer to SDSS star-forming galaxies (SDSS SFGs). SDSS QSOs at different redshift ranges are shown as light blue diamonds, and YSOs from Lupus and Sigma Orionis \citep{Rigliaco:2012, Alcala:2014} are represented by salmon stars. Blue stars refer to B[e] stars from \citet{Lamers:1998}. In the inset plot, the synthetic main sequence and giant stars loci from the library of stellar spectral energy distributions (SEDs, \citealp{Pickles:1998}) are represented by the purple and brown symbols, and the loci of white dwarf stars observed by S-PLUS (DR1, \citealp{Mendes:2019}) are represented by the light green symbols. The limiting region applied in the candidate selection is shown by black lines for hPNe. The arrow indicates the reddening vector with A$_{\mathrm{V}} \simeq$ 2 mag. It was estimated by comparing the locus of unreddened models of PNe with the reddened ones.} 
  \label{fig:Viironen}
\end{figure}

\citet{Drew:2005} designed the IPHAS survey (see Section \ref{sec:into}) showing that the $(r' - \ha{})$ vs $(r' - i)$ diagram is an optimal tool to select strong \ha{} emission-line sources \citep{Witham:2008}.
 IPHAS PNe candidates were selected using the above colour-colour diagram as well as the common $(J - H)$ vs ($H - K_s)$ 2MASS diagram. Their true nature was confirmed later on, spectroscopically \citep{corradi:2008, Viironen:2009a, Rodriguez:2014, Sabin:2014}.
 
Given that the J-PLUS and S-PLUS filter systems include the three IPHAS filters, we reconstructed the IPHAS equivalent diagram in Figure~\ref{fig:Viironen}. The $(r - J0660$) colour clearly indicates an increasing excess of the \ha{}~emission line, while the $(r - i)$ one increases with the reddening, as previously stated in \citet{corradi:2008}. Modelled PNe are found to display $(r-J0660)>0.6$ due to their strong \ha{} emission lines as well as $(r-i)<0$. It should be noted that Figure 1 of \citet{Viironen:2009a} shows that Galactic disc PNe have $(r-i)<1$, so they are redder than the hPNe. The empirical (black) lines delimit the selection criteria for being PNe (PNe zone) and are meant to include most of the modelled hPNe and exclude the majority of the contaminants. The definition of the selection criteria or zones has been made visually. This holds for the IPHAS equivalent as well as the new colour-colour diagrams of the present study. The PNe zone is contaminated by extragalactic H~{\sc ii} regions, especially for moderately lower $(r - J0660)$ colour indices or, equivalently, low-excitation nebulae. Therefore, our lists of PN candidates are not characterised by high purity. The distribution of main sequence (MS) and giant stars (purple and brown symbols in the inset figures) in the ($r$-$J$0660) vs ($r$-$i$) diagnostic diagram is such that both groups of stars occupy regions with lower ($r$-$J$0660) index (<1.2) and higher ($r$-$i$) index (>0) than the PNe zone (\citealp{Drew:2005, corradi:2008}). White dwarf (WD) stars (light green symbols in inset figures) also present low ($r$-$J$0660) index (>0), however exhibit ($r$-$i$) colour $>-1$. Thus, there is no mixing between PNe and the bulk of MS, giant and WD stars. 
The observed PNe are well located in the PNe zone with high $(r-J0660)>1$ (see also \citealp{Viironen:2009a, Viironen:2009b}).  

\subsection{New J-PLUS and S-PLUS colour-colour diagrams}
\label{sec:new}

Studying the photo-spectra of the PNe  and their contaminants (Section~\ref{sec:syn}), four new colour-colour diagrams were defined
($J0515 - J0660$) vs ($J0515 - J0861$), ($g - J0515$) vs ($J0660 - r$), ($z - J0660$) vs ($z - g$), and 
($J0410 - J0660$) vs. ($g - i$ ) to search for PN candidates in the J-PLUS and S-PLUS catalogues. 

In the first diagnostic diagram, high- and low-excitation PNe occupy  distinct regions within the PNe zone. For instance, H~4-1, a moderate high-ionisation PN, exhibits $(g - J0515)\approx-2$ , with T$_{eff}$ around 132$\times 10^3$ K (\citet{Henry:1996}), and MWC~574, a very low-excitation PN with $(g - J0515)\approx0.0$, exhibits an \oiii{}  emission even weaker than H$\beta$ \citep{Sanduleak:1972, Pereira:2007}. The ($J0515 - J0861$) colour index is associated with the continuum of the sources. Neither of these two filters contains strong emission lines. The ($J0515 - J0660$) colour provides an estimation of the \ha{} excess.

\begin{figure*}
\setlength\tabcolsep{\figstampcolsep}
\centering
\begin{tabular}{l l}
 \BowshockFig{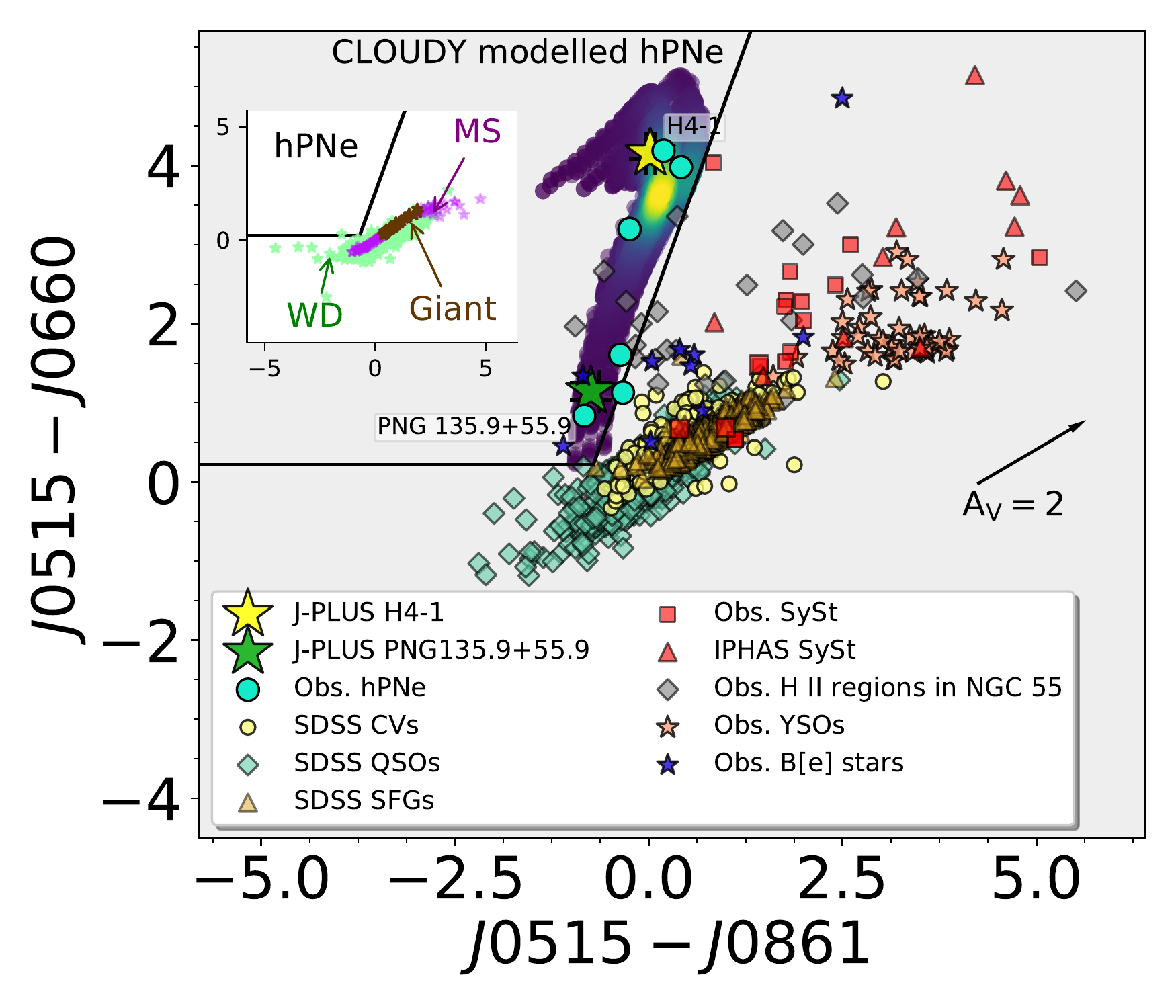} & \BowshockFig{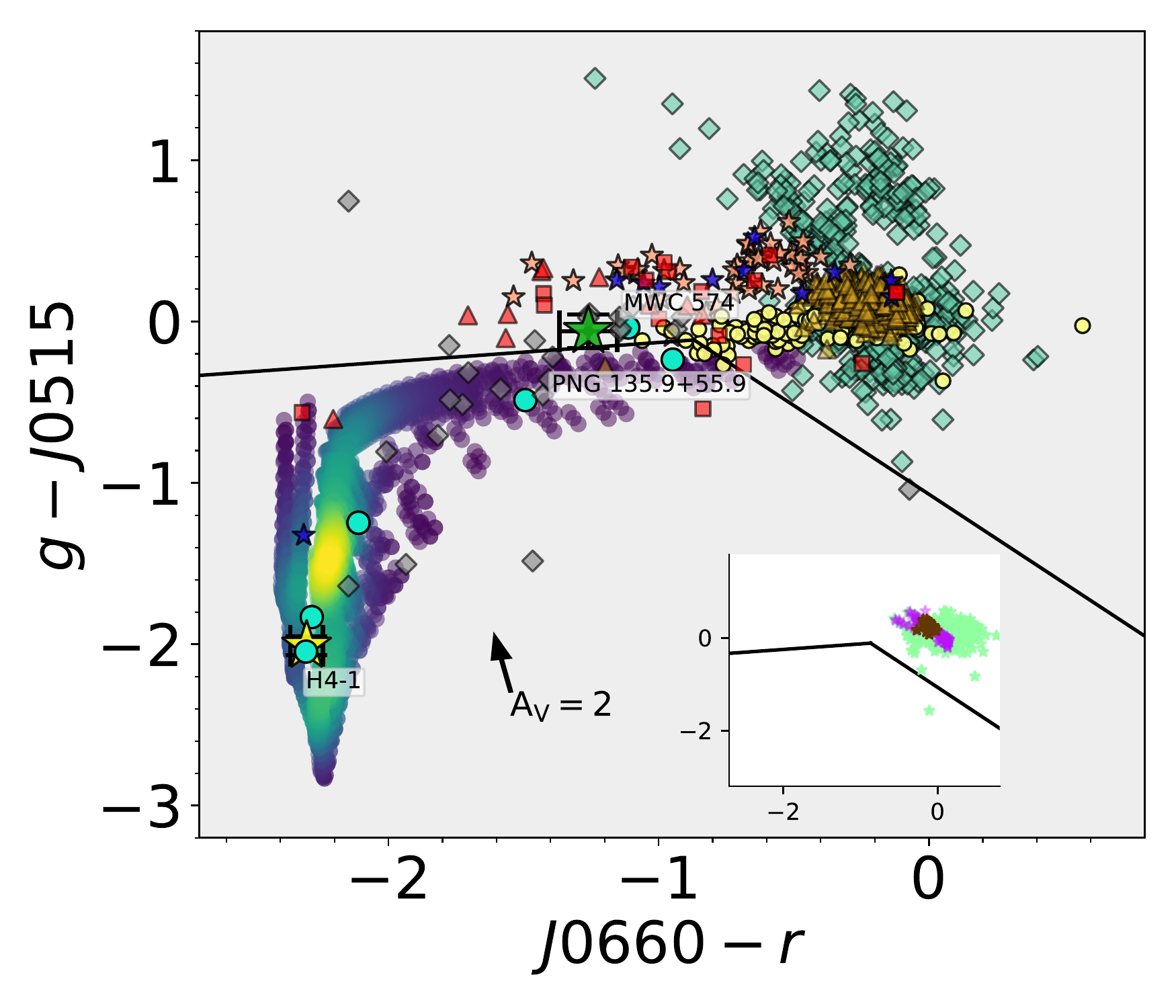} \\
\raiselabel{(\textit{a})} & \raiselabel{(\textit{b})}\\
\BowshockFig{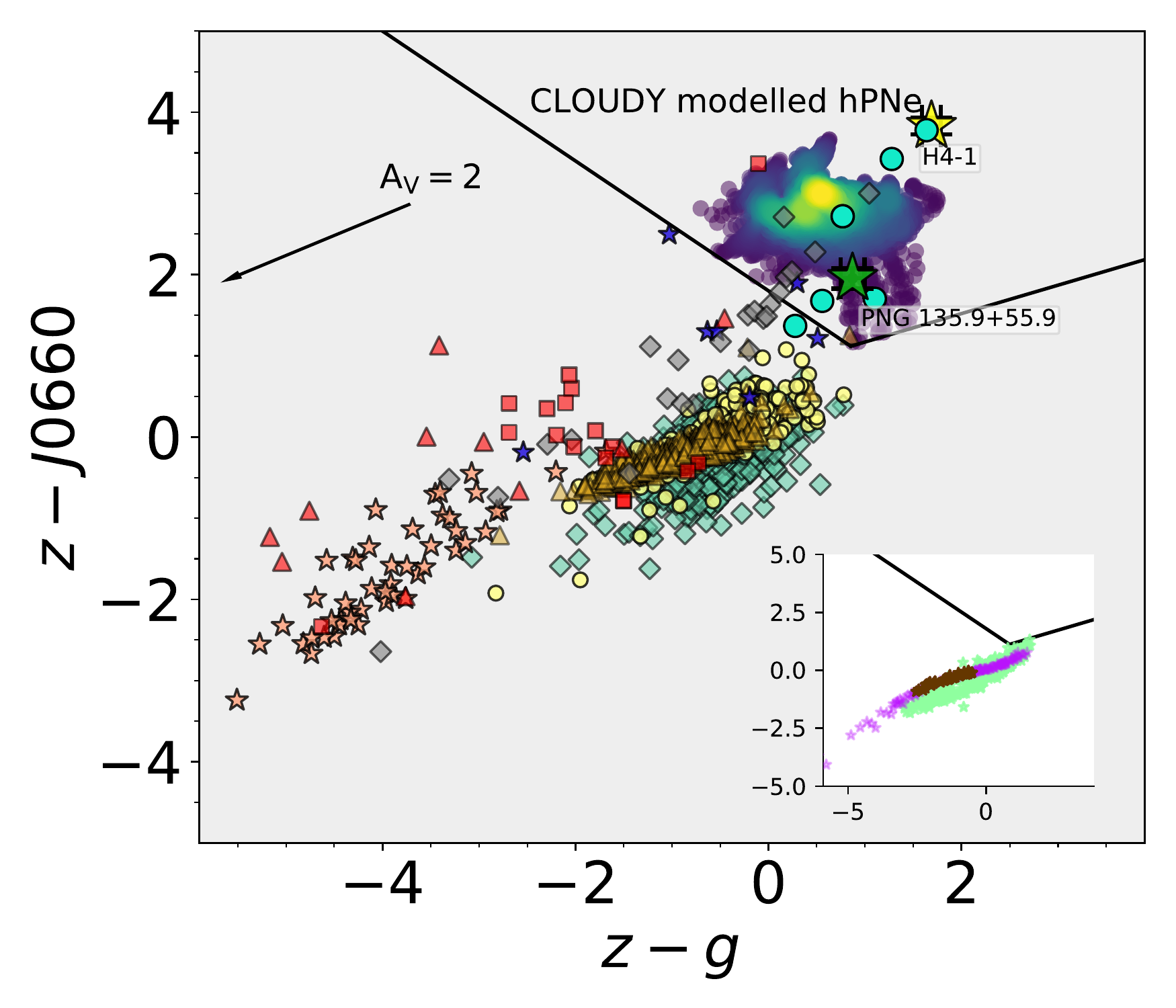} & \BowshockFig{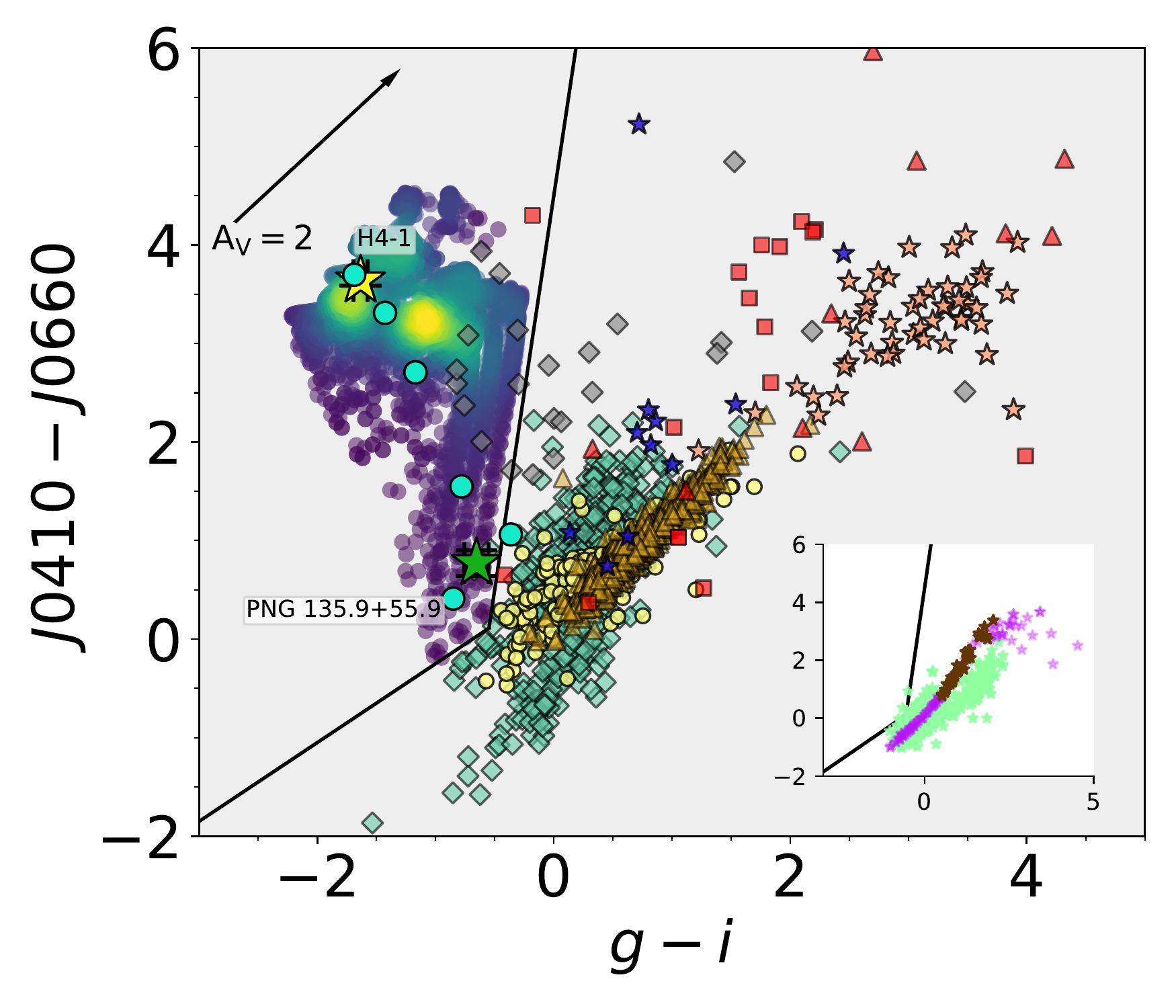} \\
\raiselabel{(\textit{c})} & \raiselabel{(\textit{d})}
  
  \end{tabular}
  \caption{J-PLUS and S-PLUS (\textit{a}) ($J0515 - J0660$) vs ($J0515 - J0861$), (\textit{b}) ($g - J0515$) vs ($J0660 - r$), (\textit{c}) ($z - J0660$) vs ($z - g$), and (\textit{d}) ($J0410 - J0660$) vs ($g - i$) colour-colour diagrams. The error bars are smaller than the symbols. The arrows  indicate the reddening vectors with A$_{\mathrm{V}} \simeq$ 2 mag. The symbols are the same as in Figure \ref{fig:Viironen}.}
  \label{fig:new-colours}
\end{figure*}

Regarding the second diagnostic diagram, the ($g - J0515$) colour index yields the excess of the \oiii{} lines. Therefore, the higher the ($g - J0515$) colour, the higher the excitation of the nebula. This can be seen from the photo-spectra of two {\sc cloudy} models with different stellar temperatures (panels $c$ and $d$ in Figure \ref{fig:phot-spect}). We remind the reader, however, that the \textit{g} broad-band filter is also affected by the H$\beta$ and other emission lines down to about 4,000\AA. The ($J0660 - r$) colour is to illustrate the \ha{} emission as it has been explained in the previous section.  

The ($z - J0660$) colour index also gives us an estimation of the \ha{} excess, by taking into account the continuum at the very red part of the spectrum, while the ($z - g$) colour is directly related to the shape of the continuum from the blue to the red part of the spectrum. The QSOs and SFGs are concentrated in a very small area ($-2 < (z - J0660) < 1$ and $-2 < (z - g) < 1$), well separated from PNe.  

Finally, ($g - i$) is another colour index associated with the shape of the continuum, and the ($J0410 - J0660$) colour reflects the \ha{} excess, though using different parts of the spectrum in contrast with the previous diagrams. From panel $d$ in Figure~\ref{fig:new-colours}, we find that QSOs and SFGs show an increase in the ($J0410 - J0660$) colour as a function of the ($g - i$) colour, while PNe appear to have a more restricted range of ($g - i$) values between -2 and 0, and ($J0410 - J0660$) values from 0 to 4. Hence, the ($J0410 - J0660$) vs ($g - i$) colour-colour diagram can distinguish PNe from galaxies. The SySt and YSOs are found to lie in the right part of the plot ($(g - i) > 2$), which is indicative of their stronger emission in the near-IR. This last diagnostic diagram turns out to be very useful to assure low contamination by B[e] stars, differing from the first three diagrams.  
 
Figure \ref{fig:new-colours} and the colour-colour diagrams also show the loci of main sequence, giant and white dwarf stars, which clearly do not affect the PN selection criteria. Leaving aside the H~{\sc ii} regions, all the colour-colour diagrams turned out to be very useful for identifying good PN and SySt candidates (a detailed analysis for SySt is to be presented in a forthcoming paper). The combination of several narrow- and broad-band filters to construct colour-colour diagrams allows characterising the whole optical spectrum of every source type. If a source simultaneously satisfies all the criteria of being a PN, the possibility of false positive identification, though not negligible, is small.

\section{Validation of the colour-colour diagrams}
\label{sec:valid}

Aiming to validate the tools to search for PNe in J-PLUS and S-PLUS, we followed two different methods. The first is to use the J-PLUS SVD (Section~\ref{sec:svd}) and the second is by applying the previously presented selection criteria (Section~\ref{sec:colour}) to the DR1 (very limited in area) of each of these surveys. 

\subsection{SVD validation}
Two hPNe, H~4-1 and PNG~135.9+55.9, were observed during the scientific verification phase of the J-PLUS, as described in Section \ref{sec:svd}. The instrumental magnitudes were calculated for each object using IRAF\footnote{Image Reduction and Analysis Facility; \url{http://iraf.noao.edu/}}. In order to find the calibrated magnitudes, we used the zero point provided by the unit of processing and data analysis (UPAD) inside the J-PLUS collaboration. The resulting photo-spectra are shown in Figure~15 of the J-PLUS presentation paper \citep{Cenarro:2019}. In Table~\ref{tab:magnitudes}, we compare the synthetic and observed magnitudes of these hPNe. For H~4-1, the two magnitudes are very compatible, with a difference smaller than 0.07, except for the \textit{u} and \textit{J}0378 filters, whose differences are 0.4 and 0.23, respectively. As for PNG~135.9+55.9, the synthetic and observed magnitudes are found to be in very good agreement, with, on average, a difference smaller than 0.1 mag. The only discrepant magnitude is the one of the \textit{J}0660 filter, for which the  difference reaches 0.35\,mag.  

\begin{table}
  \caption{Synthetic and observed magnitudes of H~4-1 and PNG 135.9+55.9, for the 12 J-PLUS filters.}  
  \label{tab:magnitudes}
  
  \centering 
  \begin{tabular}{l l l l l}
\hline\hline 
\multicolumn{1}{l}{} & \multicolumn{2}{l}{H 4-1}& \multicolumn{2}{l}{PNG 135.9+55.9} \\ 
\hline\hline  

   Filter & Synt. & Obs. &  Synt. & Obs.\\ 
   \hline
   \textit{u} & 15.74  & 16.14 $\pm$ 0.13& 17.37 & 17.22 $\pm$ 0.10\\ 
  \textit{J}0378 & 15.54 & 15.77 $\pm$ 0.02 & 17.36 & 17.31  \\ 
  \textit{J}0395 & 17.01 & 16.97 $\pm$ 0.06& 17.35 & 17.28 $\pm$ 0.12 \\ 
  \textit{J}0410 & 17.14 & 17.11 $\pm$ 0.03& 17.43 & 17.43  $\pm$ 0.10\\
  \textit{J}0430 & 16.61 & 16.68 $\pm$ 0.07& 17.48 & 17.48 $\pm$ 0.13\\
\textit{g} & 15.59 & 15.61 $\pm$ 0.04& 17.62  & 17.74 $\pm$ 0.07\\
\textit{J}0515 & 17.63 & 17.62 $\pm$ 0.04 &17.86 & 17.80  $\pm$ 0.07\\
\textit{r} & 15.75 & 15.77 $\pm$ 0.04 &17.97 & 17.92 $\pm$ 0.07\\
\textit{J}0660 & 13.45 & 13.47 $\pm$ 0.04 & 17.02 & 16.66 $\pm$ 0.08 \\
\textit{i} & 17.28 &17.24 $\pm$ 0.05 & 18.47 &  18.40 $\pm$ 0.07\\ 
\textit{J}0861 & 17.44 & 17.60  $\pm$ 0.03 & 18.69 & 18.55 $\pm$ 0.07\\ 
\textit{z} & 17.23& 17.30 $\pm$ 0.09 & 18.72  & 18.61 $\pm$ 0.10\\
  \hline                                   
\end{tabular}
\end{table}

The positions of H~4-1 and PNG~135.9+55.9 in the colour-colour diagrams --based on their J-PLUS observations-- are shown in Figures \ref{fig:Viironen} and \ref{fig:new-colours} as yellow and green stars, respectively. H~4-1 satisfies all five criteria of being a PN (see Figures \ref{fig:Viironen} and \ref{fig:new-colours}), while PNG~135.9+55.9 passes four of them, violating the (\textit{g}-\textit{J}0515) colour criterion given in panel \textit{(b)} of Figure~\ref{fig:new-colours}. We note, however, that this PN is located close to the border of the PN zone. This border was defined in a conservative way, to avoid more contamination of H~{\sc ii} regions and SySt, since low-excitation PNe and H~{\sc ii} regions are extremely hard to differentiate.

\subsection{DR1 validation}
\label{sec:result}

In order to apply the colour-colour diagrams to the J-PLUS and S-PLUS data releases, 
there are a few cleaning instructions (FLAGS) that need to be considered, since they avoid artefacts. For J-PLUS, we used two FLAGS. One responsible for excluding the objects that have neighbours, bright and close enough to significantly affect the photometry, or that have bad pixels (more than 10$\%$ of the integrated area affected). The other FLAG allows us to exclude objects originally blended with another one. In the case of S-PLUS, only one FLAG was applied, which avoids the objects for which the aperture photometry is likely biased by neighbour sources or by more than 10\%\ of bad pixels. Note that not considering the sources flagged with these known issues affects the selection, because of the quality of the photometry. 

To encounter the best way of applying all five selection criteria described in Section~\ref{sec:colour}, we follow a few validating steps, using the J-PLUS data set. To ensure the recovery of the largest amount of potential candidates due to lack of detection in some filters, we explore how the limiting magnitude errors for different sets of filters affect the results: (i) by limiting the error associated with the magnitudes corresponding to IPHAS-like magnitudes ($i$, $r$ and $J$0660), and admitting any error for the other 9 photometries, (ii) by limiting the magnitude errors of the $J$0660 and broad-band filters, and finally, (iii) by limiting the magnitude errors of all narrow- and broad-band filters. Whenever it applies, the limiting magnitude error is  $\pm$0.2. All the procedure was performed with the 6~$\arcsec$ aperture.

Since option (i) is the least restrictive of all, it returned a higher number of sources, nine in total. With this option, we are allowing sources that are faint in any other band (high uncertain photometry) or even not detected at all, except in $i$, $r,$ and $J$0660 bands. Among the selected sources, which are known objects and classified either by the Set of Identifications, Measurements and Bibliography for Astronomical Data (SIMBAD) or the NASA/IPAC Extragalactic Database (NED), there are four \hii~regions, two \hii~galaxies, one WR star, and one PN. An additional source (NED/SDSS \lq extended or galaxy') that also matches all criteria, but for which a detailed search in the literature does not show any further information (not even redshift or recession velocity), was identified as a PN candidate. 

To test whether further limitations would minimise the number of contaminants, we use step (ii) of the validation process. This option accounts for the fact that the continuum of PNe is usually faint, with the consequence of uncertain photometry (large errors) in the narrow-band filters. Therefore, this time we limited the errors in the \ha\ and broad-band magnitudes. Four sources are returned in this case, they are the two known \hii~galaxies, the known PN, and the PN candidate described in step (i). We note that this time no \hii~regions were selected.

As a final test, we try the most restrictive way of applying the selection criteria, as in step (iii), for which all 12 magnitude errors are limited. The result is shown in Figures \ref{fig:Viironen-apply} and \ref{fig:new-colours-applied}. By applying the selection criteria in this way, we identify the same four sources as in (ii). Thus, this last step emphasises the robustness of the results obtained in step (ii), and proves that the imposed error limitation does not compromise the analysis. 

No PN candidate is identified in S-PLUS DR1. We find that one true, one likely, and three possible PNe from the HASH catalogue were observed by either J-PLUS or S-PLUS. All these objects are presented in the selection diagrams (Figure~\ref{fig:Viironen-apply} and Figure~\ref{fig:new-colours-applied}), and discussed in the rest of this section as well as in Section~\ref{sec:hash}.

\subsection{Photometry and spectroscopy of the PN candidate}
\label{sec:photometry}

\begin{figure}
  \includegraphics[width=\linewidth, trim=10 10 10 10]{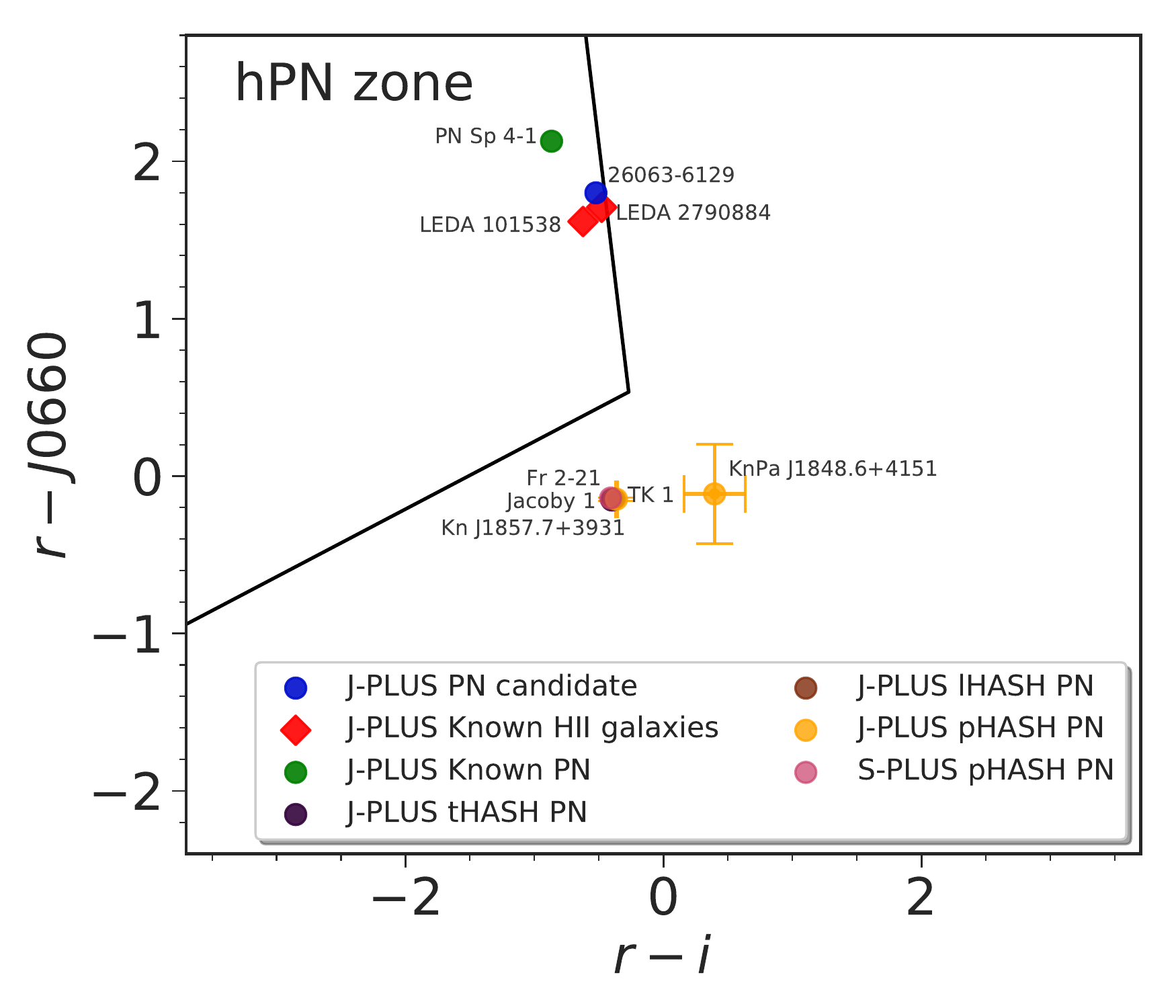}
  \caption{J-PLUS $(r-J0660)$ vs $(r - i)$ colour-colour diagram with the objects selected from J-PLUS DR1. The blue circle represents the PN candidate with ID 26063-6129, the green circle is the known PN Sp 4-1, the red diamonds are the H~{\sc ii} galaxies LEDA 2790884 and LEDA 101538. The matches of the J-PLUS and S-PLUS data with the HASH catalogue are also presented in the diagrams. In agreement with the classification of HASH PN database. The purple circle is the true PN (tHASH PN) Jacoby~1, the brown circle is the likely PN (lHASH PN) TK 1 and the orange circles represent the possible PNe (pHASH PN) Kn J1857.7+3931, and KnPa J1848.6+4151. These objects were observed by J-PLUS. The pHASH PN Fr 2-21 is indicated by the magenta circle. This source was found in S-PLUS DR1. See text for further details. The error bars are smaller than the symbols.}
  \label{fig:Viironen-apply}
\end{figure}

\begin{figure}
\setlength\tabcolsep{\figstampcolsep}
\centering
\begin{tabular}{l l}
 \BowshockFigg{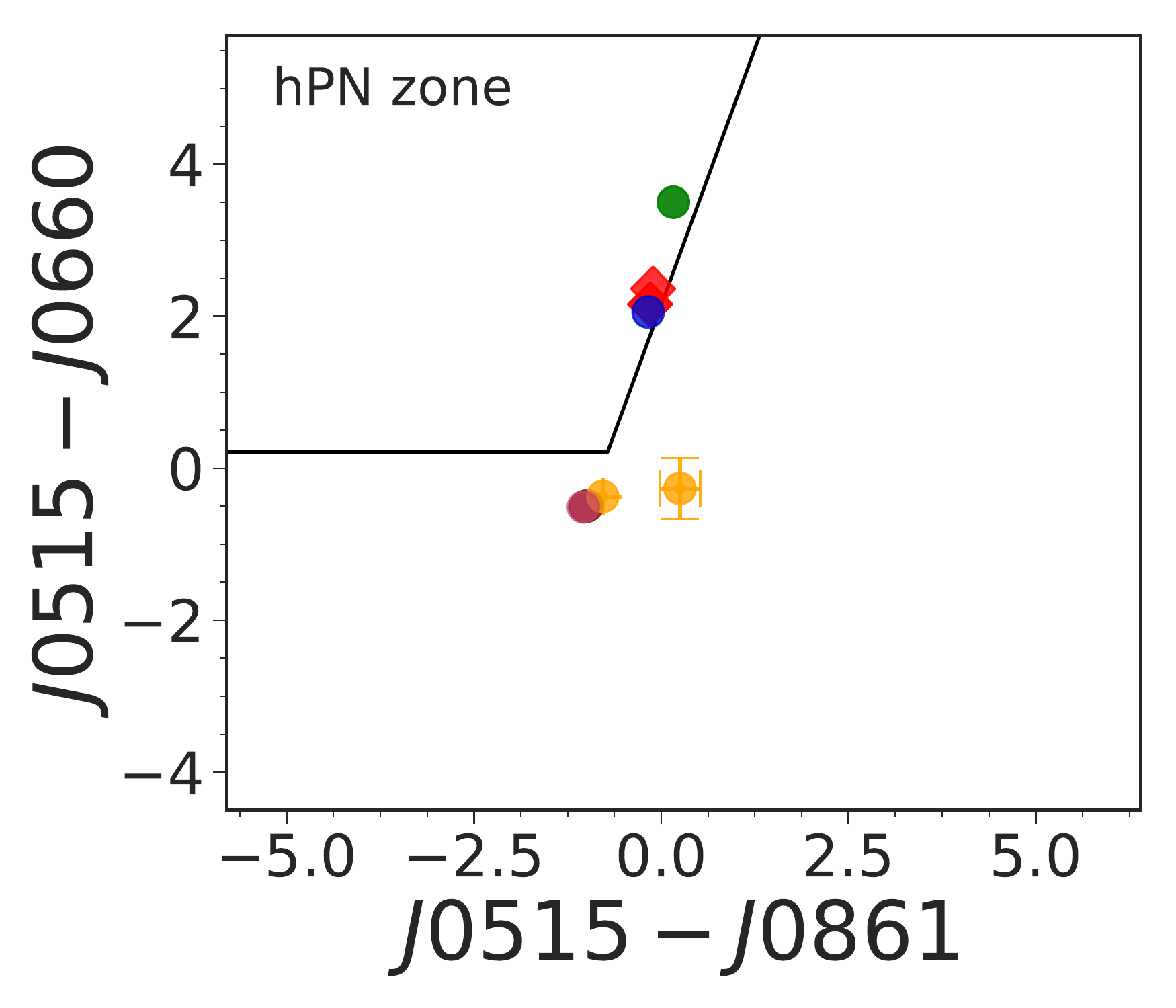} & \BowshockFigg{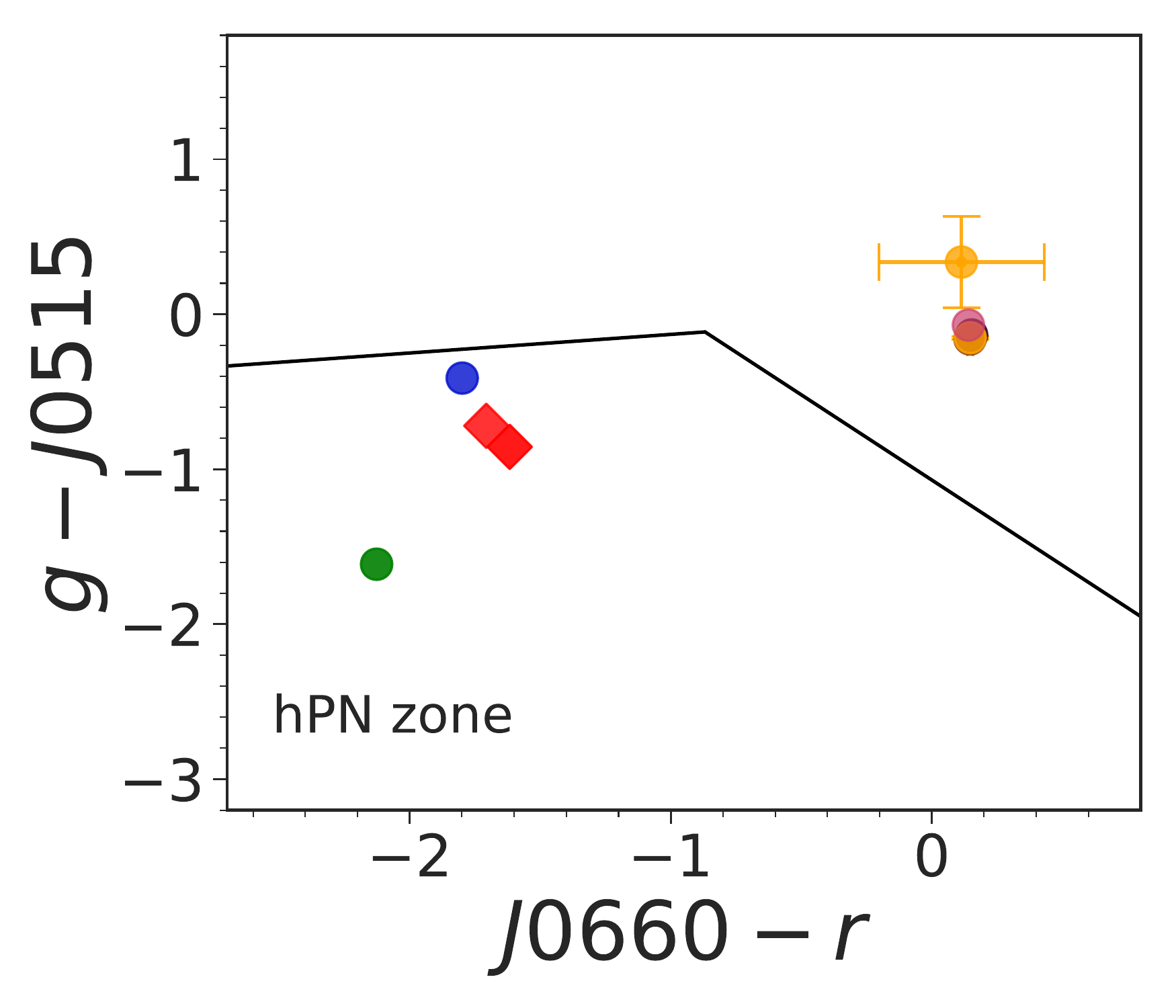} \\
\raiselabell{(\textit{a})} & \raiselabell{(\textit{b})}\\
\BowshockFigg{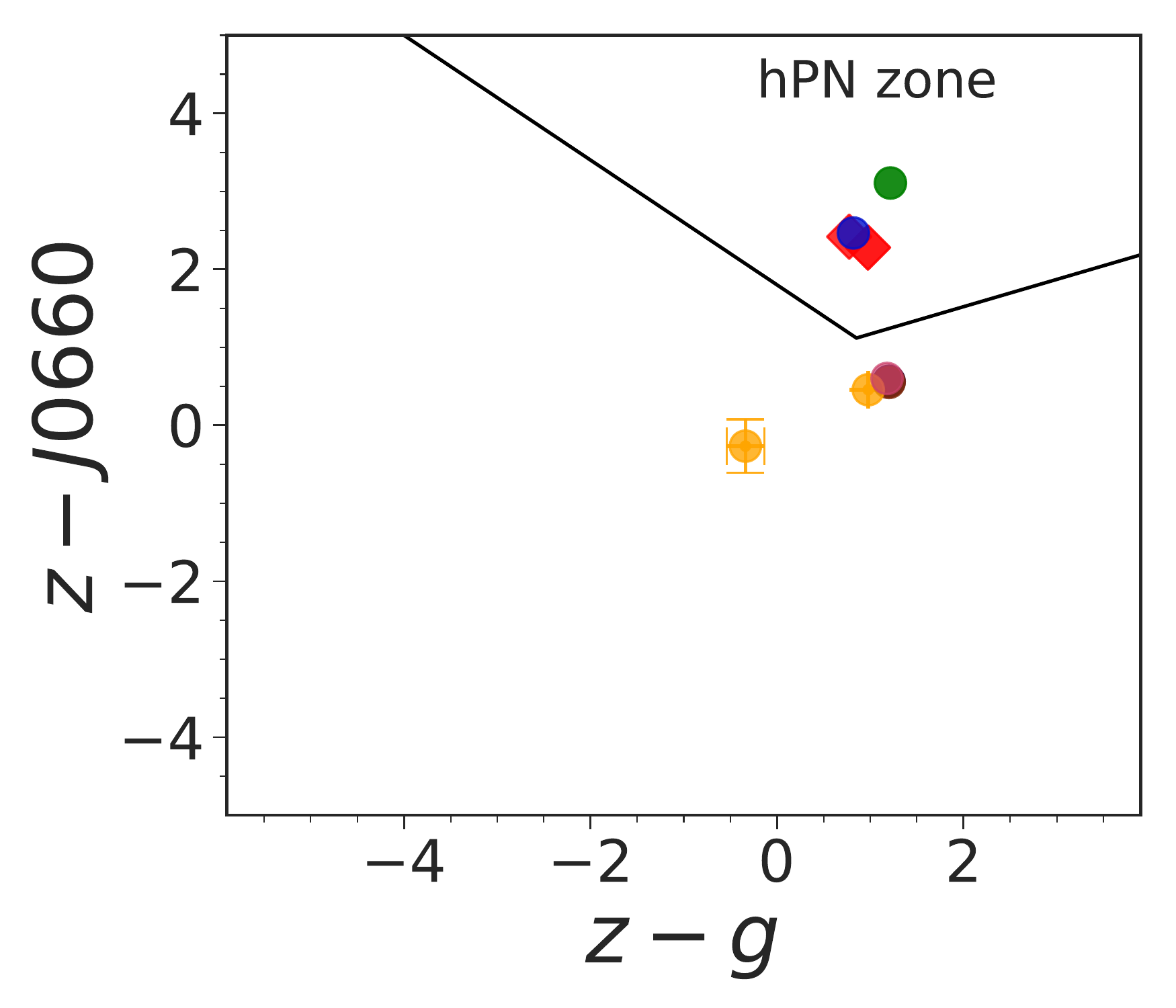} & \BowshockFigg{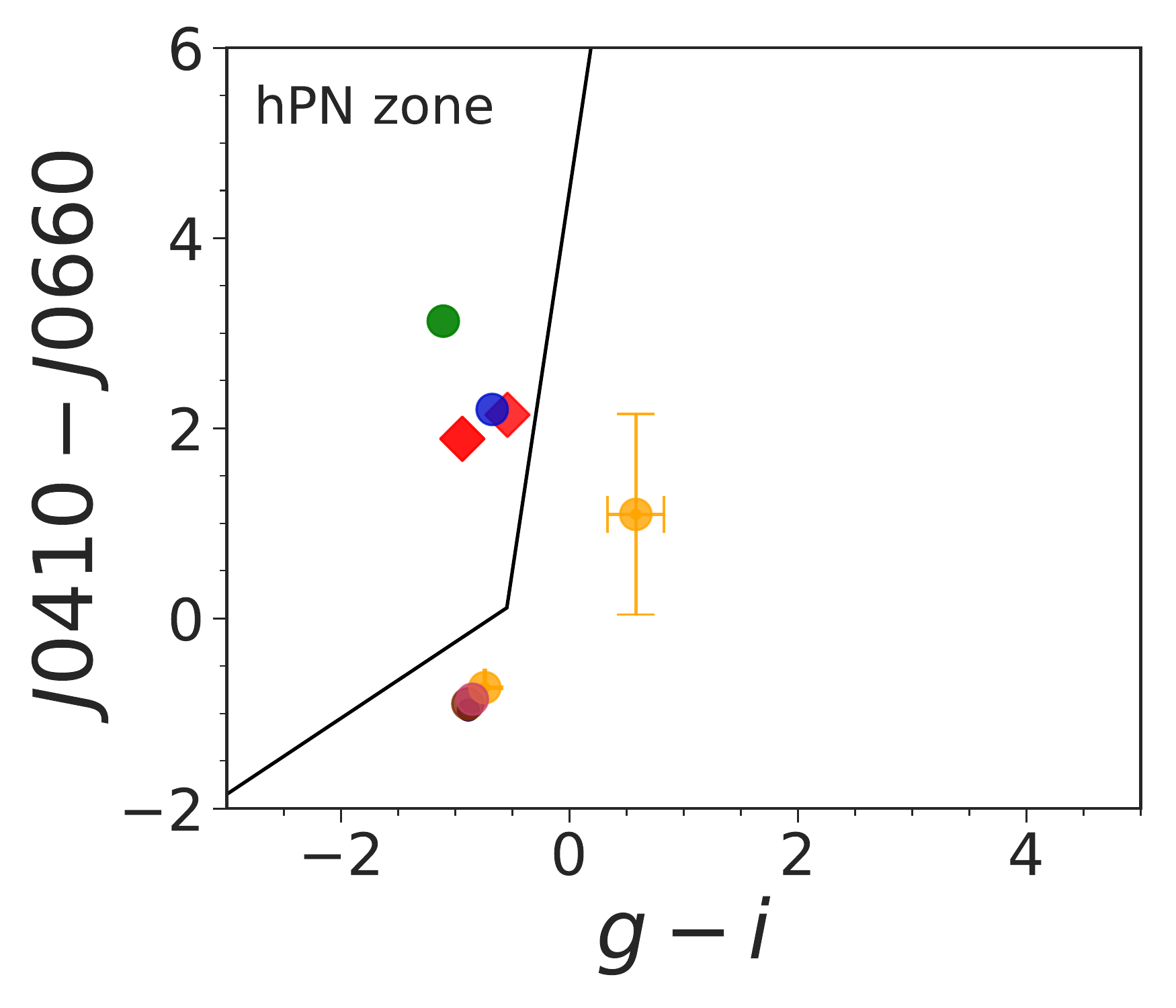} \\
\raiselabell{(\textit{c})} & \raiselabell{(\textit{d})}
  
  \end{tabular}
  \caption{ J-PLUS and S-PLUS (\textit{a}) ($J0515 - J0660$) vs ($J0515 - J0861$), (\textit{b}) ($g - J0515$) vs ($J0660 - r$), (\textit{c}) ($z - J0660$) vs ($z - g$), and (\textit{d}) ($J0410 - J0660$) vs ($g - i$) colour-colour diagrams. The symbols are the same as in Figure \ref{fig:Viironen-apply}.}
  \label{fig:new-colours-applied}
\end{figure}

\begin{figure}
  \centering
  \includegraphics[width=0.9\linewidth]{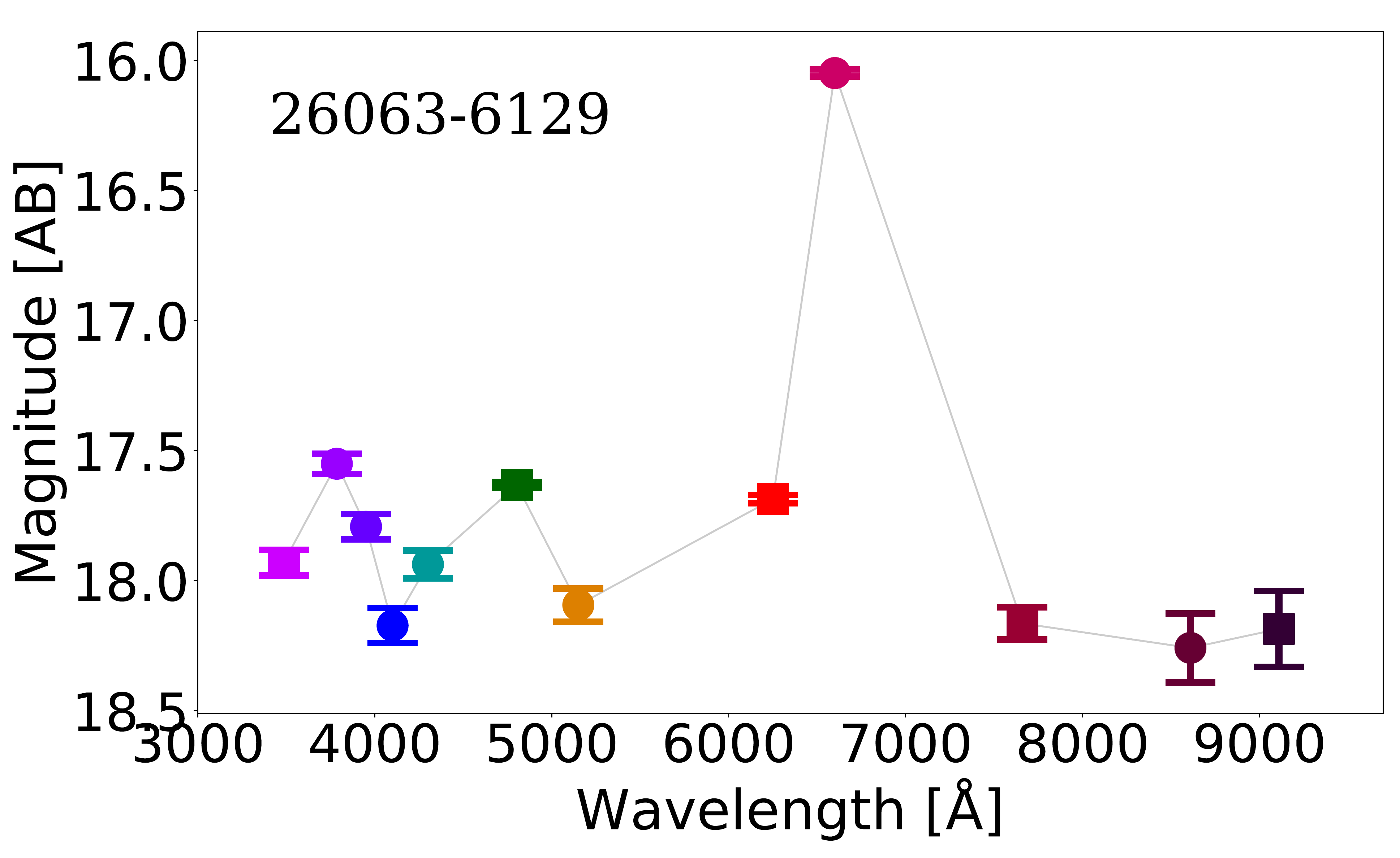}
  \llap{\raisebox{2.35cm}{\includegraphics[width=0.29\linewidth, trim=10 10 -40 0]{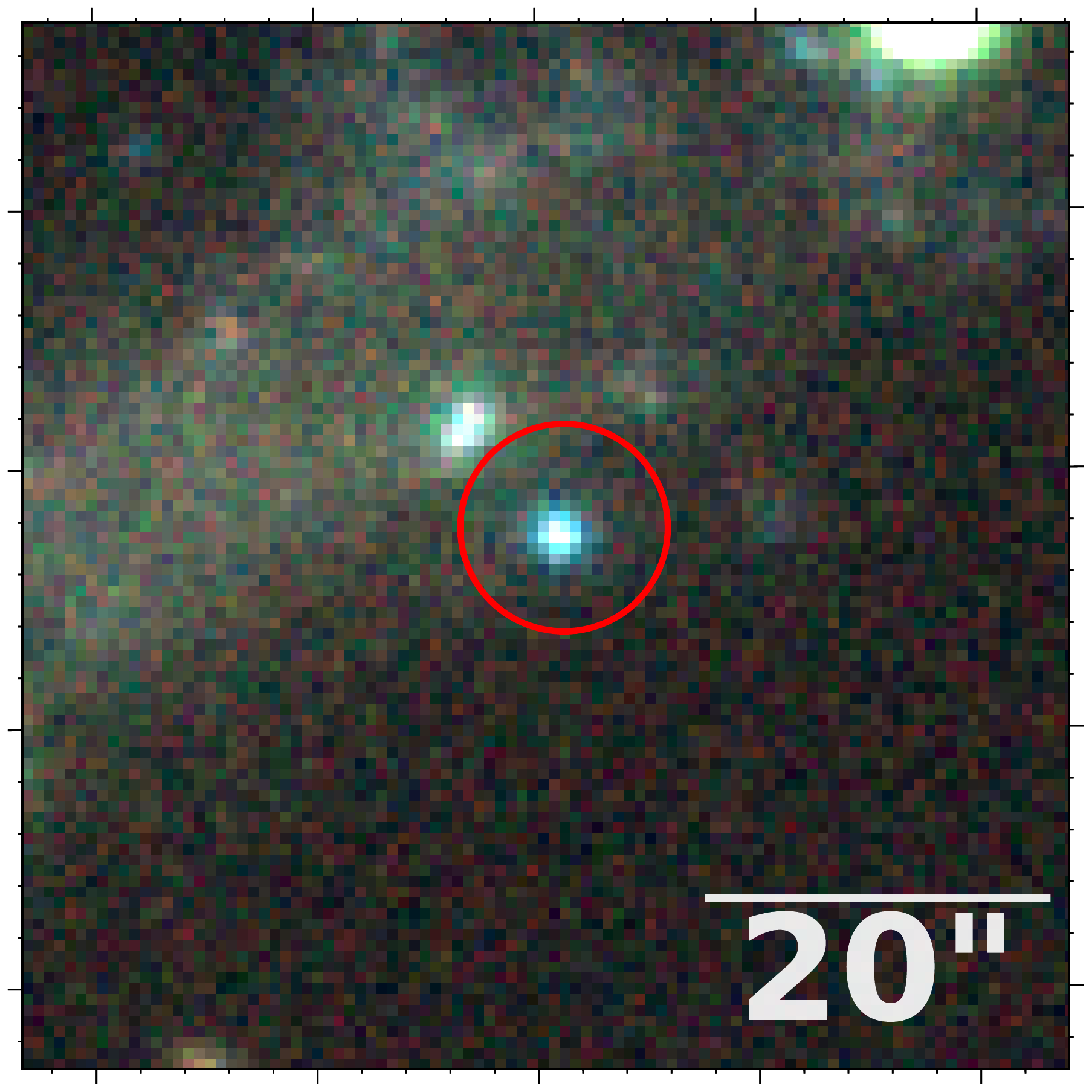}}}
  
  \caption{Photo spectrum and a combination of the $g$, $r,$ and $i$ broad-band images of the J-PLUS PN candidate.}
  \label{fig:cand-photo}
\end{figure}

The location of the PN candidate, with ID in J-PLUS DR1 26063-6129, (J2000 RA: J09 50 20.92, DEC: 31 29 11.02) is represented by the blue circle in Figures~\ref{fig:Viironen-apply} and \ref{fig:new-colours-applied}. The source displays a clear \ha{} excess (see Figure \ref{fig:cand-photo}). Also, from the \textit{(b)}, \textit{(c),} and \textit{(d)} diagrams of Figure \ref{fig:new-colours-applied}, it is possible to argue that this candidate has moderate \oiii{} and/or H${\beta}$ emission. In fact, the $(g - J0515$), $(z - g$) and $(z - i$) colours have approximate values of -0.5, 0.5, and -0.6, respectively, indicating low to moderate contribution of the \oiii{} and/or H${\beta}$ lines to the \textit{g}-band magnitude. 

Figure \ref{fig:cand-photo} displays the PN candidate photo-spectrum, whose  shape is very similar to that of typical PNe in the J-PLUS and S-PLUS configurations (see, for instance, panels \textit{(a)} and \textit{(b)} of Figure \ref{fig:phot-spect}), with strong emission lines and relatively flat continuum. From its J-PLUS \ha{} image, we derived an angular radius of $\sim$1.5~arcsec. This figure also presents a  composite \textit{g}, \textit{r,} and \textit{i} image centred on the candidate. It clearly shows another object, a diffuse emission, which we identified as the UGC~5272 galaxy, at 7~Mpc  (\citealp{Garrido:2004}; \citep{Karachentsev:2014}). If the candidate was at the distance of UGC\,5272, it would have a size of 40-50\,pc, and it would be a H~{\sc ii} region. It could also be a Milky Way halo source, at 30 to 40~kpc, for instance, displaying a 0.2 to 0.3~pc size.

\begin{figure}
  \centering
  \includegraphics[width=0.9\linewidth]{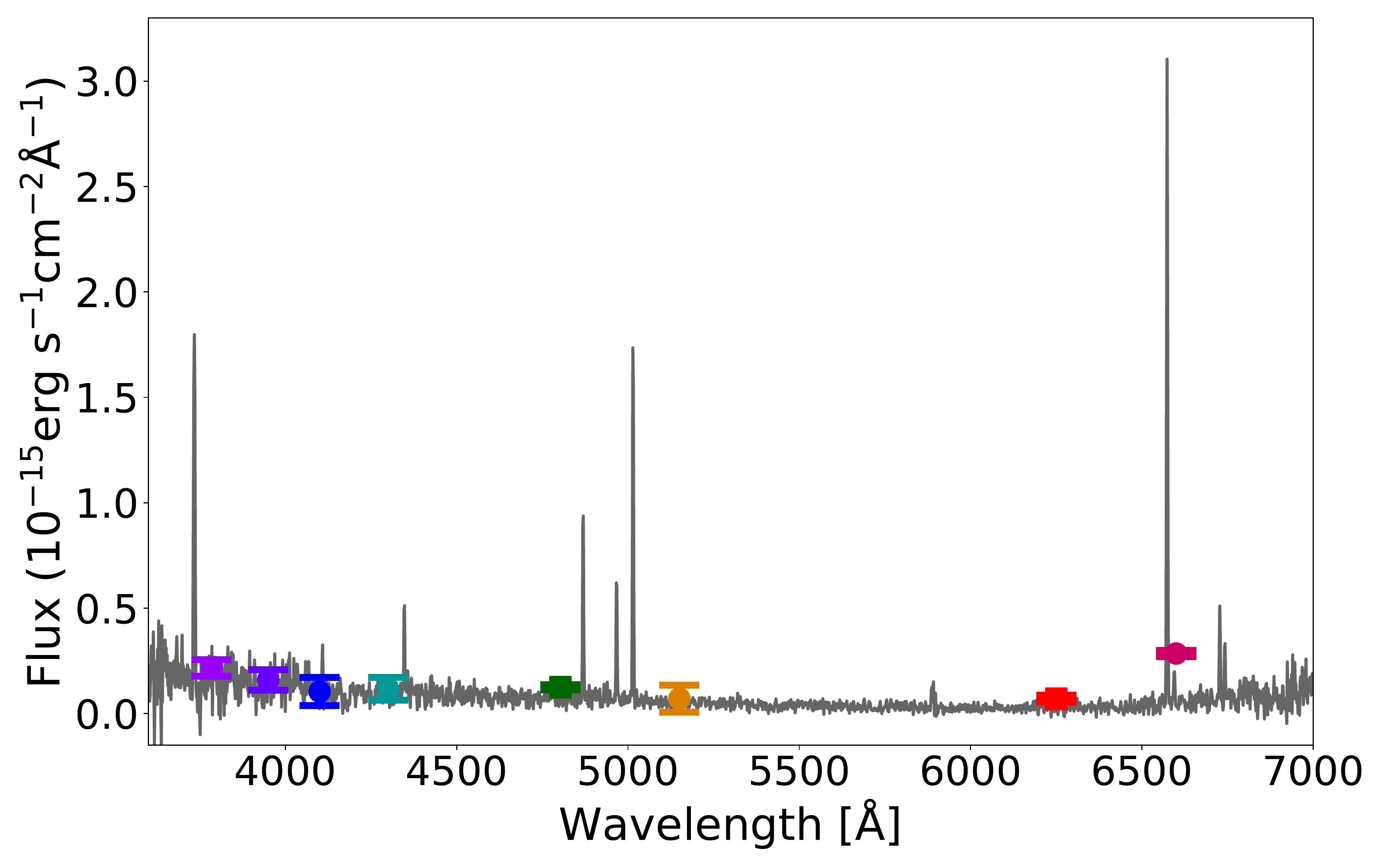}
  \caption{INT IDS spectrum (not corrected by extinction) of the J-PLUS PN source. The coloured points represent the J-PLUS photometry in flux units.}
  \label{fig:spectrum}
\end{figure}

 The PN candidate was observed with the 2.54-m INT at Roque de los Muchachos observatory in La Palma (Spain), on November 15 2018, using the Intermediate Dispersion Spectrograph (IDS). The spectrum of the candidate, together with the J-PLUS photometry, is presented in Figure \ref{fig:spectrum}. The fluxes of the emission lines detected in our spectrum are presented in Table~\ref{tab:emis-lines}. Based on the H$\alpha$, H$\beta$, and H$\gamma$ lines, its extinction is 0.35. From the observed central wavelength of the detected lines, we also derived the heliocentric velocity of the source, 515~km~s$^{-1}$. This value agrees with the heliocentric velocity of the galaxy (513$\pm$2~km~s$^{-1}$ \citealp{Garrido:2004}), and it confirms that our J-PLUS PN candidate belongs to UGC~5272. The \sii~$\lambda\lambda$6716/6731 diagnostic ratio of 1.75 suggests a very tenuous H~{\sc ii} region.

  \begin{table}
  \caption{Emission line fluxes of J-PLUS PN candidate. The five columns show, respectively, the line identification, the rest frame and observed central wavelengths, the observed fluxes, and the extinction corrected intensities.}  
  \label{tab:emis-lines}
  
  \centering 
  \begin{tabular}{l c c c c}

\hline\hline  

   Line & $\mathrm{\lambda_{\mathrm{rest}} (\AA)}$ & $\mathrm{\lambda_{\mathrm{Obs}} (\AA)}$ & F\tablefootmark{a} &  I\tablefootmark{a} \\ 
   \hline
  \oii\tablefootmark{b}         & 3727.0   & 3734.3    & 290.6  & 357.8 \\
  $\mathrm{H\delta}$ & 4101.7   & 4108.4    & 24.8  & 28.7 \\ 
  $\mathrm{H\gamma}$ & 4340.5   & 4347.3    & 41.7  & 46.2 \\ 
  $\mathrm{H\beta}$  & 4861.4   & 4869.1    & 100.0  & 100.0 \\ 
  \oiii              & 4958.9   & 4966.8    & 67.9  & 66.6 \\
  \oiii              & 5006.8   & 5014.8    & 199.4  & 193.7 \\
  \hei               & 5875.6   & 5888.9    & 29.6  & 24.9 \\
   \ha{}             & 6562.8   & 6573.5    & 373.7  & 288.5 \\
   \nii{}            & 6583.4   & 6594.4    & 22.3  & 17.2 \\
\sii                 & 6716.4   & 6727.3    & 53.2  & 40.3 \\
\sii                 & 6730.8   & 6742.0    & 30.4  & 23.1 \\ 
\hline
  F($\mathrm{H\beta}$)\tablefootmark{c} & $-$    & $-$     & 3.246 & $-$ \\
  \hline                                   
\end{tabular}
\tablefoot{ \tablefoottext{a}{In units of $\mathrm{H\beta}$ = 100.}
 \tablefoottext{b}{Blended line.}
 \tablefoottext{c}{In units of $10^{-15}$~erg~cm$^{-2}$~s$^{-1}$}.}
\end{table}

\subsection{Recovering DR1 known H${\alpha}$ emitters}
\label{sec:recover}

\indent One PN 
and two H~{\sc ii} galaxies were recovered applying our selection methodology to the J-PLUS catalogue.

\begin{figure}
\setlength\tabcolsep{\figstampcolsep}
\centering
\begin{tabular}{l l}
\includegraphics[width=0.9\linewidth, trim=10 105 10 0]{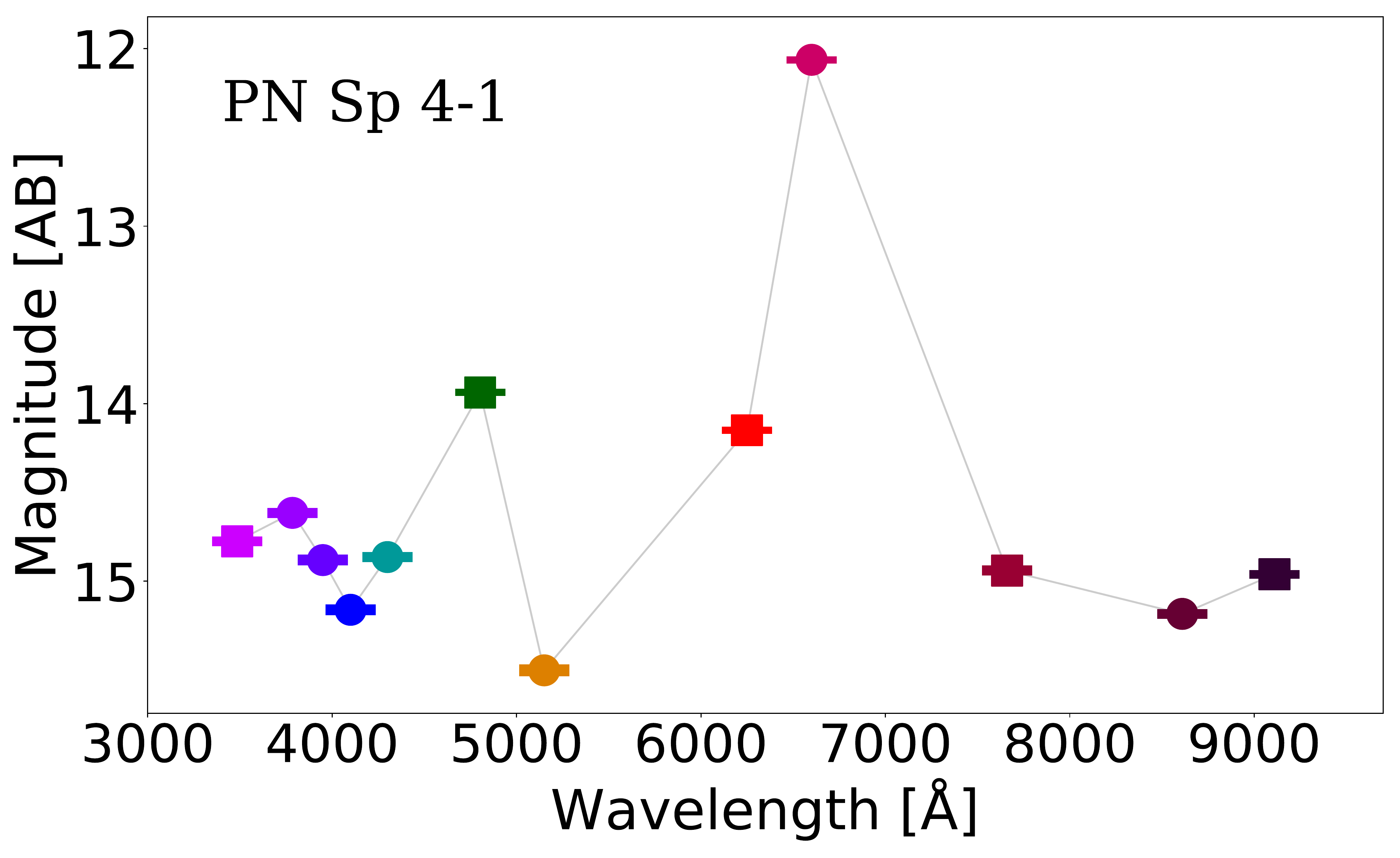}  \llap{\raisebox{1.75cm}{\includegraphics[width=0.29\linewidth, trim=10 10 -40 0]{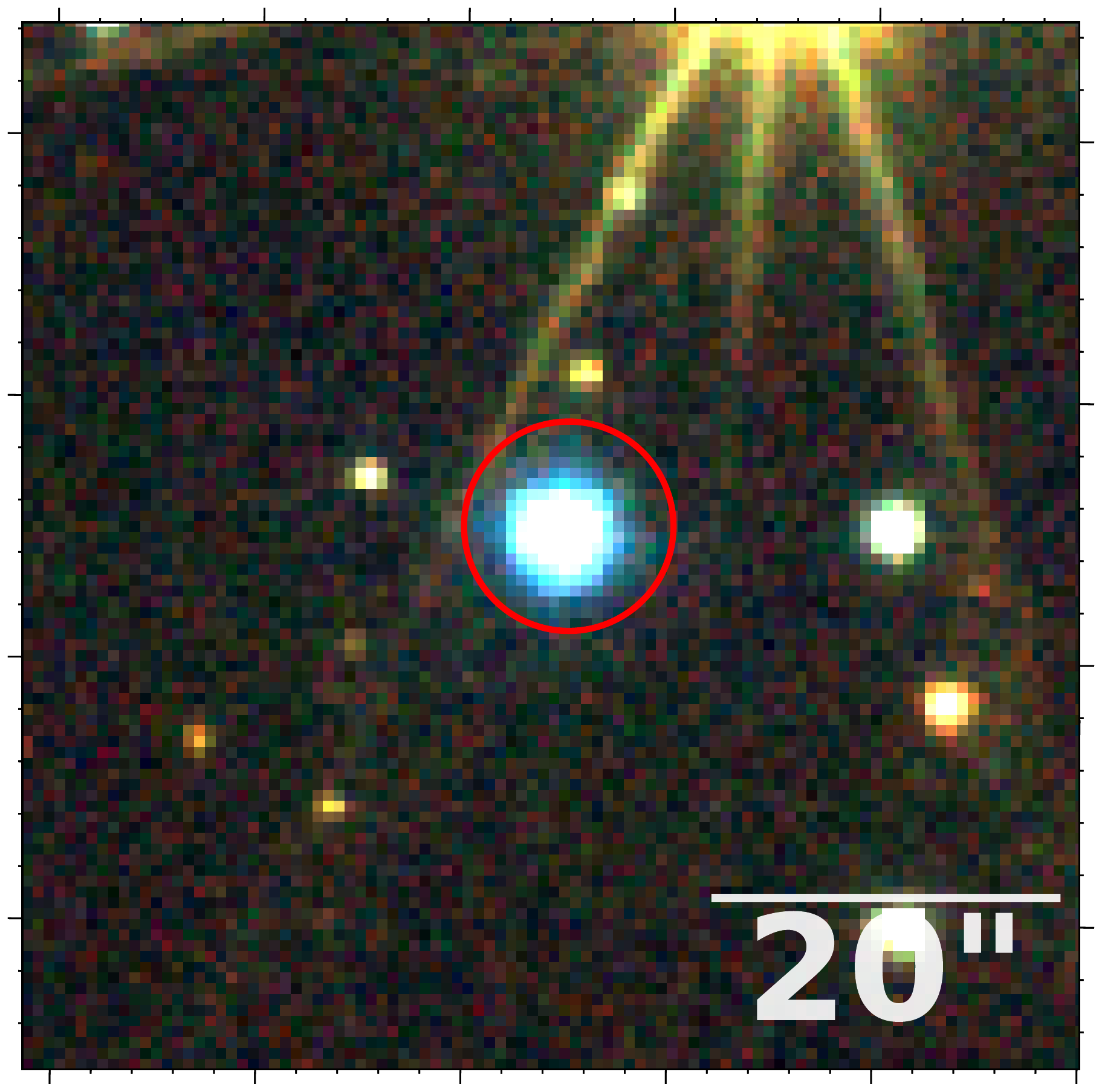}}}\\ \includegraphics[width=0.9\linewidth, trim=50 102 10 30]{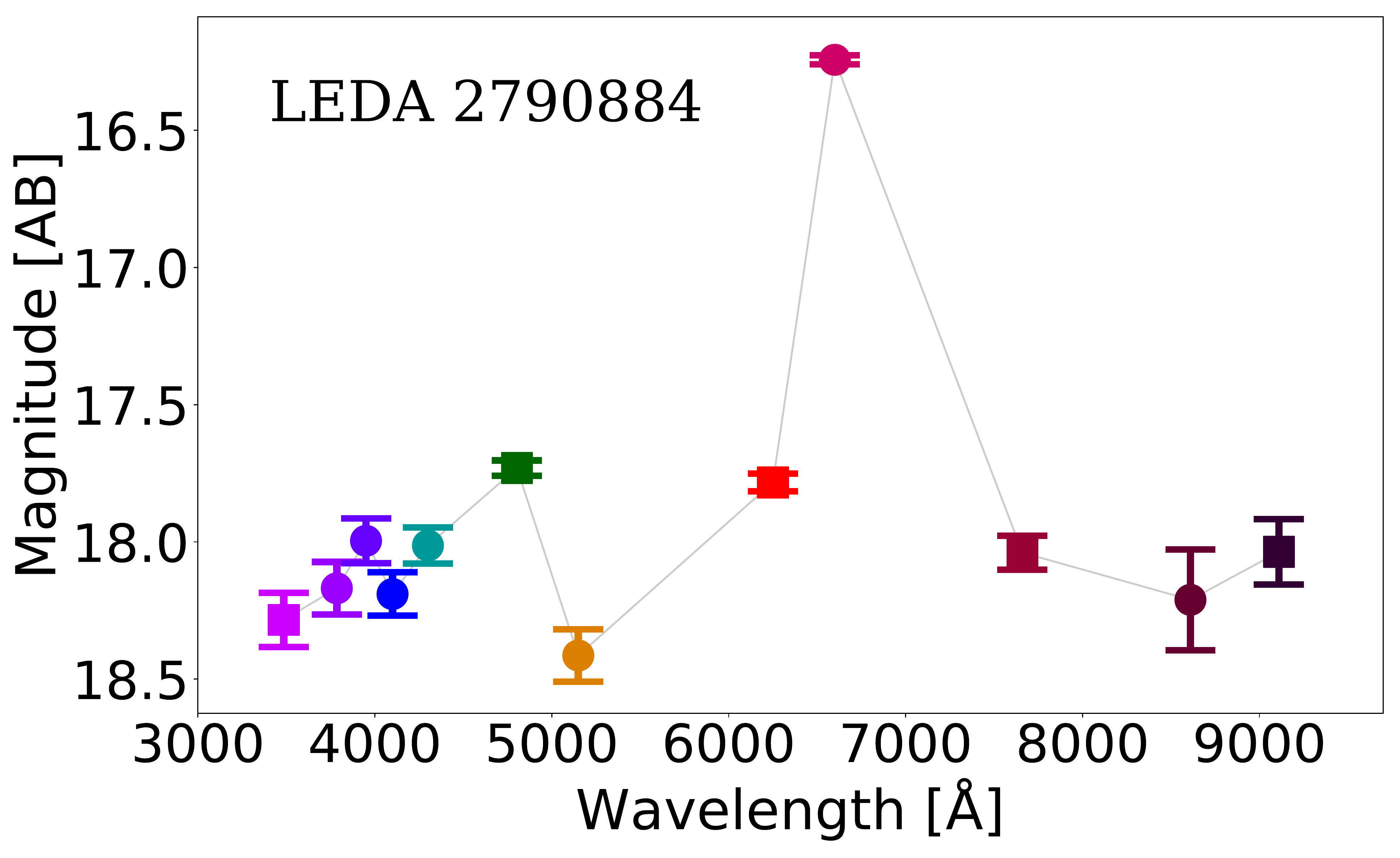}  \llap{\raisebox{1.87cm}{\includegraphics[width=0.29\linewidth, trim=10 10 -40 0]{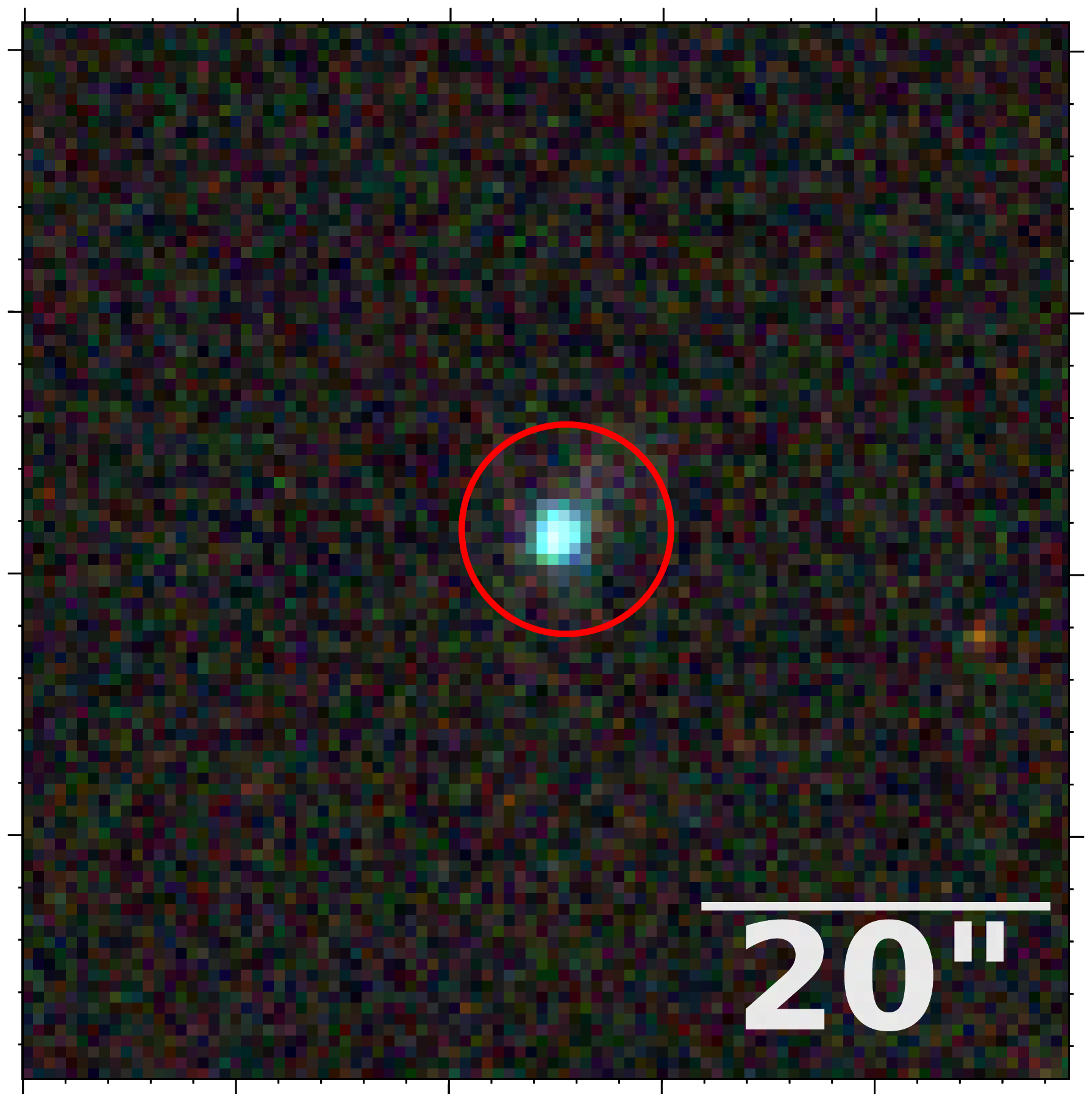}}} \\ \includegraphics[width=0.9\linewidth, trim=50 15 10 30]{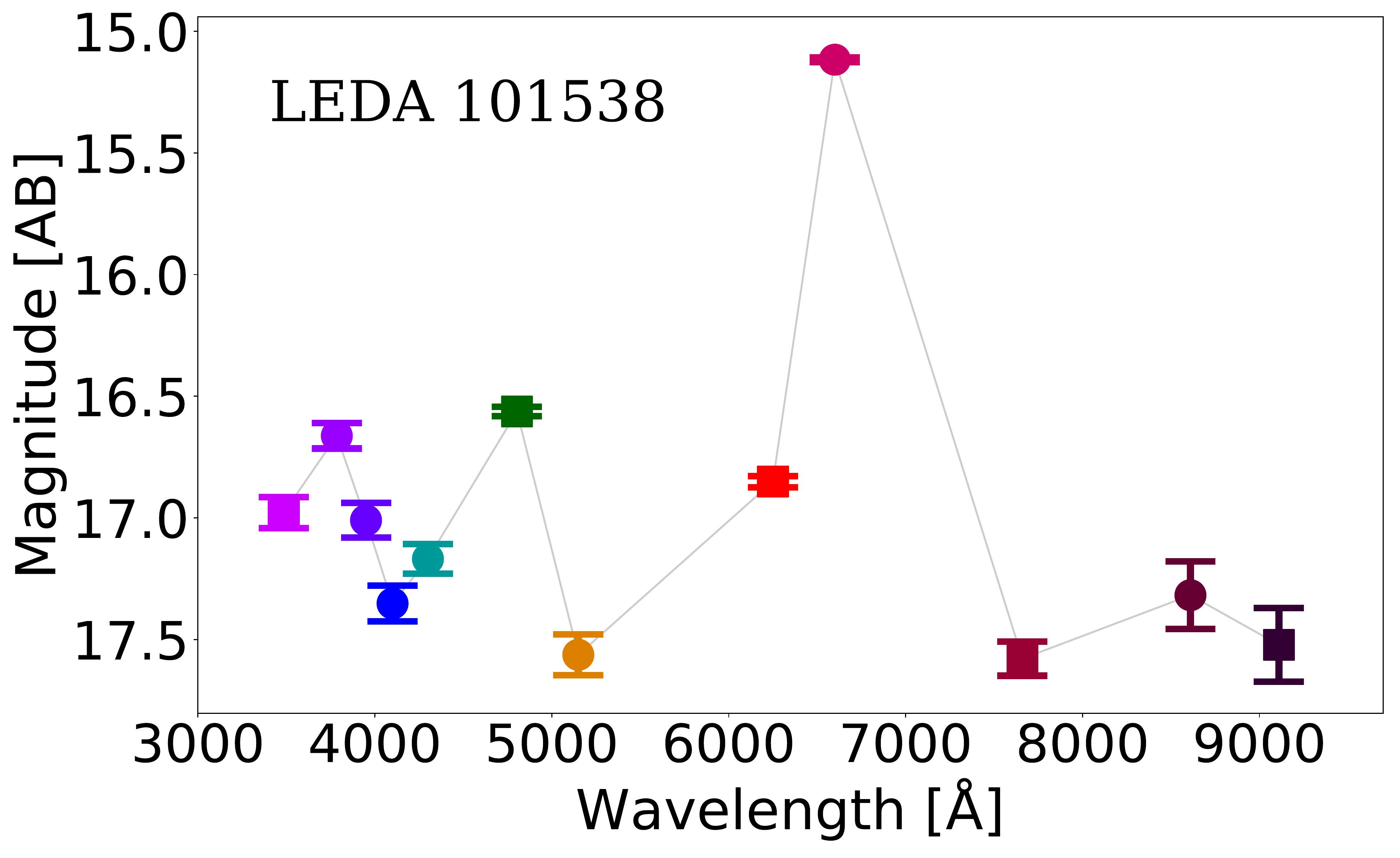}  \llap{\raisebox{2.5cm}{\includegraphics[width=0.29\linewidth, trim=10 10 -30 0]{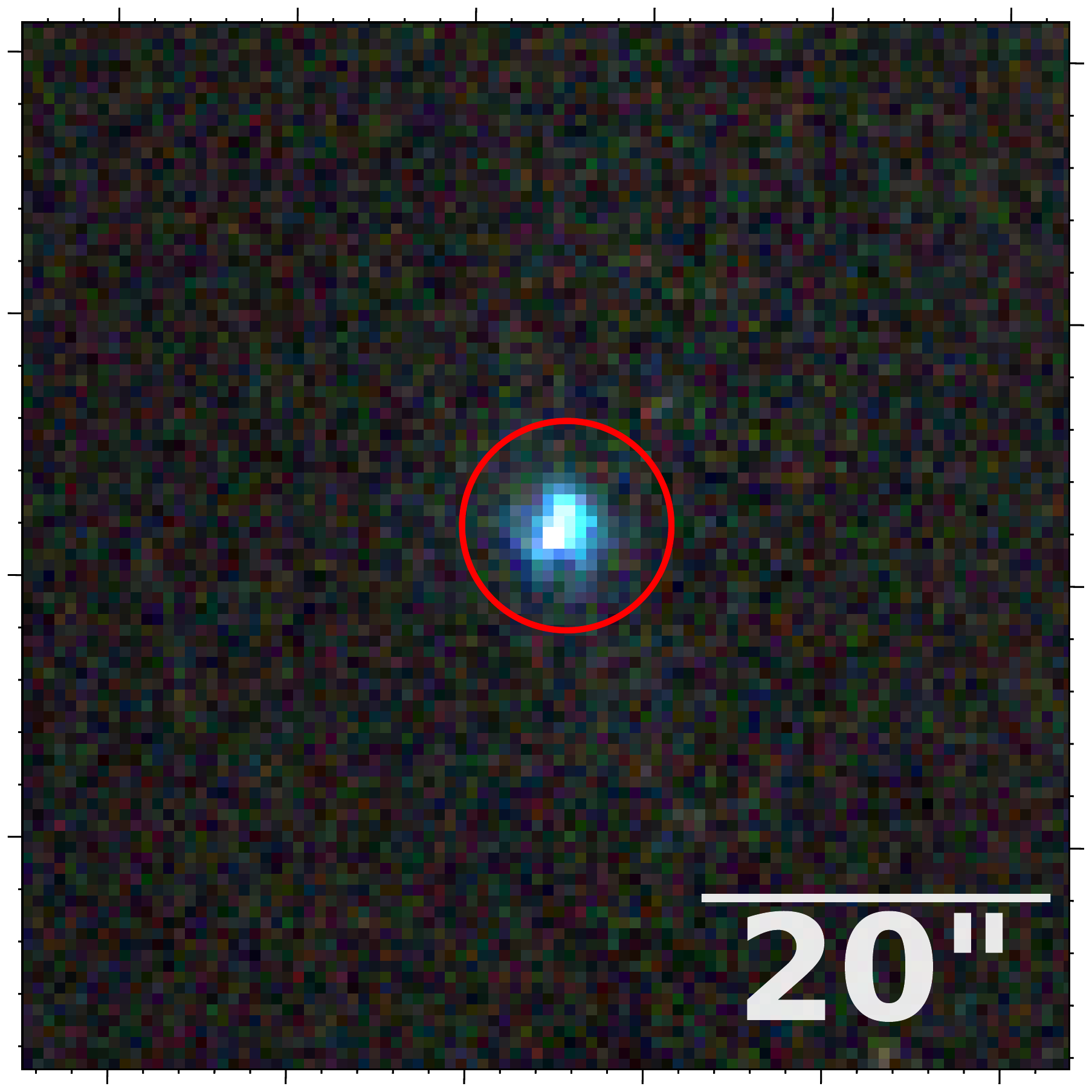}}}
  \end{tabular}
  \caption{J-PLUS photo-spectra and corresponding images of disc Galactic PN Sp 4-1 (\textit{upper panel}), H~{\sc ii} galaxy LEDA 2790884 (\textit{middle panel}) and H~{\sc ii} galaxy LEDA 101538 (\textit{lower panel}). These objects were recovered by applying the colour-colour diagrams. The images resulted from the combination of the J-PLUS broad-band filters: $g, r,~\mathrm{and}~i$.}
  \label{fig:photo-image}
\end{figure}

 \textit{PN Sp 4-1} (J2000 RA: 19 00 26.5, DEC: 38 21 07.99), also called PN G068.7+14.8, is a previously confirmed compact Galactic PN \citep{Acker:1992, Moreno:2016}. In the diagrams of Figures \ref{fig:Viironen-apply} and \ref{fig:new-colours-applied}, Sp~4-1 is represented by the green circle. The colours $(J0515 - J0660)$ and $(g - J0515)$ are estimated to be 3.4 and -1.6, respectively. The former colour clearly indicates a strong \ha{} emitter, while the latter 
implies a significant contribution from the \oiii{} and $H\beta$ lines to the $g$-band. The upper panel of Figure~\ref{fig:photo-image} displays its photo-spectrum and corresponding image. There is no doubt that these data correspond to those of a typical PN photo-spectrum, of a roundish and compact object with full width at half maximum (FWHM)$\sim$1.3~arcsec.

 \textit{LEDA 2790884} (J2000 RA: 8 25 55.50, DEC: 35 32 32.01) is a H~{\sc ii} galaxy (blue compact galaxy) at $z \sim 0.0024$ \citep{Moiseev:2010}. 
The loci where the system is represented in the diagrams of Figures~\ref{fig:Viironen-apply} and \ref{fig:new-colours-applied} (one of the red diamonds) corresponds to that of a low-excitation PN, in agreement with the expectations from the  H~{\sc ii} regions. The photo-spectrum and image of this source are shown in the middle panel of Figure~\ref{fig:photo-image}. 

 \textit{LEDA 101538} (J2000 RA: 16 16 23.51, DEC: 47 02 02.64), the red diamond in Figure~\ref{fig:new-colours-applied}, is a H~{\sc ii} galaxy (blue compact dwarf galaxy) at $z \sim 0.002$, with a very blue colour, which indicates that it is likely a starburst \citep{Ann:2015}. The photo-spectrum of this galaxy is presented in the lower panel of Figure \ref{fig:photo-image}. The spectra of the H~{\sc ii} galaxies are very similar to those of giant extragalactic H~{\sc ii} regions \citep{Sargent:1970}. For this reason, LEDA 101538 lies in the PN zone, even though it is not a PN.

\section{DR1 vs. the complete PNe catalogue - HASH}
\label{sec:hash}

\indent We cross-matched the J-PLUS and S-PLUS data with the HASH catalogue \citep{Parker:2016}. We remind the reader that this catalogue contains all known PNe, which are classified as $true$, $likely$ and $possible$ PNe. 
We found four matches with J-PLUS, and one with S-PLUS. 
The J-PLUS ones appear in the colour-colour diagrams as purple (1 true PN), brown (1 likely PN), and orange (2 possible PNe) circles, while the S-PLUS match is the dark magenta (1 possible PN) 
circle in Figures~\ref{fig:Viironen-apply} and \ref{fig:new-colours-applied}. It is simple to see from these diagnostic diagrams that, following our selection criteria, none of these sources are  classified as PNe, all being located outside the PN zone. Four of these HASH sources are located in the zone of very blue WDs. Why does this happen?  
Below, we describe each source, thus highlighting that their sizes preclude their automatic recovery by our criteria.

\subsection{J-PLUS}
\label{sec:obs-jplus}

\begin{figure}
\setlength\tabcolsep{\figstampcolsep}
\centering
\begin{tabular}{l}
\includegraphics[width=0.9\linewidth, trim=50 115 10 8]{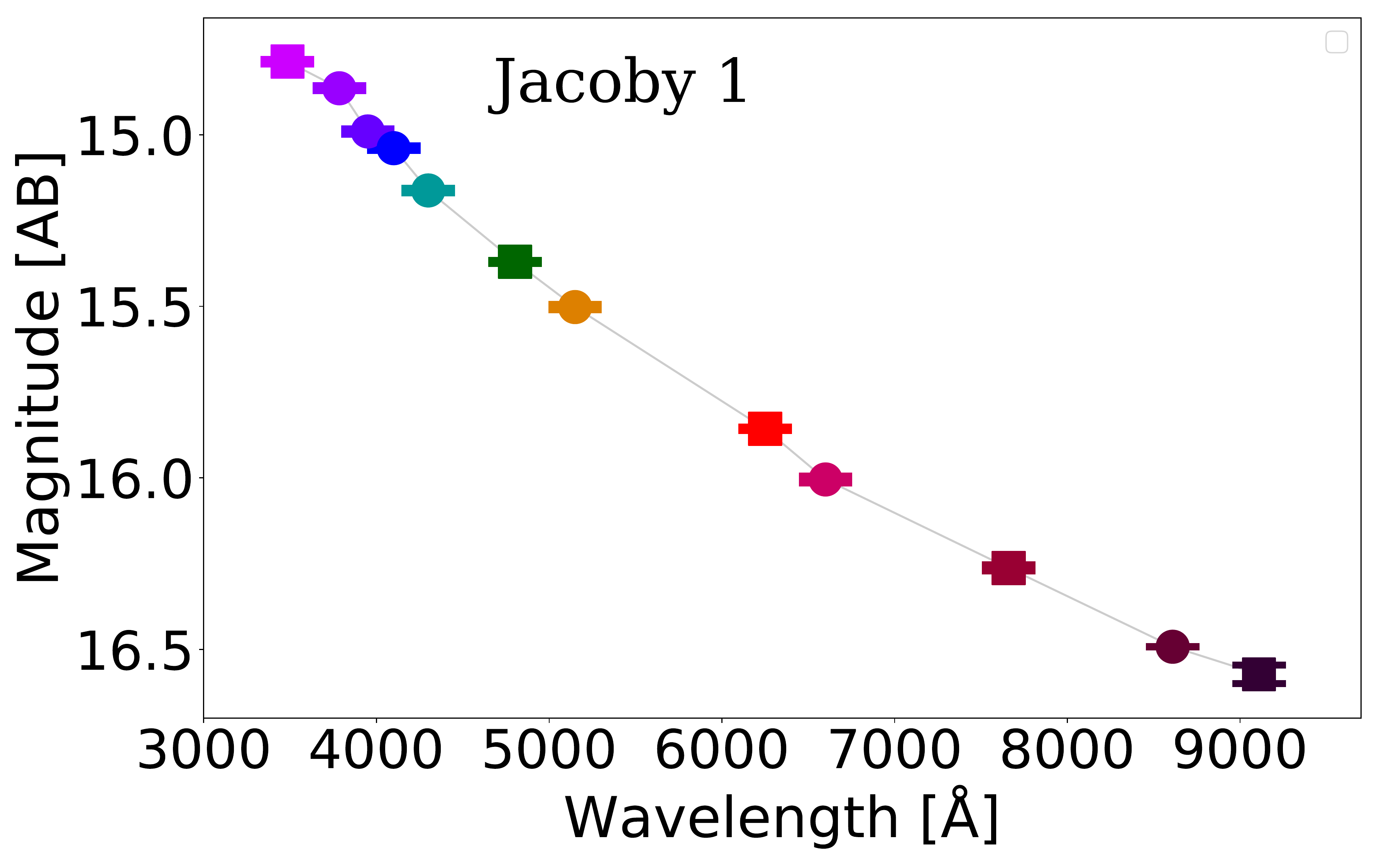}  \llap{\raisebox{1.65cm}{\includegraphics[width=0.31\linewidth, trim=10 10 -30 0]{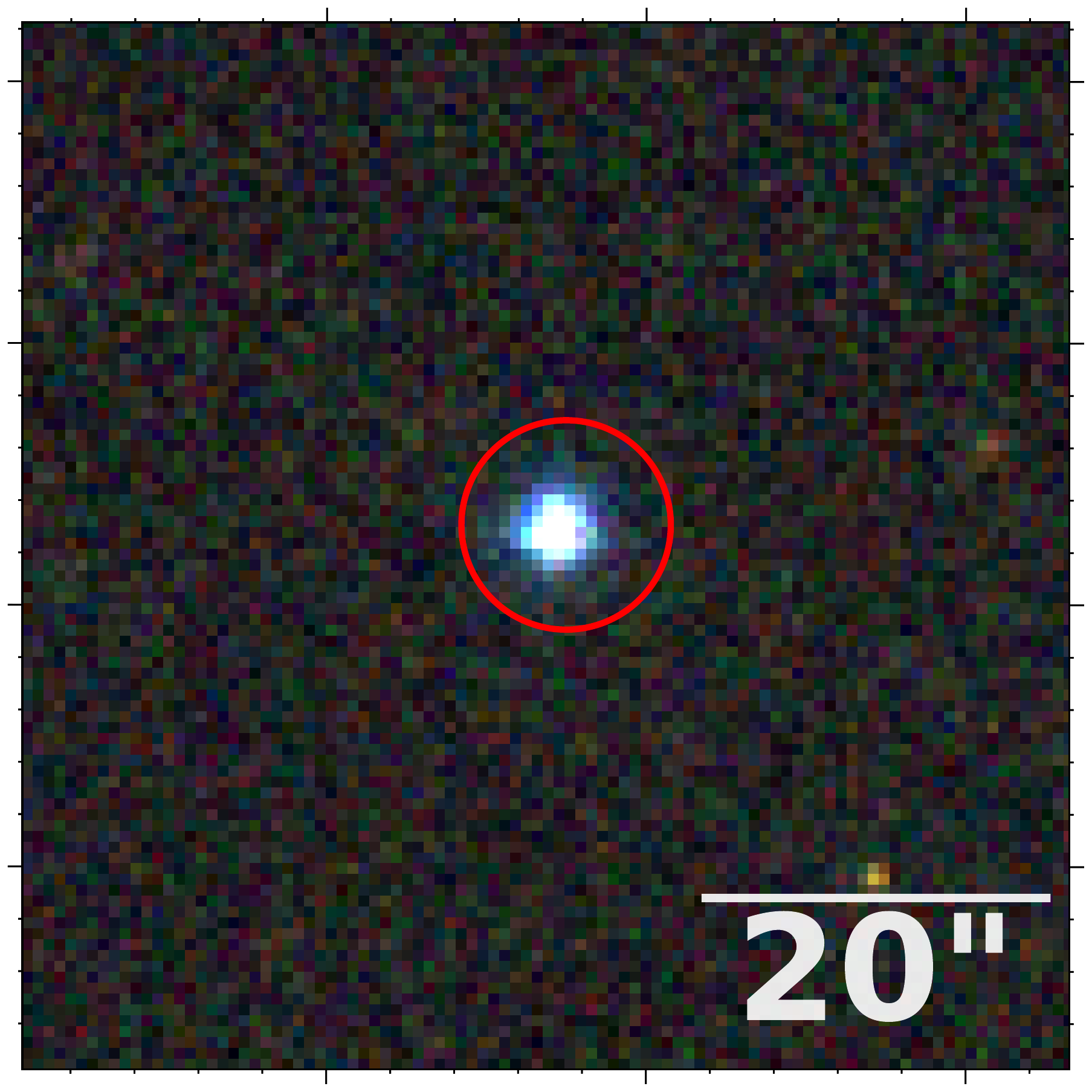}}} \lapbox[\width]{-21.em}{\raisebox{0.05cm}{\includegraphics[width=0.26\linewidth, trim=52 52 8 8, clip]{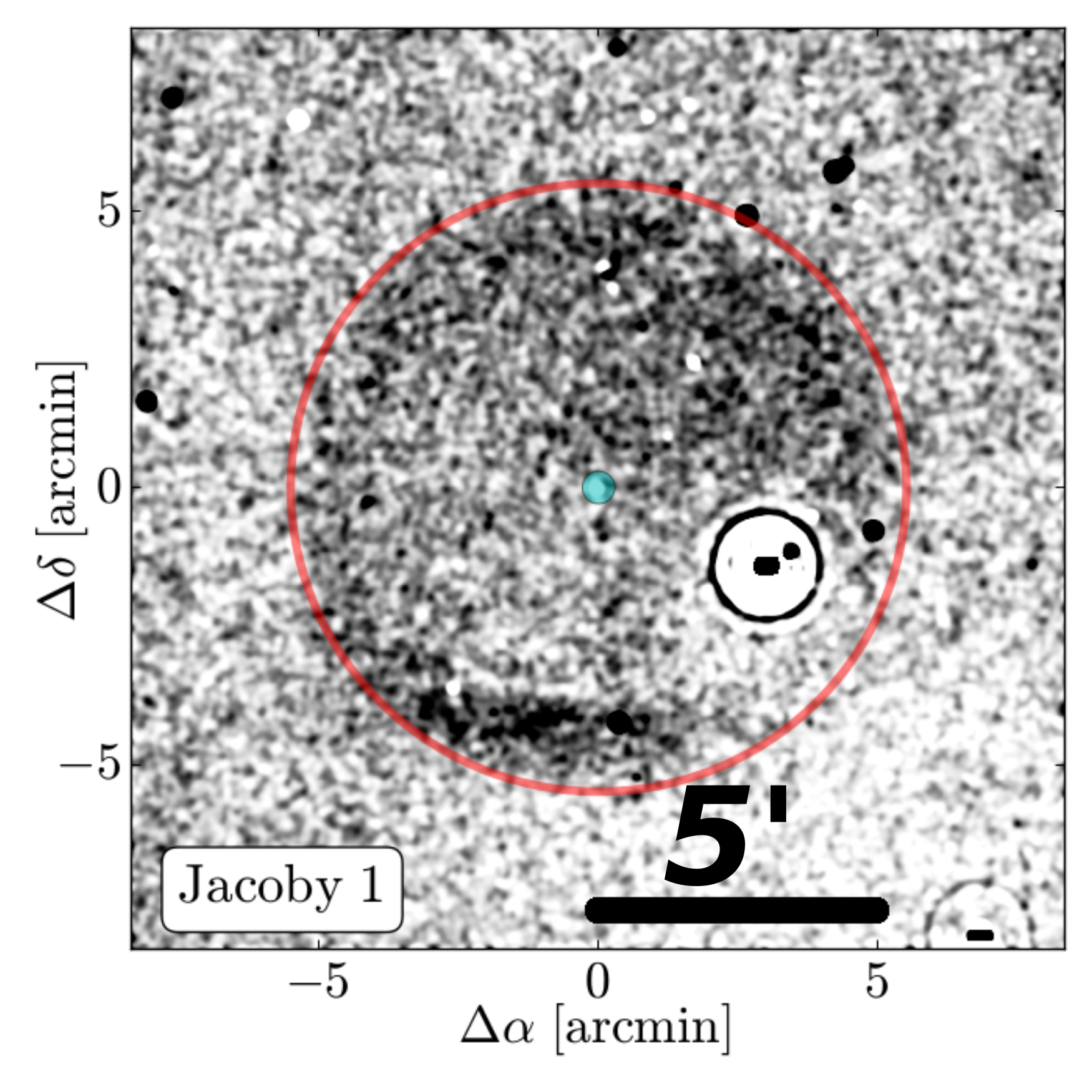}}} \\ \includegraphics[width=0.9\linewidth, trim=50 115 10 8]{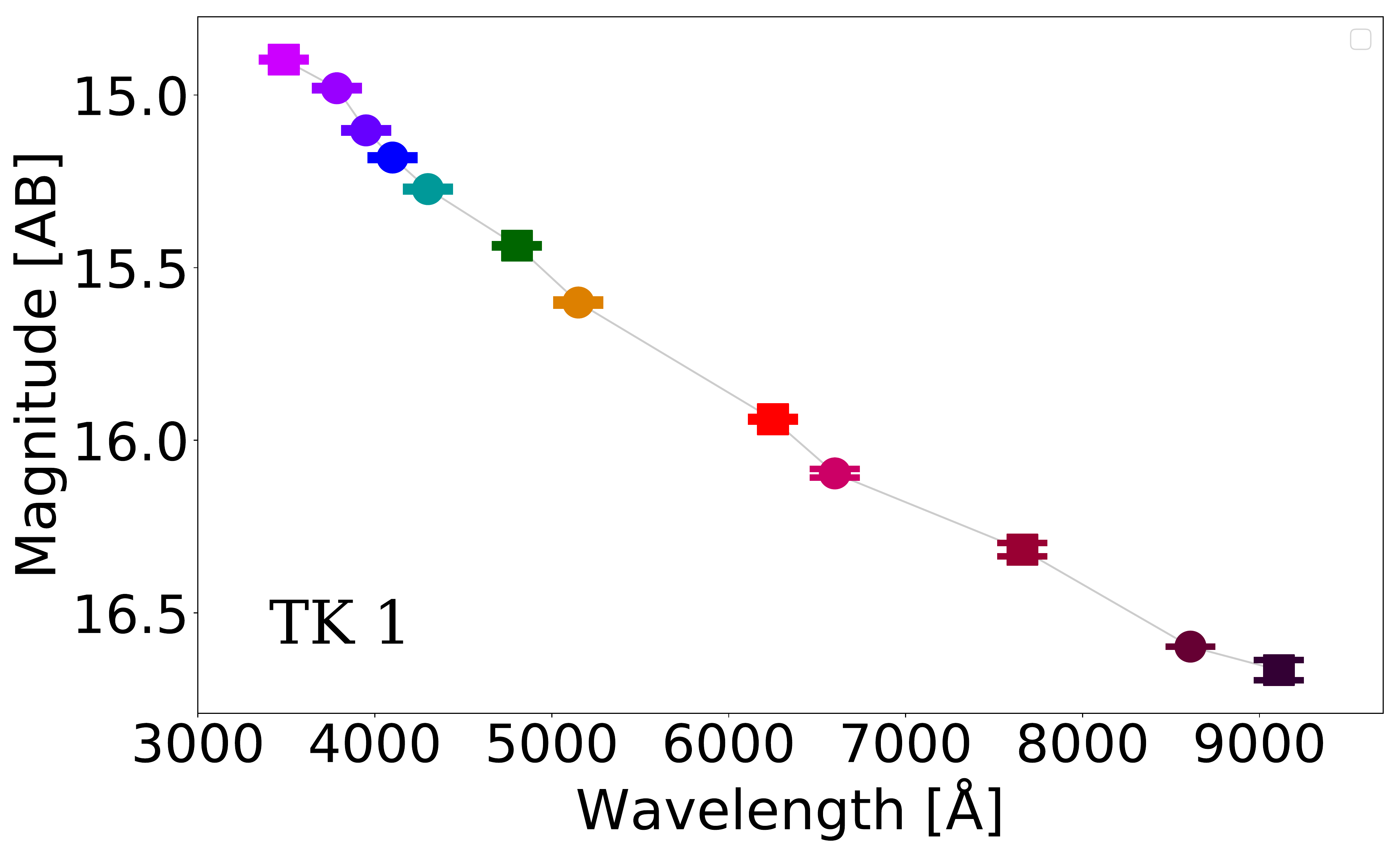} \llap{\raisebox{1.6cm}{\includegraphics[width=0.31\linewidth, trim=10 10 -30 0]{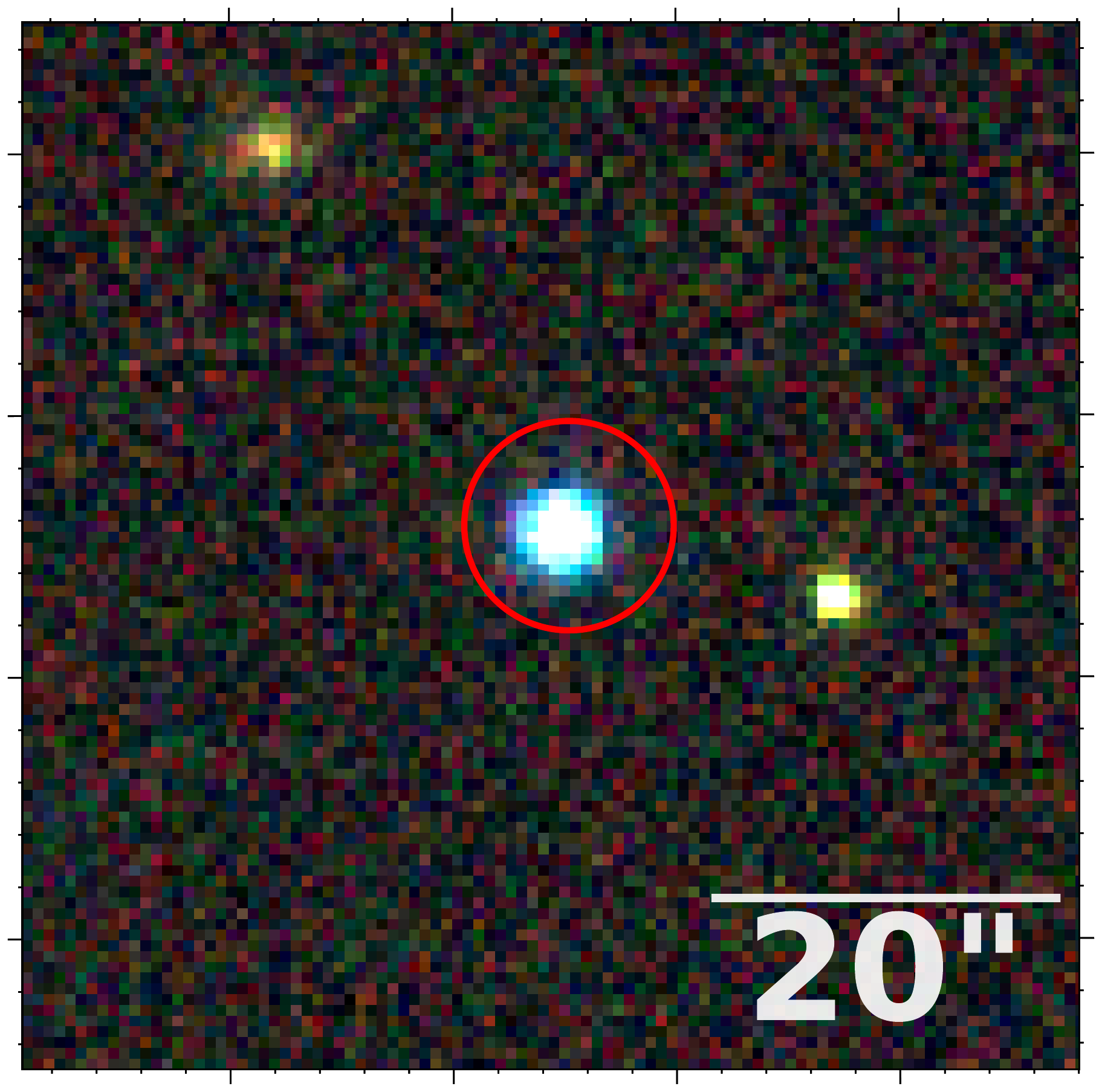}}} \\ \includegraphics[width=0.9\linewidth, trim=50 120 10 8]{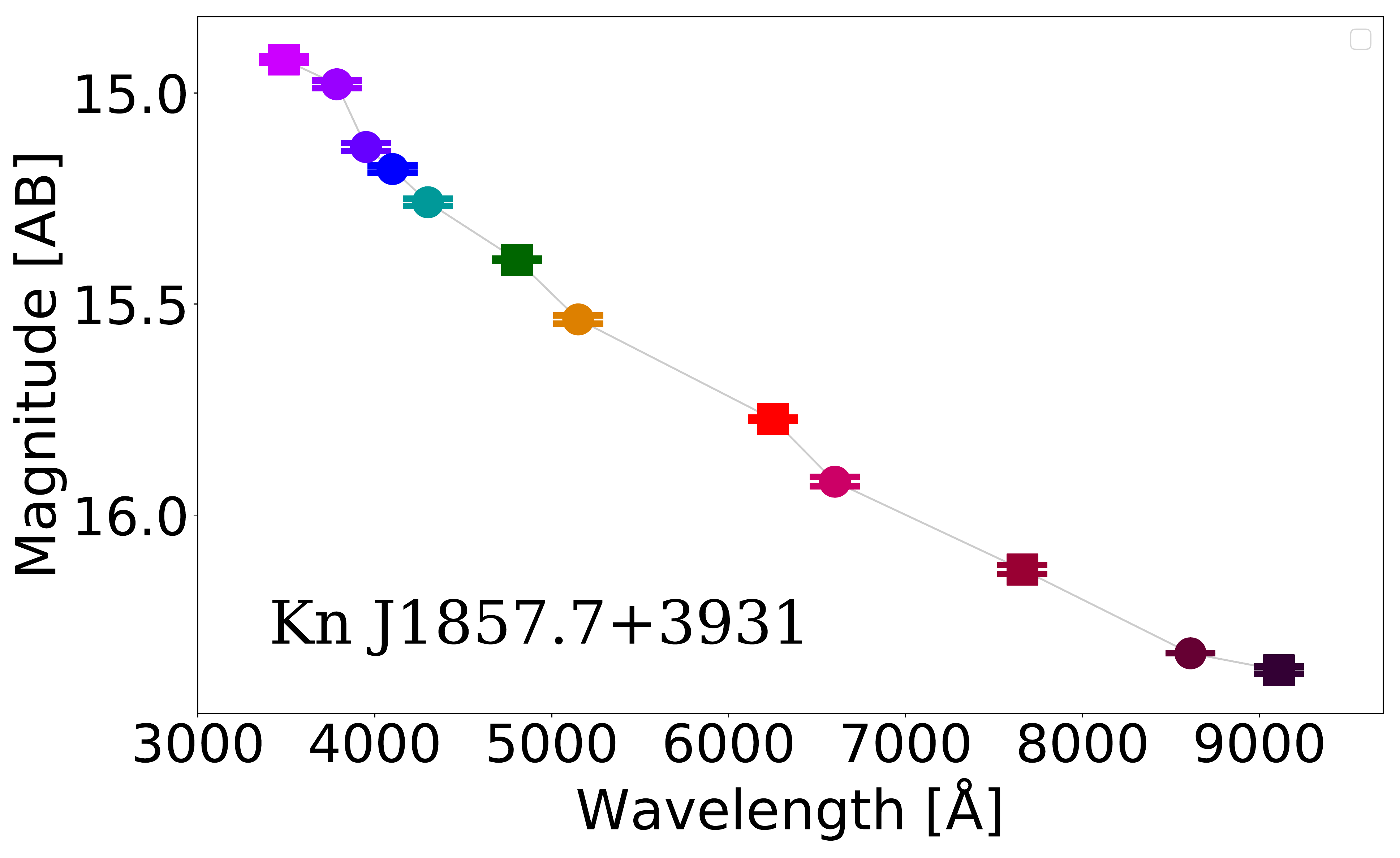} \llap{\raisebox{1.55cm}{\includegraphics[width=0.31\linewidth, trim=10 10 -30 0]{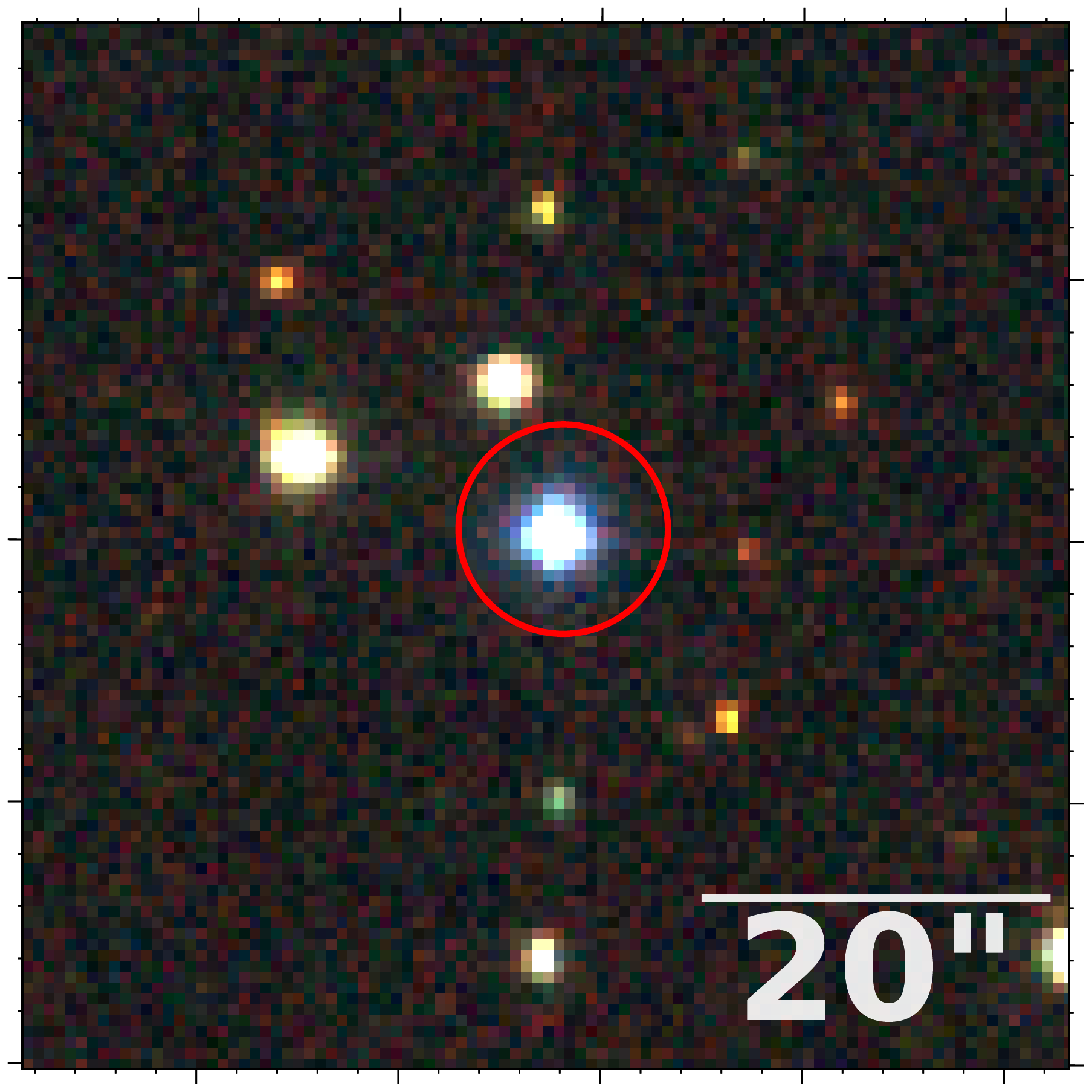}}} \\ \includegraphics[width=0.9\linewidth, trim=10 15 10 8]{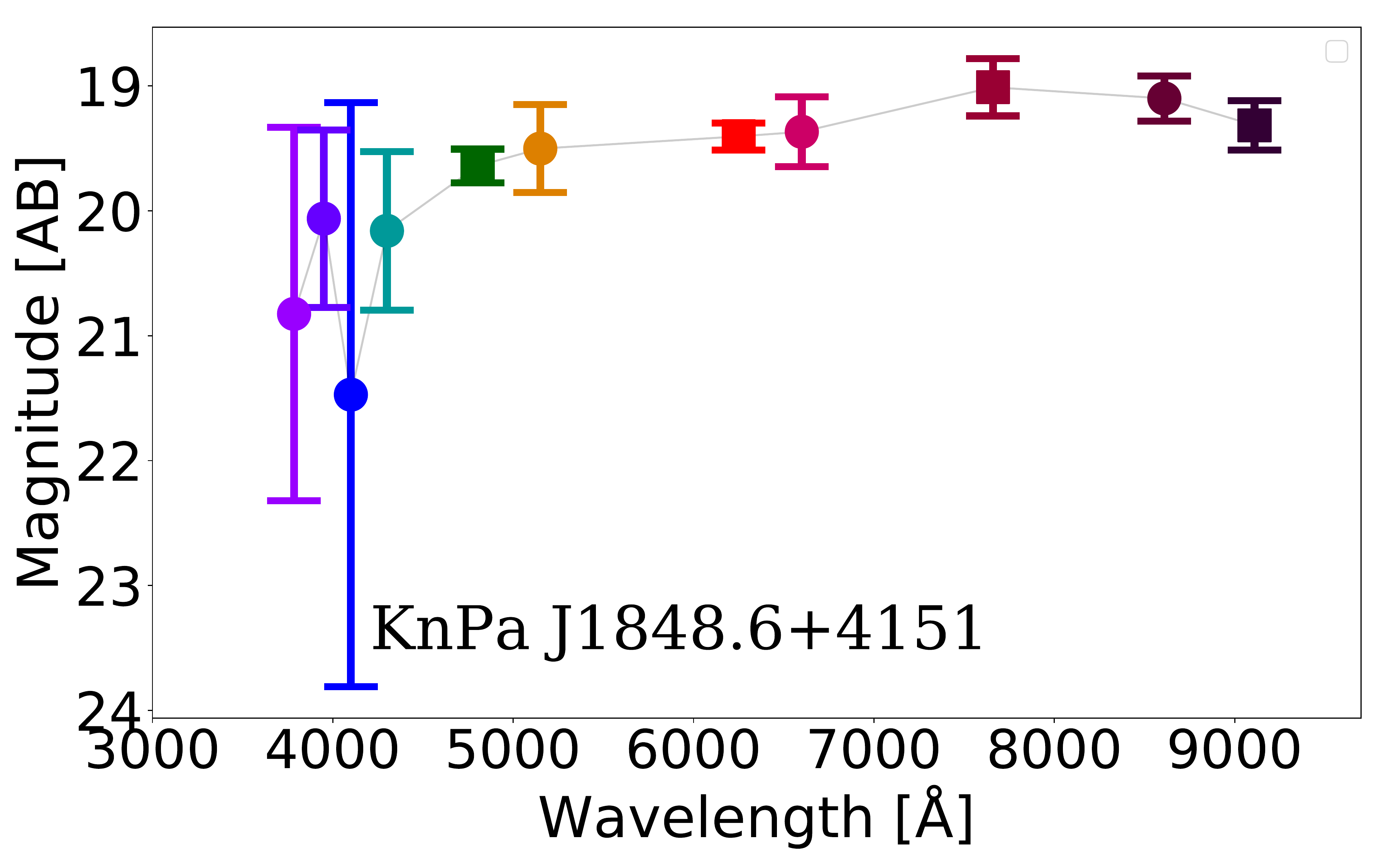}  \llap{\raisebox{1.6cm}{\includegraphics[width=0.31\linewidth, trim=10 10 -30 0]{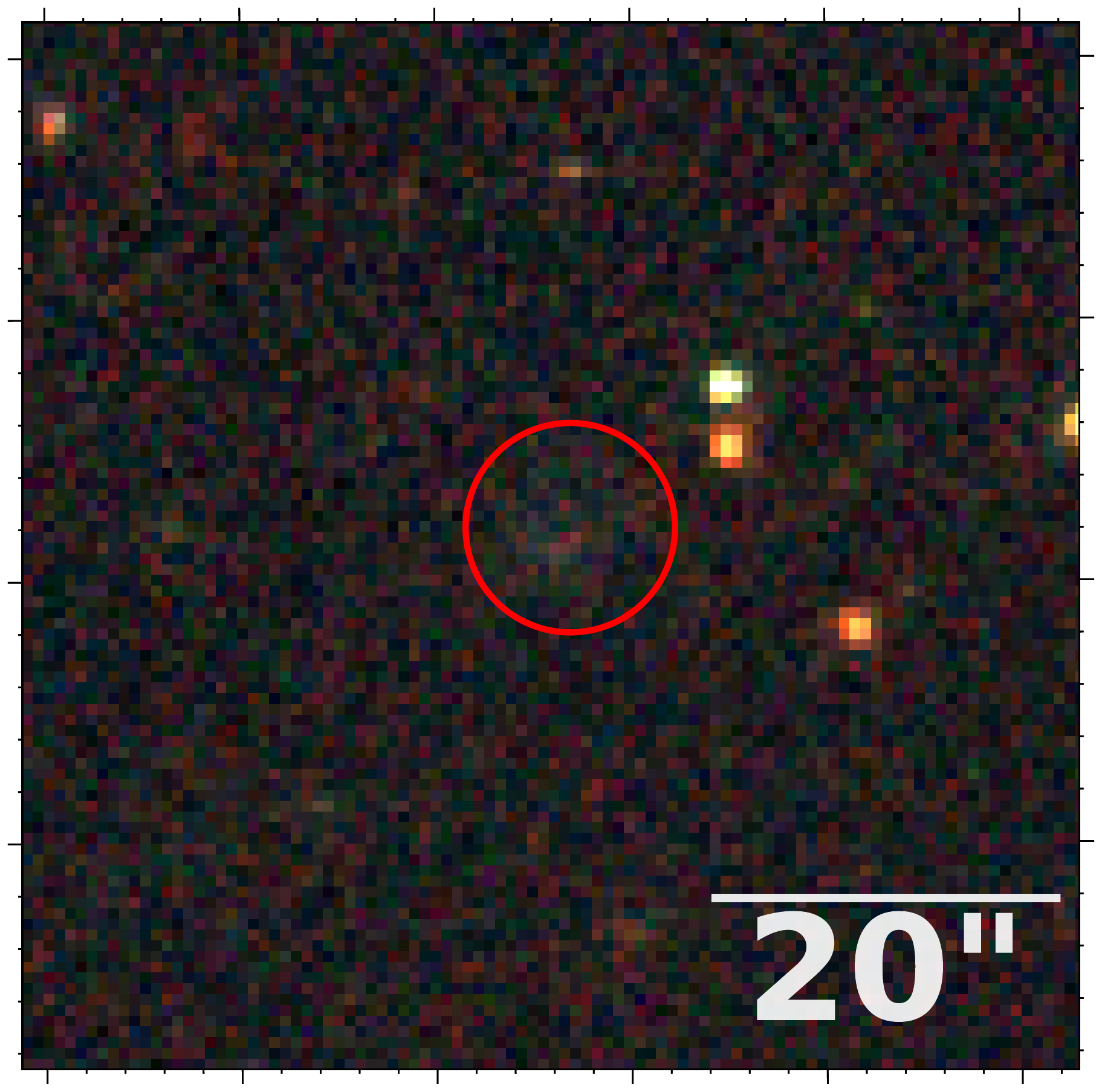}}}
  \end{tabular}
  \caption{(\textit{First panel}) J-PLUS photo-spectra, corresponding composite images (upper-right inset image), as in Figure \ref{fig:photo-image} and composite -- $J0660$, $r$, and $i$ -- image  (lower-left inset image) of  tHASH PN Jacoby~1. The extended PN is clearly visible, as indicated by the red circle. The \ha{} emission is detected  by applying a Gaussian smoothing filter of 10 pixels. Photo-spectra and corresponding images of lHASH PN TK 1 (\textit{second panel}), pHASH PNe Kn J1857.7+3931 (\textit{third panel}), and KnPa J1848.6+4151 (\textit{fourth panel}) of the HASH PN catalogue.}
  \label{fig:photo-image-PN}
\end{figure}

 \textit{Jacoby~1} (J2000 RA: 15 21 46.56, DEC: 52 22 04.05) is classified as a true (tHASH) PN, previously reported by \citet{Jacoby:1995}. The low $(r - J0660$) and  $(J0515 - J0660)$ colour indices are indicative of very weak, or totally absent, \ha{} line-emission. The first panel of Figure \ref{fig:photo-image-PN} displays the photo-spectrum, which turns out to be typical for white dwarfs (e.g. Figure 14 of \citealp{Cenarro:2019}) and its combined image (right inset image of the figure). Following \citet{Tweedy:1996}, Jacoby~1 is a highly evolved PN of $\sim$ 11~arcmin in size. Such large sizes can not be recovered/found by automatic photometric criteria, this is only possible by visual inspection of the fields (e.g. \citealp{Sabin:2014}). We managed to detect the \ha{} emission from the PN by applying a Gaussian smoothing filter of 10 pixels to the combined RGB -- $J0660, r,$ and $i$ -- as shown in the left inset in Figure~\ref{fig:photo-image-PN}.

 \textit{TK 1} (J2000 RA: 08 27 05.52, DEC: 31 30 08.10) is a likely (lHASH) PN, of up to 15~arcmin \citep{Tweedy:1996}. Its position in our colour-colour diagrams is very close to Jacoby~1 suggesting similar spectral characteristics (see photo-spectrum in the second panel of Figure~\ref{fig:photo-image-PN}). So, 
we suggest that the recovered emission is that of a WD star, in agreement with \citet{Rebassa:2015}. The above technique that recovered the nebular emission of Jacoby~1 was used, though it could not confirm the nebula that is possibly surrounding this white dwarf.

 \textit{Kn J1857.7+3931} (J2000 RA: 18 57 42.24, 39 31 00.13) is a possible (pHASH) PN. It shares the same position in the diagnostic diagrams, as well as an akin photo-spectrum, with Jacoby~1 and TK~1 (see Figure \ref{fig:photo-image-PN}). Nevertheless, when we looked for the extended nebula, this was not detected. We argue that \textit{Kn J1857.7+3931} is also a WD star, possibly associated with a nebular emission too faint  to be detected by J-PLUS.
 
 \textit{KnPa J1848.6+4151} (J2000 RA: 18 48 38.36, DEC: 41 51 02.52), another pHASH PN, is located far from the other HASH objects in our colour-colour diagrams (orange circle with large error bar), and again with no \ha{} excess. \textit{KnPa J1848.6+4151} lies in the regime of SDSS SFG and QSOs (fourth panel of Figure~\ref{fig:photo-image-PN}). It is a very faint object with $r \sim 19.5$ and 
its estimated angular size is of $\sim$~10.3$''$. \textit{KnPa J1848.6+4151} is very likely a galaxy (e.g. Figure~16 of \citealp{Cenarro:2019, Alam:2015, Greiss:2012}), and less likely to be a genuine PN.

\newcommand*{\shifttext}[2]{%
  \settowidth{\@tempdima}{#2}%
  \makebox[\@tempdima]{\hspace*{#1}#2}%
}

\subsection{S-PLUS}
\label{sec:obs-splus}

\begin{figure}
\centering
\begin{tabular}{l l}
  \includegraphics[width=0.9\linewidth, trim=10 20 0 8]{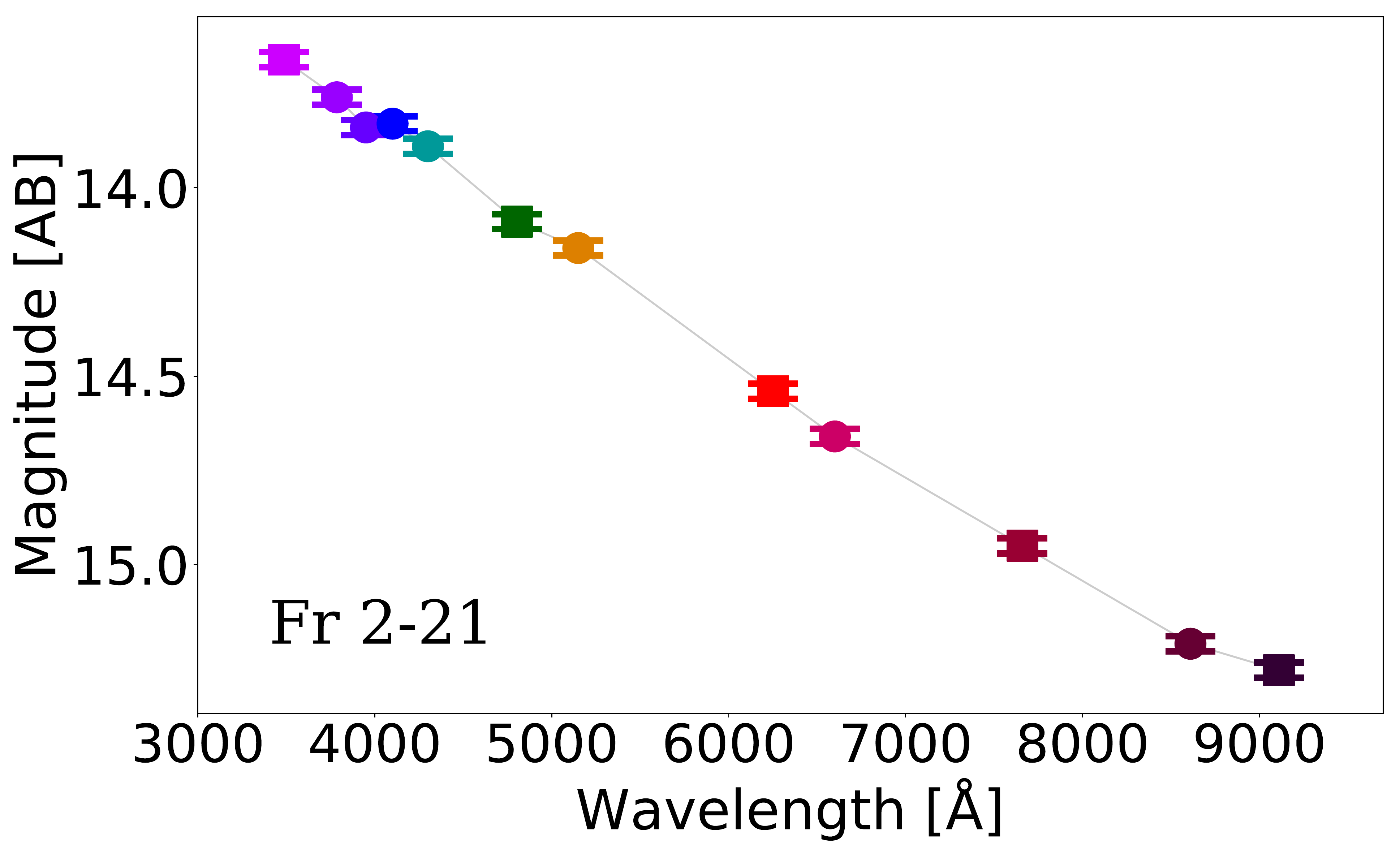}
  \llap{\raisebox{2.2cm}{\includegraphics[width=0.3\linewidth, trim=10 10 -40 0]{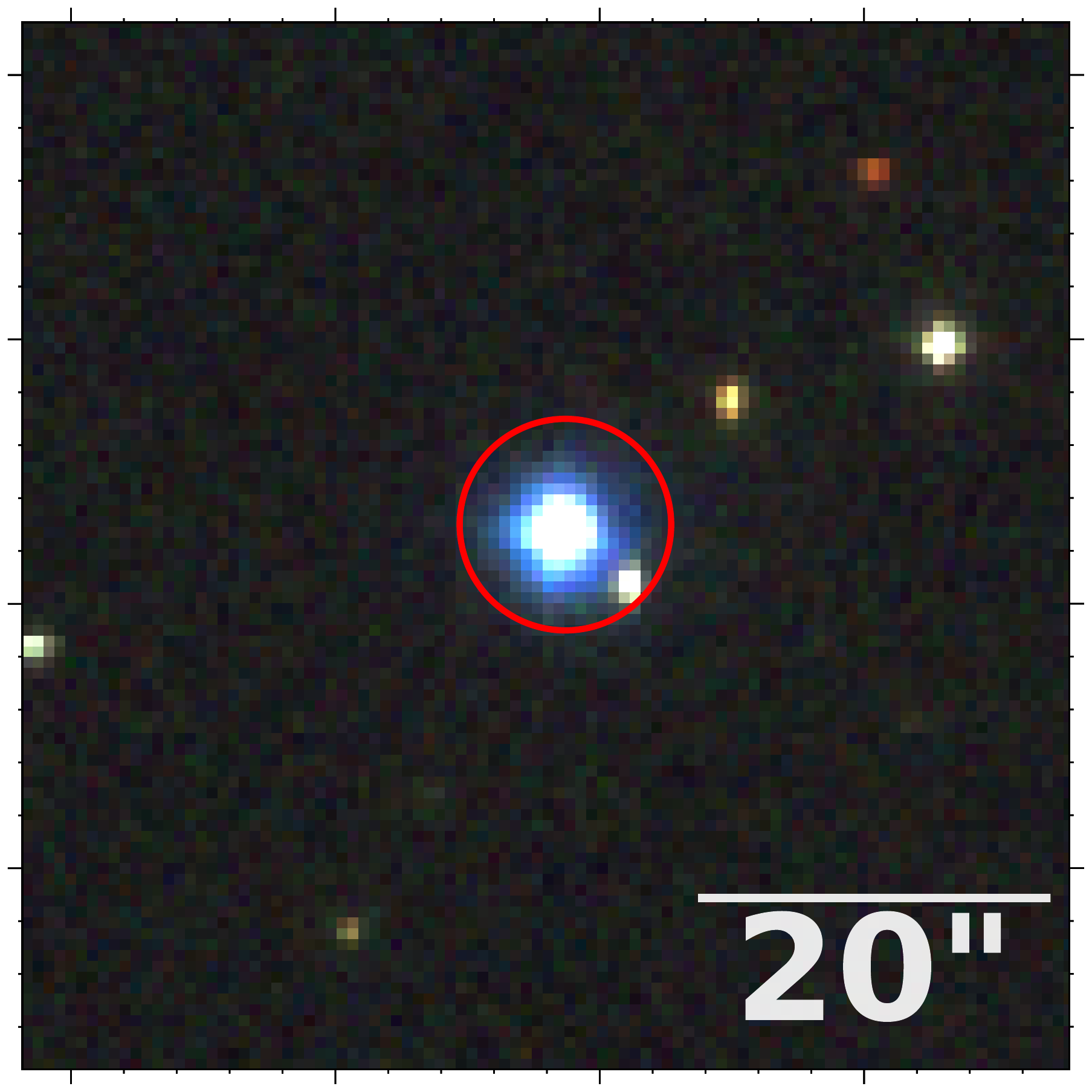}}}
  \end{tabular}  
  \caption{S-PLUS photo-spectrum of Fr 2-21 and corresponding image. This object is classified as possible PN in the HASH catalogue.}
\label{fig:fr221}
\end{figure}

 \textit{Fr 2-21} (J2000 RA: 21 26 21.17, DEC: 00 58 34.22) is also called PHL~4, and previously identified as a hot subdwarf \citep{Kilkenny:1984}, in addition to being a pHASH PN. It is located very close to the other HASH objects in the diagnostic diagrams. Its properties (Figure~\ref{fig:fr221}) suggest another WD. No extended nebula was found combining J0660, r and i images.

\section{Discussion and Conclusions}
\label{sec:Dis}

\indent The results described above highlight the potential of the data provided by J-PLUS and S-PLUS to explore the population of planetary nebulae in the Galactic halo. The developed photometric tools used to identify strong emission-line sources were applied to the very limited area surveyed up to now -- in both J-PLUS and S-PLUS -- returning promising results.  

Our selection criteria were very successful in avoiding all the strong emission-line emitters present in the fields, apart from the compact H~{\sc ii} regions (H~{\sc ii} galaxies), which means that J-PLUS and S-PLUS can really do a good job in terms of searching for PNe, at the same time minimising the list of spectroscopic follow-up candidates. As anticipated in Section~\ref{sec:syn}, the ability of photometrically distinguish PNe and H~{\sc ii} regions is strongly hampered by the similarities between low-excitation PNe and H~{\sc ii} regions. Therefore, our only way of avoiding lots of H~{\sc ii} regions among the candidates is by limiting the automatic search to compact PN. For this reason, we adopt apertures up to 6~arcsec. Unfortunately, the first PN candidate identified in these surveys turned out to be a H~{\sc ii} region located in the UGC\,5271 galaxy, which resembles a compact PN.

The larger planetary nebulae with lower density and corresponding lower surface brightness are more difficult to discover. These objects are essentially undetectable by survey methods. The cross-match between the HASH and J-PLUS/S-PLUS catalogues made this point very clear. The magnitudes of the resultant PNe matches correspond only to the photometry of the central WDs, as can be seen from their photo-spectra with no H$\alpha$ emission.

As in the case of IPHAS, a parallel study needs to be done in order to find extended PNe \citep{Sabin:2014}. It is necessary to bin the images in order to increase the surface brightness of the extended nebula and make their recovery feasible by visually inspecting one by one the images of the surveys. 


In conclusion, we considered the photometric systems of two twin imaging surveys (J-PLUS and S-PLUS), by using the synthetic photometry of different types of strong emission-line sources, 
to select the best combination of filters and construct diagnostic colour-colour diagrams to isolate hPNe. We reproduced the equivalent  IPHAS and proposed four new diagnostic colour-colour diagrams to separate hPNe from SySt, CVs, B[e] stars, QSOs, extragalactic H~{\sc ii} regions, star-forming galaxies, and young stellar objects. We find that the there is a high probability that a candidate found by our criteria will end up being a genuine PN if it is located well within the PN zones in the IPHAS equivalent and the other four colour-colour diagrams we proposed. It is also true that J-PLUS and S-PLUS are not completely successful at discriminating low-excitation PNe from compact H~{\sc ii} regions (H~{\sc ii} galaxies). We have validated our colour-colour diagrams through the photometry of the known hPNe H~4-1 and PNG~135.9+55.9 observed by J-PLUS during the SVD phase, and by  applying them to the J-PLUS DR1 and S-PLUS DR1, which make up an observed area of the sky of $\sim$1,190~deg$^2$.      

Having proven that our selection criteria are very effective in selecting PN candidates, even considering the contamination by H~{\sc ii} regions/galaxies, 
we intend to continuously apply these photometric tools to the forthcoming J-PLUS and S-PLUS data releases, to provide a more complete list of hPN candidates for spectroscopic follow-up.  

\begin{acknowledgements}
The authors acknowledge anonymous referee for very insightful
comments and for helping us to significantly improve our paper. We thank J. A. Caballero and R. Lopes de Oliveira for their useful comments and suggestions. LAGS acknowledges the support of CAPES -the Brazilian Federal Agency for Support and Evaluation of Graduate Education within the Ministry of Education of Brazil. DRG thanks the partial support of CNPq (grant 304184/2016-0) and S.A. acknowledges CAPES for a fellowship from the National Postdoctoral Program (PNPD). L.G. was funded by the European Union's Horizon 2020 research and innovation programme under the Marie Sk\l{}odowska-Curie grant agreement No. 839090. R.L.O. was partially supported by the Brazilian agency CNPq (PQ 302037/2015-2).
This work is based on observations made with the JAST/T80 telescope at the Observatorio Astrof\'isico de Javalambre (OAJ), in Teruel, owned, managed and operated by the Centro de Estudios de F\'isica del Cosmos de Arag\'on. Funding for the J-PLUS Project has been provided by the Governments of Spain and Aragon throug the Fondo de Inversiones de Teruel, the Spanish Ministry of Economy and Competitiveness (MINECO: under grants AYA2015-66211-C2-1-P, AYA2015-66211-C2-2, AYA2012-30789 and ICTS-2009-14), and European FEDER funding (FCDD10-4E-867, FCDD13-4E-2685). The Brazilian agencies FAPESP and the National Observatory of Brazil have also contributed to this project. We acknowledge the OAJ Data Processing and Archiving Unit (UPAD) for reducing and calibrating the OAJ data used in this work. This study used data collected at the T80-South, a new 0.826 meter telescope carried out by S-PLUS project. The T80-South robotic telescope \citep{Mendes:2019} was founded as a partnership between the S\~ao Paulo Research Foundation (FAPESP), the Observat\'orio Nacional (ON), the Federal University of Sergipe (UFS) and the Federal University of Santa Catarina (UFSC), with important financial and practical contributions from other collaborating institutes in Brazil, Chile (Universidad de La Serena) and Spain (CEFCA). We want to thanks to the S-PLUS team for the reducing and calibrating of the data. This research has made use of the HASH PN database at hashpn.space. Funding for the Sloan Digital Sky Survey IV has been provided by the Alfred P. Sloan Foundation, the U.S. Department of Energy Office of Science, and the Participating Institutions. SDSS-IV acknowledges
support and resources from the Center for High-Performance Computing at
the University of Utah. The SDSS web site is www.sdss.org.

SDSS-IV is managed by the Astrophysical Research Consortium for the 
Participating Institutions of the SDSS Collaboration including the 
Brazilian Participation Group, the Carnegie Institution for Science, 
Carnegie Mellon University, the Chilean Participation Group, the French Participation Group, Harvard-Smithsonian Center for Astrophysics, 
Instituto de Astrof\'isica de Canarias, The Johns Hopkins University, Kavli Institute for the Physics and Mathematics of the Universe (IPMU) / 
University of Tokyo, the Korean Participation Group, Lawrence Berkeley National Laboratory, 
Leibniz Institut f\"ur Astrophysik Potsdam (AIP),  
Max-Planck-Institut f\"ur Astronomie (MPIA Heidelberg), 
Max-Planck-Institut f\"ur Astrophysik (MPA Garching), 
Max-Planck-Institut f\"ur Extraterrestrische Physik (MPE), 
National Astronomical Observatories of China, New Mexico State University, 
New York University, University of Notre Dame, 
Observat\'ario Nacional / MCTI, The Ohio State University, 
Pennsylvania State University, Shanghai Astronomical Observatory, 
United Kingdom Participation Group,
Universidad Nacional Aut\'onoma de M\'exico, University of Arizona, 
University of Colorado Boulder, University of Oxford, University of Portsmouth, 
University of Utah, University of Virginia, University of Washington, University of Wisconsin, 
Vanderbilt University, and Yale University.

\end{acknowledgements}

\bibliography{ref-pne}

\end{document}